\documentclass[fleqn,usenatbib]{mnras}

\usepackage{newtxtext,newtxmath}

\usepackage[T1]{fontenc}
\DeclareUnicodeCharacter{0327}{\c{}}
\DeclareRobustCommand{\VAN}[3]{#2}
\let\VANthebibliography\thebibliography
\def\thebibliography{\DeclareRobustCommand{\VAN}[3]{##3}\VANthebibliography}

\usepackage{graphicx}	
\usepackage{amsmath}	
\usepackage{multicol}
\usepackage{multirow}

\newcommand{\photflam}{erg/cm$^2$/s/\AA}
\newcommand{\fluxcgs}{erg~s$^{-1}$~cm$^{-2}$}
\newcommand{\luxcgs}{erg~s$^{-1}$}
\newcommand{\sfr}{$M_\mathrm{\odot}$ yr$^{-1}$}
\newcommand{\chisq}{$\chi^2$}

\newcommand{\mout}{$\dot{m}_\text{ejec}$}
\newcommand{\afunnel}{$\theta_\mathrm{f}$}
\newcommand{\msun}{M$_\odot$}

\newcommand{\rin}{R$_\text{in}$}
\newcommand{\rsph}{R$_{\text{sph}}$}

\newcommand{\rout}{$R_{\text{out}}$}

\newcommand{\tbabs}{\texttt{tbabs}}
\newcommand{\chandra}{\textit{Chandra}}

\newcommand{\hst}{\textit{HST}}
\newcommand{\nustar}{\textit{NuSTAR}}
\newcommand{\xmm}{\textit{XMM-Newton}}
\newcommand{\swift}{\textit{Swift}}

\newcommand{\cloudy}{\texttt{Cloudy}}
\newcommand{\nh}{$N_{\text{H}}$}
\newcommand{\ebv}{$E(B-V)$}

\newcommand{\fout}{$f_\mathrm{out}$}
\newcommand{\met}{$12 + \log$(O/H)}
\newcommand{\ulx}{NGC 1313~X--1}
\newcommand{\theulx}{NGC 1313~X--1}

\newcommand{\ha}{H$\alpha$}
\newcommand{\hb}{H$\beta$}
\newcommand{\heii}{He{\sc ii}\,$\lambda$4686}
\newcommand{\oiii}{[O~{\sc iii}]\,$\lambda$5007}
\newcommand{\oi}{[O~{\sc i}]\,$\lambda$6300}

\title[Quasi-isotropic UV Emission in the ULX NGC~1313~X--1]{Quasi-isotropic UV Emission in the ULX NGC~1313~X--1}

\author[A.~G\'urpide \& N.~Castro~Segura]{
A.~G\'urpide$^1$\thanks{E-mail: a.gurpide-lasheras@soton.ac.uk} \&
N.~Castro~Segura$^1$
\\
\\
$^{1}$School of Physics \& Astronomy, University of Southampton, Southampton, Southampton SO17 1BJ, UK
}

\date{Accepted XXX. Received YYY; in original form ZZZ}

\pubyear{2015}

\begin{document}
\label{firstpage}
\pagerange{\pageref{firstpage}--\pageref{lastpage}}
\maketitle

\begin{abstract}
A major prediction of most super-Eddington accretion theories is the presence of anisotropic emission from supercritical disks, but the degree of anisotropy and its dependency with energy remain poorly constrained observationally. A key breakthrough allowing to test such predictions was the discovery of high-excitation photoionized nebulae around Ultraluminous X-ray sources (ULXs).
We present efforts to tackle the degree of anisotropy of the UV/EUV emission in super-Eddington accretion flows by studying the emission-line nebula around the archetypical ULX NGC~1313~X--1. We first take advantage of the extensive wealth of optical/near-UV and X-ray data from \hst, \xmm, \swift-XRT and \nustar\ observatories to perform multi-band, state-resolved spectroscopy of the source to constrain the spectral energy distribution (SED) along the line of sight. We then compare spatially-resolved \cloudy\ predictions using the observed line-of-sight SED with the nebular line ratios to assess whether the nebula `sees' the same SED as observed along the line of sight. We show that to reproduce the line ratios in the surrounding nebula, the photo-ionizing SED must be a factor $\approx 4$ dimmer in ultraviolet emission than along the line-of-sight. Such nearly-iosotropic UV emission may be attributed to the quasi-spherical emission from the wind photosphere. We also discuss the apparent dichotomy in the observational properties of emission-line nebulae around soft and hard ULXs, and suggest only differences in mass-transfer rates can account for the EUV/X-ray spectral differences, as opposed to inclination effects. Finally, our multi-band spectroscopy suggest the optical/near-UV emission is not dominated by the companion star.
\end{abstract}




\section{Introduction}

Ultraluminous X-ray sources are defined as extragalactic off-nuclear point-like sources with an X-ray luminosity exceeding the Eddington limit of a 10\,\msun\ black hole (BH) \citep[e.g. ][]{kaaret_ultraluminous_2017, king_ultraluminous_2023}. It is now established that the vast majority of these systems are powered by super-Eddington accretion onto a stellar-mass compact objects in binary configurations with a donor star. While it is speculated that the population of ULXs might be dominated by transient systems briefly reaching the ULX threshold \citep{brightman_new_2023}, most of the well-known systems shine persistently at such extreme luminosities, acting as laboratories for the study of \textit{sustained} super-Eddington accretion. However, despite decades of studies, how such extreme luminosities are produced remains a matter of debate. 

One major prediction of super-Eddington accretion theory is the presence of highly anisotropic emission \citep{shakura_black_1973,poutanen_supercritically_2007}. As the mass-transfer rates reaches or exceeds the Eddington limit, powerful radiation-driven optically-thick outflows are launched from the accretion disc, creating an evacuated cone or funnel around the rotational axis of the compact object. This causes observers at high-inclination to see the reprocessed emission of the outflow photosphere, whereas observer peering down the funnel will see the hot emission from the inner parts of the accretion flow \citep{poutanen_supercritically_2007, abolmasov_optically_2009}. Even if the quantitative details may differ, there is now a body of numerical simulations which indeed reproduce the accretion flow geometry and anisotropic emission pattern envisioned by \citet{shakura_black_1973} \citep[e.g.][]{kawashima_comptonized_2012, narayan_spectra_2017, mills_spectral_2023}, .

The discovery of neutron stars (NSs) in ULXs through X-ray pulsations \citep[][]{bachetti_ultraluminous_2014, israel_accreting_2017, castillo_discovery_2020} opened up new avenues in which super-Eddington accretion may proceed. For instance, it is known that magnetic fields reduce the cross-section for electron scattering, thereby increasing the allowed Eddington luminosities \citep{basko_limiting_1976, mushtukov_maximum_2015}. NSs are also expected to be more radiatively efficient compared to BHs, as the latter swallow the excess radiation which is emitted otherwise at the NS surface \citep{takahashi_supercritical_2018}. Additionally, at high-mass transfer rates, the NS may be engulfed in an optically-thick magnetosphere, whose spectrum is predicted to emit instead as a multi-color blackbody with a dipolar temperature dependency \citep{mushtukov_optically_2017}.

Constraining the degree of anisotropy in ULXs is thus not only key to understand the accretion flow geometry powering them and test existing theories, but also imperative to understand the effect of the ULX on its environment. For instance, the exact radiative output will inform about the role of ULXs/X-ray binaries on the epoch of re-ionization \citep{madau_radiation_2017} as well as help explain how or whether ULXs can shape some galaxy properties. In this regard, explaining the presence of bright nebular \heii\ emission line in the integrated spectra of metal-poor galaxies remains a long-standing issue, as regular stars do not produce enough photons above its ionisation potential (IP = 54 eV) to explain it \citep[][and references therein]{schaerer_x-ray_2019}. The discovery of HeIII regions around a few ULXs \citep{pakull_optical_2002, kaaret_high-resolution_2004, abolmasov_optical_2008} together with the fact that this line seems more prevalent in low-metallicity galaxies, where hard-ionising sources such as X-ray binaries and ULXs are more common \citep[e.g.][]{shirazi_strongly_2012,kovlakas_census_2020, lehmer_metallicity_2021}, has made ULXs receive attention as a potential explanation for this so-called `He{\sc ii} problem' \citep[e.g.][]{simmonds_can_2021, kovlakas_ionizing_2022}. Whether that is the case remains uncertain, mainly due the poor understanding of the ULX UV emission and anisotropy effects \citep{simmonds_can_2021, kovlakas_ionizing_2022}.

A key discovery allowing to constrain the degree of anisotropy was the observation of extended (25–80 pc) EUV/X-ray photoionized gas around a handful of ULXs \citep{pakull_optical_2002}. Such nebulae --when spatially-resolved -- effectively provide a 2D map of the ionising SED not directed onto the line of sight, allowing observers to compare the nebular emission lines with those expected from the line-of-sight SED. If the expected emission lines and the observed ones from the nebula are comparable, then the degree of anisotropy must be relatively small. 

Most works to date have focused on the HeIII region around the ULX Holmberg~II~X--1 \citep[e.g.][]{pakull_optical_2002, kaaret_high-resolution_2004, berghea_first_2010, berghea_spitzer_2012}. These works have found the nebula sees similar SED as that observed along the line of sight, arguing for isotropic emission. It must be noted however that most of these works were based on optical observations, with particular focus on the \heii\ line, which is most effective in probing the extreme-UV (EUV) emission. Instead, theoretical/numerical works suggest anisotropy must be strongest in the X-ray band \citep[e.g.][]{poutanen_supercritically_2007, narayan_spectra_2017}, although observations of the Galactic edge-on ULX-like system SS433 suggest collimation along the polar funnel may well take place in the EUV too \citep{waisberg_collimated_2019}. A crucial aspect is therefore the exact extrapolation of the X-ray spectrum to the unaccesible EUV. Evidence suggest that in general, direct extrapolation of X-ray derived models to the UV band is not accurate for ULXs \citep{dudik_spitzer_2016, abolmasov_optical_2008}. A way forward is therefore to combine broadband spectroscopy with nebular observations, of which only a few studies exist \citep{berghea_first_2010, berghea_spitzer_2012}.

In this work, we attempt to constrain the degree of anisotropy in the ULX NGC 1313 X–1 using the $\sim$200 pc photoionized nebula we discovered in our previous work \citep{gurpide_muse_2022}. Here we combine state-resolved multi-band spectroscopy – which allows to us to constrain the SED along the line of sight and reduce uncertainties related to the extrapolation to the UV – with spatially-resolved line maps from IFU spectroscopy \citep{gurpide_muse_2022} – which allow us to constrain the SED seen by the nebula along two different sight lines. 

We will show that the degree of collimation of the UV/EUV emission is small (a factor $\sim$4) in agreement with previous works \citep[e.g.][]{kaaret_high-resolution_2004}. However, we will show that unlike the ULXs Holmberg~II~X--1 or NGC~6946~X--1, \theulx\ does not produce a strong HeIII region. We will argue that the reason most likely lies in the differences in mass-accretion rate between these three sources.  

This paper is structured as follows: in Section~\ref{sec:data_reduction} we present our multi-band data reduction and its classification into the spectral states of NGC~1313~X--1 based on the long-term behaviour of the source. In Section~\ref{sec:multi_band_sed} we present spectral modelling of the multi-band spectral states of NGC~1313~X--1. Section~\ref{sec:cloudy} presents the \cloudy\ photo-ionization modelling of the nebular emission along two different sight lines and finally, Sections~\ref{sec:discussion} and \ref{sec:conclusion} present our Discussion and Conclusions. 

\section{Data Reduction}\label{sec:data_reduction}

In order to characterize the broadband SED of \theulx, we have investigated the available archival data to characterize the source temporal and spectral properties. In particular, we have considered the long-term monitoring provided by the \swift-XRT \citep{burrows_swift_2005}, optical photometry from the \textit{Hubble Space Telescope} and X-ray spectroscopy from \xmm\ \citep{jansen_xmm-newton_2001} and the \nustar\ \citep{harrison_nuclear_2013} observatories. 

\begin{figure*}
    \centering
    \includegraphics[width=0.98\textwidth]{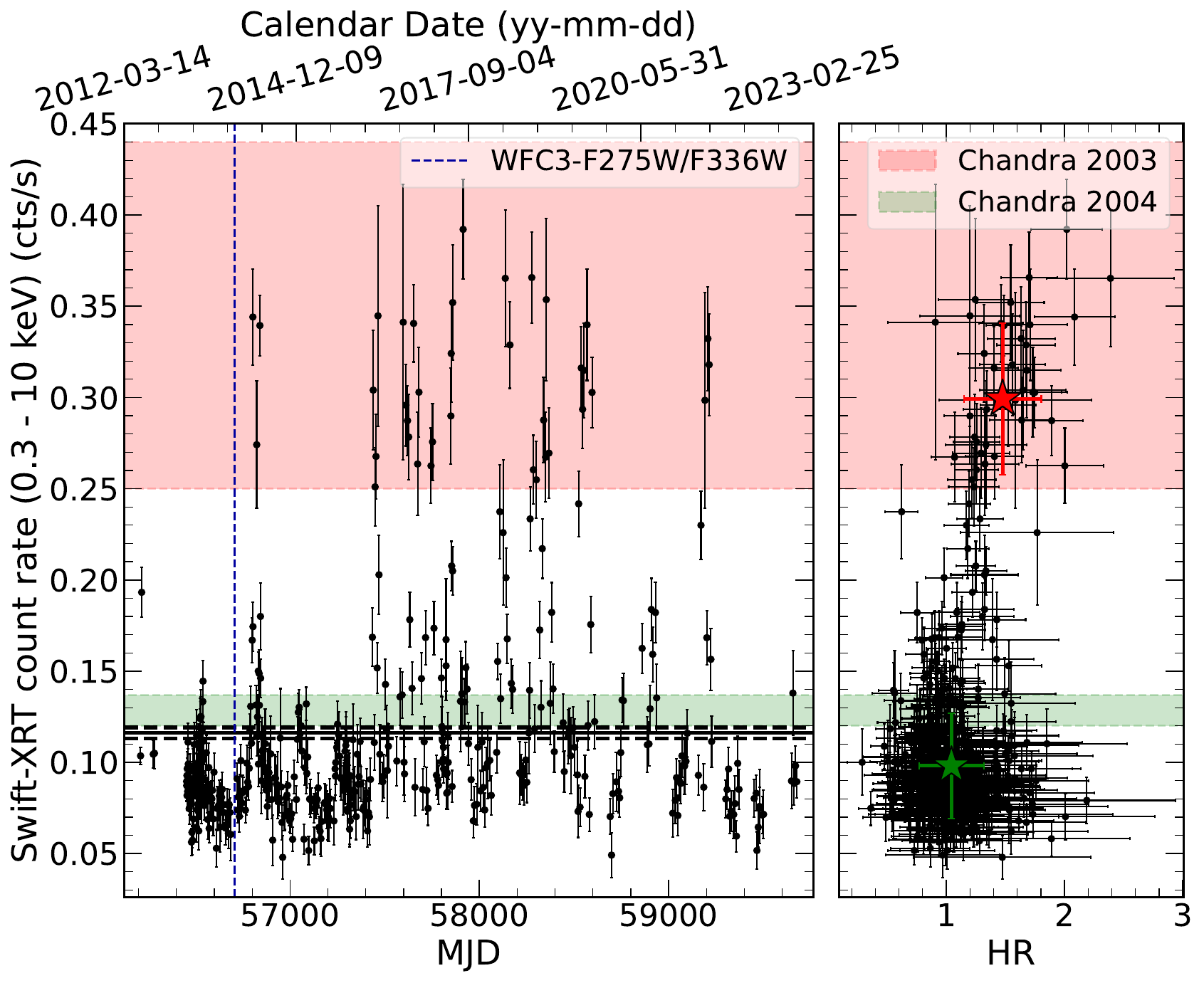}
    \caption{\swift-XRT observations showing the variability and spectral states of \theulx. (Left) \swift-XRT lightcurve. The dashed blue line shows the time when the \hst/WFC3/UVIS observations were performed (too close in time to be distinguished here). The red and green shaded areas show the equivalent \swift-XRT countrates of the 2003 and 2004 \chandra\ observations respectively. The uncertainties on the former includes any uncertainties associated with the pile up modelling (see text for details). The black solid and dashed line shows the mean \swift-XRT count-rate and its standard error. (Right) hardness ratio given as the count rate in the 1.5 -- 10 keV band over the 0.3 -- 1.5 keV band. The red and green stars mark the high and low state of NGC~1313~X--1 respectively, which were derived taking the mean and standard deviation of the snapshots above and below 0.225 ct/s respectively. }
    \label{fig:swift_hst}
\end{figure*}

The long-term lightcurve of NGC 1313 X–1 along with the hardness-intensity diagram (HID) is shown in Fig.~\ref{fig:swift_hst} and was extracted using the standard online tools \citep{evans_online_2007, evans_methods_2009}. From the Figure, it is clear how the source transits through two main states \citep[a behaviour also supported by \xmm\ observations;][]{gurpide_long-term_2021, pintore_x-ray_2012}. We refer to them as the `high' ($\gtrsim 0.25$ \swift-XRT ct/s) and `low' ($< 0.15$ ct/s) states, respectively, hereafter. Such bi-modality is commonly observed in other ULXs \citep{luangtip_x-ray_2016, amato_ultraluminous_2023} but it appears less strong in the case of \theulx\ \citep{weng_evidence_2018}. As we discuss below in Section~\ref{sec:cloudy}, our nebular modelling is insensitive to short-term changes in flux. Therefore we focused on building the broadband SED of the source for these two broadly defined spectral states. 
\subsection{X-rays}
X-ray spectral products from \xmm\ and \nustar\ were taken from \cite{gurpide_long-term_2021}. In particular, based on the recurrent behaviour of \theulx\ from the \swift-XRT lightcurve (Fig.~\ref{fig:swift_hst}) and the analysis from \cite{gurpide_long-term_2021}, we selected \xmm\ and \nustar\ observations taken in 2012 (corresponding to 10XN and 11XN in the notation used by \citet{gurpide_long-term_2021} or XN1 in \citet{walton_unusual_2020}), March of 2017 (14XN or XN3 respectively) and August/September 2017 (17XXN or XN5 respectively) to characterise the low state. For the characterisation of the high state, we extracted the first 10\,ks of \xmm\ obsid 0803990101 together with the first 20\,ks (owing to the lower number of counts) of \nustar\ obsid 30302016002 taken in June of 2017 (15XN or XN4). More details on the characterization of the high state are provided in the Appendix~\ref{sec:chandra}. 

We noticed recent important calibration updates for \nustar, so we reduced the data using \texttt{nuproducts} version 2.1.2 with the most recent calibration files as of February of 2023. Source and background regions were extracted following \citet{gurpide_long-term_2021}. All \xmm\ (EPIC-PN and the MOS cameras) and the \nustar\ spectra were rebinned using the scheme proposed by \citet{kaastra_optimal_2016} and fitted over the band where the source dominated above the background (this was the $\sim$ 0.3--10 keV band for the \xmm\ data and typically up to 20--25\,keV for the \nustar\ data). All spectra had sufficient number of counts per bin to use the $\chi^2$ statistic.

\subsection{UV, Optical and IR data}

In order to characterise the IR/optical/UV emission from \theulx, we used all archival \textit{Hubble Space Telescope} (\hst) images available from the MAST portal\footnote{\url{https://mast.stsci.edu/portal/Mashup/Clients/Mast/ Portal.html}}, which provides calibrated, geometrically-corrected drizzled images. We found a total of 14 observations taken over 5 different epochs. The details of these observations are given in Table~\ref{tab:hst_data}. The first 4 epochs were presented in \citet{yang_optical_2011} and covered the optical/IR range. The last epoch comprised 4 observations taken in 2014 with the WFC3/UVIS and covered instead the UV.

In order to extract fluxes in the different filters from the optical counterpart identified by \citet{yang_optical_2011}, we performed aperture photometry using a 0.2"-radius circular aperture in the ACS/WFC and WFC3/UVIS detectors (corresponding to 4 and 5 pixels respectively) and 0.15" (6 pixels) for the ACS/HRC. The aperture centroid was determined using 2D Gaussian fitting. We then corrected the counts for the finite aperture, following a two step process, as recommended\footnote{\url{https://www.stsci.edu/hst/instrumentation/wfc3/data-analysis/photometric-calibration/uvis-encircled-energy}}: first, because the aperture correction at small scales is known to vary with time and location on the detector, we estimated the aperture correction to 10 pixels using isolated bright stars in the field, typically selecting 1 to 4 stars (depending on the availability of isolated stars in the field) and averaging the results when possible. Next, we corrected the 10-pixel fluxes to infinity using the tabulated values for each combination of detector and filter\footnote{\url{https://www.stsci.edu/hst/instrumentation/acs/data-analysis/aperture-corrections}}. In instances where it was not possible to find any suitable star to estimate the aperture correction, we relied on the tabulated values for the full correction.

The background level was determined by taking the 3-$\sigma$-clipped median count rate in a concentric annulus around the source region. For the UVIS filters these regions contained several bright stars, so we instead used a nearby $\sim$0.7" circular region relatively free of stars but still containing some of the galaxy diffuse emission. The uncertainties on the final background-subtracted count rates where derived assuming Poisson statistics for the source and background regions and accounting for the uncertainty on the aperture correction.

In order to derive extinction-corrected fluxes, we first defined a likelihood function as the sum of the uncertainty-weighted residuals between the background-subtracted count rates and the predicted count rates by a model when convolved with the corresponding \hst\ filter. Along with the source model parameters (described below), we modelled extinction using two components: a Galactic component using \citet{cardelli_relationship_1989} reddening law, with $R_\mathrm{V} =3.1$  and an additional, extragalactic component with $R_\mathrm{V} =4.05$ using the extinction curve from \citet{calzetti_dust_2000}, which may be appropriate for a star-forming galaxy such as NGC~1313. The galactic component was fixed to the value along the line of sight \citep[\ebv$_\mathrm{G}$ = 0.11;][]{schlafly_measuring_2011} while the extragalactic component was included as a Gaussian ($\mu_\mathrm{E(B-V)}$ = 0.15, $\sigma_\mathrm{E(B-V)}$ = 0.03) prior on \ebv. The constraints on the extragalactic extinction come from our MUSE spectrum and are justified later in Section~\ref{sub:extinction}, where we show there is additional extinction ($A_\mathrm{v}$ = 0.59$\pm$0.11 mag) towards NGC~1313~X--1. By including \ebv\ in this manner we were able to potentially constrain it further based on the \hst\ filters or in the worst case scenario, propagate its uncertainties to the final deabsorbed-fluxes. However, in all instances we could not constrained further and obtained our prior back in the posteriors of \ebv.

For the source model, we assumed the same absorbed powerlaw ($F_\lambda = N\lambda^\alpha$ where $F_\lambda$ are the fluxes in \photflam\ and $\lambda$ are the pivot wavelengths of the filters) for all filters in a given epoch. Exceptions to this were epoch 2004-02-22 where the fluxes clearly deviated from a single powerlaw \citep[$\chi^2 > 20$ for 1 degree of freedom; see also figure~4 in][]{yang_optical_2011} and epochs for which only one measurement was available. In these cases we assumed a flat spectrum in \photflam, parametrised by its amplitude $N$. 

To estimate the best-fit model parameters and their uncertainties, we drew parameter samples (either $N, \alpha$, \ebv\ or $N$,\ebv) to sample the posteriors using 32 Markov-Chain Monte Carlo (MCMC) chains using the \texttt{emcee} package \citep{foreman-mackey_emcee_2013}. The chains were run until a) the number of steps reached 100 times the integrated autocorrelation time ($\tau$), which was estimated on the fly every 800 samples, and b) $\tau$ changed less than 1\% compared to the previous estimate. We then burned the first 30$\tau$ samples and thinned the chains by $\tau$/2. The final intrinsic fluxes and its uncertainties were estimated by drawing 2,000 realizations of $N, \alpha$, (or $N$ in cases of a flat spectrum) from the posteriors to estimate the distribution of best-fit intrinsic fluxes in each filter. The mean and the 1$\sigma$ confidence interval of the distribution were taken as our final estimates (these posteriors were symmetric and Gaussian-like in all instances). The final extinction-corrected ST magnitudes and fluxes are reported in Table~\ref{tab:hst_data}. 

\begin{table*}
   \begin{center}
        \caption{Derived \hst\ fluxes for the counterpart of NGC~1313~X--1. }
    \label{tab:hst_data}
     \resizebox{\textwidth}{!}{\begin{tabular}{@{\extracolsep{4pt}}ccccccccc} 
    \hline
    \hline
    \noalign{\smallskip}
   Obs id & Date & Instrument & Filter & Exp & Ac$^a$ & ST & F$_\lambda$ & State\\
 & & & & ks & & mag & 10$^{-18}$ \photflam\\
 & & & &  &  & \multicolumn{2}{c}{$A_\mathrm{V} = 0.93\pm0.11$$^b$ mag}  & \\
 \cline{7-8}
 \noalign{\smallskip}
 \hline
 \noalign{\smallskip}
9796\_a1\_acs\_hrc\_f330w & \multirow{4}{*}{2003-11-17} & ACS/HRC & F330W & 2.76 & 0.69$\pm$0.05 & 21.82$\pm$0.09 & 6.8$\pm$0.5 & \multirow{4}{*}{High}\\ 
9796\_01\_acs\_wfc\_f435w &  & \multirow{3}{*}{ACS/WFC}  & F435W & 2.52& 0.839 & 22.67$\pm$0.07 & 3.1$\pm$0.2 &  \\ 
9796\_01\_acs\_wfc\_f555w & &  & F555W & 1.16& 0.78$\pm$0.06 & 23.38$\pm$0.06 & 1.61$\pm$0.08 & \\ 
9796\_01\_acs\_wfc\_f814w & & & F814W & 1.16 & 0.830 & 24.74$\pm$0.05 & 0.46$\pm$0.02  & \\ 
 \noalign{\smallskip}
9796\_05\_acs\_wfc\_f555w & 2004-02-22 & ACS/WFC & F555W & 2.40 & 0.841 & 23.42$\pm$0.06 & 1.56$\pm$0.08 & Low\\ 
 \noalign{\smallskip}
9774\_05\_acs\_wfc\_f435w & \multirow{3}{*}{2004-07-17} & \multirow{3}{*}{ACS/WFC} & F435W & 0.68 &0.822$\pm$0.001 & 22.51$\pm$0.08 & 3.6$\pm$0.3  & --\\ 
9774\_05\_acs\_wfc\_f555w & & & F555W & 0.68 &0.79$\pm$0.03 & 23.55$\pm$0.08 & 1.38$\pm$0.10 &  --\\ 
9774\_05\_acs\_wfc\_f814w &  & & F814W & 0.68 & 0.78$\pm$0.01 & 24.61$\pm$0.06 & 0.52$\pm$0.03 & --\\ 
 \noalign{\smallskip}
10210\_06\_acs\_wfc\_f606w & \multirow{2}{*}{2004-10-30} & \multirow{2}{*}{ACS/WFC} & F606W & 1.06& 0.779$\pm$0.001 &  23.72$\pm$0.06 & 1.18$\pm$0.06 & --\\ 
10210\_06\_acs\_wfc\_f814w &  &  & F814W & 1.38&0.694$\pm$0.001 & 24.29$\pm$0.13 & 0.70$\pm$0.08 &  --\\ 
 \noalign{\smallskip}
13364\_98\_wfc3\_uvis\_f275w & \multirow{2}{*}{2014-02-16} & \multirow{4}{*}{WFC3/UVIS} & F275W & 2.53 & 0.78$\pm$0.05 &20.9$\pm$0.1 & 15.3$\pm$1.7 & \multirow{4}{*}{Low}\\ 
13364\_98\_wfc3\_uvis\_f336w &  &  & F336W & 2.41 & 0.78$\pm$0.06 &21.65$\pm$0.12 & 7.9$\pm$0.9 & \\ 
 \noalign{\smallskip}
13364\_85\_wfc3\_uvis\_f275w & \multirow{2}{*}{2014-02-19} &  & F275W & 2.53 & 0.8$\pm$0.02 & 20.91$\pm$0.10 & 15.7$\pm$1.5 &  \\ 
13364\_85\_wfc3\_uvis\_f336w &  & & F336W & 2.41&  0.76$\pm$0.02 & 21.65$\pm$0.09 & 8.0$\pm$0.7 & \\ 
\noalign{\smallskip}
\hline
\hline
\end{tabular} 
}
\end{center}
\begin{minipage}{\linewidth}
\textbf{Notes.} Magnitudes have been corrected for Galactic extinction along the line of sight and for additional extinction towards \theulx. Uncertainties represent the 1$\sigma$ confidence level and take into account the uncertainties on the aperture correction, the best-fit model parameters and the extinction used to derive the intrinsic fluxes. 
`$-$' Indicates there is no X-ray simultaneous information to categorise the source in one of its two states.\\
$^a$Aperture correction.\\
$^b$Total extinction used for the correction estimated from the Balmer decrement.
\end{minipage}
\end{table*}

While the exact values obviously differ from those reported by \citet{yang_optical_2011} due to the different treatment of the extinction, our analysis is in agreement with the variability reported in the F555W filter by \citet{yang_optical_2011}. The rest of the filters for which multi-epoch data exist are approximately consistent within uncertainties. We verified that correcting only for foreground extinction as in \citet{yang_optical_2011} yielded fluxes in good agreement with their results.

We also attempted to supplement the multi-band data with UV fluxes from the optical monitor (OM) onboard \xmm. We ran \texttt{omichain} on obsid 0803990101 and found \theulx\ is detected with a significance above the 8$\sigma$ level in the UV filters, while it is undetected in the optical filters. We used the default aperture of 6" in radius and converted the instrumental-corrected count-rates to fluxes using the average tabulated values\footnote{\url{https://www.cosmos.esa.int/web/xmm-newton/sas-watchout-uvflux}}. However, upon comparison of the OM fluxes with those from \hst\ in similar bands we found the OM fluxes ($\gtrsim$10$^{-16}$\, erg/cm$^2$/\AA/s) to be 2 order of magnitude overestimated, owing to the large aperture which likely contains significant stellar contribution. We concluded that the contribution of \theulx\ to the detection must be minimal and discarded the OM data from further analysis.
\subsection{Source spectral states}\label{sub:spectral_states}

In order to place the \hst\ observations in context with respect to the X-ray spectral states, we looked in the archives for X-ray data taken simultaneously with the \hst\ observations. We found three \chandra\ observations simultaneous with the first two sets of \hst\ observations (taken in November 2003 and February 2004), which were presented also in \citet{yang_optical_2011}. The details of the \chandra\ observations are given in Table~\ref{tab:chandra_data}. The 2014 WFC3/UVIS observations were instead covered by the \swift-XRT long-term monitoring (Fig.~\ref{fig:swift_hst}).

The \swift-XRT places the \hst\ 2014 observations when \theulx\ was in the low state. The \chandra\ data simultaneous with the other \hst\ observations requires more detailed modelling, particularly given the presence of pile-up in some observations. We therefore relegate the full analysis to the Appendix (Section~\ref{sec:chandra}) but briefly, we found that despite the presence of pile up, we can confidently place \theulx\ in the high state during the November 2003 observations, and in the low state during the February 2004 observations. The unabsorbed fluxes and luminosities for the best fits for both \chandra\ observations are reported in Table~\ref{tab:chandra_data} along with the determined spectral state. Fig.~\ref{fig:swift_hst} shows the two equivalent \swift-XRT count rates determined from the \chandra\ observations on the \swift-XRT lightcurve.     

\begin{table*}
    \centering
    \caption{Fits to the \chandra\ data used to determine the spectral state of NGC~1313~X-1 during the simultaneous \hst\ observations.}
    \label{tab:chandra_data}
    \begin{tabular}{ccccccccc} 
    \hline
    \hline
    \noalign{\smallskip}
   Obs id & Date & Detector & Exp & $P_\mathrm{f}$ & $\Gamma /kT^a$ & $F_\mathrm{X}^b$ & $L_\mathrm{X}^c$ & State\\
 & & & ks & \%  & \-/keV & 10$^{-12}$\,erg/s/cm$^2$ & 10$^{40}$\,erg/s & \\
 \noalign{\smallskip}
 \hline
 \noalign{\smallskip}
4747 & 2003-11-17 & ACIS-S & 5.28&$>$20& $0.55_{-0.07}^{+0.01}$ & 13.6$^{+1.8}_{-2}$ &  2.9$\pm$0.4 & High\\ 
4748$^d$& 2004-02-22 & ACIS-S & 5.05 & $>$20&  &  &  & Low\\ %
4750 & 2004-02-22 & ACIS-S  & 4.65 & $\sim$2.5--3 & $2.0\pm$0.1 &6.9$\pm$0.3 & 1.48$\pm0.06$ & Low\\ 
\noalign{\smallskip}
\hline
\hline
\multicolumn{9}{l}{In all fits \nh\ was frozen to the value derived by \citet{gurpide_long-term_2021}. Uncertainties are given at the 90\% confidence level for one parameter of interest.} \\
\multicolumn{9}{l}{$^a$Photon index of the \texttt{powerlaw} or temperature of the \texttt{diskbb} component.}\\
\multicolumn{9}{l}{$^b$Unabsorbed flux over the 0.3--10 keV band.}\\
\multicolumn{9}{l}{$^c$Unabsorbed luminosity over the same band.}\\
\multicolumn{9}{l}{$^d$This observation was not used as it was simultaneous with obsid 4750, which instead was free of pile up.}\\
    \end{tabular}
\end{table*}

As a summary, Table~\ref{tab:ulx_states} presents all the data gathered to characterized the broadband SED for the low and high states, respectively. We proceed to characterise the state-resolved broadband SED of the source in the following Section.

\begin{table*}
    \centering
        \caption{Datasets used to characterised the state-resolved broadband SED of \theulx.}
    \label{tab:ulx_states}
    \begin{tabular}{ccccc}
    \hline 
    \hline 
    \noalign{\smallskip}
   State &  Telescope & Obs. ID &  Detector/Filter & Net Exposure   \\
         &            &         &          &           ks         \\
     \noalign{\smallskip}
     \hline 
  \multirow{9}{*}{Low} &  \xmm\  & 0693850501 & pn/MOS1/MOS2 & 90.9/112.0/115.1 \\
                    &  \xmm\  & 0693851201 & pn/MOS1/MOS2 & 85.4/120.8/122.5 \\
                      & \nustar\ & 30002035002 &  FPMA/FPMB                       & 100.9/100.8      \\
                      & \nustar\ & 30002035004 &  FPMA/FPMB                       & 127.0/127.0      \\
                      & \hst\    & 9796\_05\_acs\_wfc\_f555w & ACS/WFC/F555W      & 2.40 \\
                      & \hst\    & 13364\_98\_wfc3\_uvis\_f275w & WFC3/UVIS/F275W & 2.53 \\
                      & \hst\    & 13364\_98\_wfc3\_uvis\_f336w & WFC3/UVIS/F336W & 2.41 \\
                      & \hst\    & 13364\_85\_wfc3\_uvis\_f275w & WFC3/UVIS/F275W & 2.53 \\
                      & \hst\    & 13364\_85\_wfc3\_uvis\_f336w & WFC3/UVIS/F336W & 2.41 \\
  \noalign{\smallskip}
  \hline
  \noalign{\smallskip}
        \multirow{7}{*}{High}  &  \xmm\  & 0803990101 & pn/MOS1/MOS2 & 10/10/10 \\
                      & \nustar\ & 30302016002 &  FPMA/FPMB & 20/20     \\
                      & \hst\    & 9796\_a1\_acs\_hrc\_f330w & ACS/HRC/F330 & 2.76 \\
                      & \hst\    & 9796\_01\_acs\_wfc\_f435w & ACS/WFC/F435W & 2.52 \\
                      & \hst\    & 9796\_01\_acs\_wfc\_f555w & ACS/WFC/F555W & 1.16 \\
                      & \hst\    & 9796\_01\_acs\_wfc\_f814w & ACS/WFC/F814W & 1.16 \\
   \noalign{\smallskip}
        \hline 
    \hline 
     \end{tabular}

\end{table*}

\section{Multi-band SED modelling: What we see} \label{sec:multi_band_sed}
Our aim here is mostly focused on constraining the EUV emission along the line of sight by testing different spectral models. We will then test these models using \cloudy\ photo-ionization modelling of the emission-line nebula. Additionally, we wish to determine whether the optical emission is dominated by the companion star or reprocessing in the outer disc, which remains unclear \citep[e.g.][]{grise_optical_2012} and can only be tackled with strictly simultaneous broadband data, of which only a few studies exist \citep[e.g.][]{soria_birth_2012, sathyaprakash_multi-wavelength_2022}.

\subsection{Extinction} \label{sub:extinction}
Modelling of the optical data requires knowledge of the level of extinction towards \theulx. We used the MUSE data presented in \citet{gurpide_muse_2022} to estimate the level of extinction from the ratio of the Balmer lines (\hb\ and \ha). To this end, we extracted an average spectrum from cube 1 in that work from a circular region of 1" (corresponding roughly to the PSF FWHM) around the optical counterpart. We then corrected the spectrum from Galactic extinction using the \citet{cardelli_relationship_1989} extinction curve with $R_\mathrm{V} = 3.1$. From this foreground-extinction-corrected average spectrum we measured $F$(H$\beta)$ = 4.2$\pm$0.1  and $F$(H$\alpha$) = 14.2$\pm$0.1, both in units of 10$^{-18}$ erg/s/cm$^{2}$, using a simple constant model for the local continuum around each line and a Gaussian for the line itself. The Balmer decrement $F$(\ha)/$F$(\hb) = 3.39$\pm$0.11 suggests there is additional extinction towards NGC~1313~X--1. The extinction correction requires knowledge of the intrinsic Balmer decrement, which depends on the electron density ($n_\text{e}$) and temperature of the gas ($T$) \citep{osterbrock_astrophysics_2006}. While we do not have access to temperature-sensitive line diagnostics, the typical values we found in these regions for the electron-density sensitive line ratio were [S~{\sc ii}]$\lambda$6716 / [S~{\sc ii}]$\lambda$6731 $> 1.3$. These values correspond to the low-density regime which is nearly temperature-insensitive, indicating $n_\text{e} < 100$ cm$^{-2}$. Assuming case B recombination and $T = 10000$ K (broadly consistent with our nebular modelling; Section~\ref{sec:cloudy}) the intrinsic ratio is \ha/\hb = 2.863 \citep[noting the likely absence of shocks in this region also constrains $T< 20,000$\,K;][]{osterbrock_astrophysics_2006}. Therefore we found $E(B-V) = 0.15\pm0.03$ adopting the \citet{calzetti_dust_2000} extinction curve with $R_\mathrm{V}$ = 4.05 as stated in Section~\ref{sec:data_reduction}. The total absorption is thus $A_\mathrm{v} = 0.93\pm0.11$ mag.

Another argument supporting the additional level of extinction can be made considering the relationship between the neutral hydrogen absorption column (\nh) and reddening found by \citet{guver_relation_2009} using simultaneous X-ray and optical observations of Galactic supernova remnants:
\begin{equation}
    N_{\text{H}} = (2.21\pm0.09) \times 10^{21} A_\text{V}
\end{equation}
Using the Galactic value along the line of sight $E(B-V)$ = 0.11 \citep{schlafly_measuring_2011} with $R_\text{v}$ = 3.1 would suggest \nh\ $\sim 7.5\times10^{20}$\,cm$^{-2}$, which is an order of magnitude below the value derived from X-ray spectral fitting \cite[e.g. from][(2.6$\pm$0.03) $\times$10$^{21}$ \text{cm}$^{-2}$]{gurpide_long-term_2021}. We will additionally show that all models used below also require \nh\ $> 2\times10^{21}$\,cm$^{-2}$. Instead, using the total $A_\mathrm{V}$ derived above suggests \nh\ $= (2.1\pm0.3)\times10^{21}$\,cm$^{-2}$, which is consistent within the order of magnitude with the values derived from X-ray spectral fitting and the values derived below. This suggest our estimate for the amount of reddening is reasonable.

\subsection{Spectral fits}

 To model the spectra we decided to use two physically motivated models along with a third phenomenological model typically used to describe the 0.3--25\,keV emission in ULXs, which we summarize below. The low-state presented the well-known strong residuals at around 1\,keV \citep{middleton_diagnosing_2015} which have been associated with radiatively-driven relativistic outflows \citep{pinto_resolved_2016, pinto_xmmnewton_2020}. These residuals have a minor impact on the continuum estimation but they can affect the estimation of the level of neutral absorption. Moreover, because we wanted our $\chi^2$ to reflect closely the goodness of fit, we modelled the residuals with a Gaussian at $E \sim1$keV with $\sigma \sim$0.09 keV, which resulted in $\Delta \chi^2$ improvements upwards of 100 (depending on the exact model) for 3 degrees of freedom. 
 
 Neutral X-ray absorption was modelled with a \tbabs\ model component as explained in Appendix~\ref{sec:chandra}. Whether \nh\ is variable in ULXs remains a contested issue virtually unstudied owing to the lack of physically motivated models. In \theulx, \cite{gurpide_long-term_2021}, using phenomenological models, showed \nh\ varies only in an unusual state of the source -- termed `obscured state' in that work \citep[see also][]{middleton_diagnosing_2015}. Since this state is not considered here, we decided to tie \nh\ between the high and low states, while leaving the rest of the parameters free to vary unless stated otherwise.
 
\begin{figure*}
    \centering
    \includegraphics[width=0.8\textwidth]{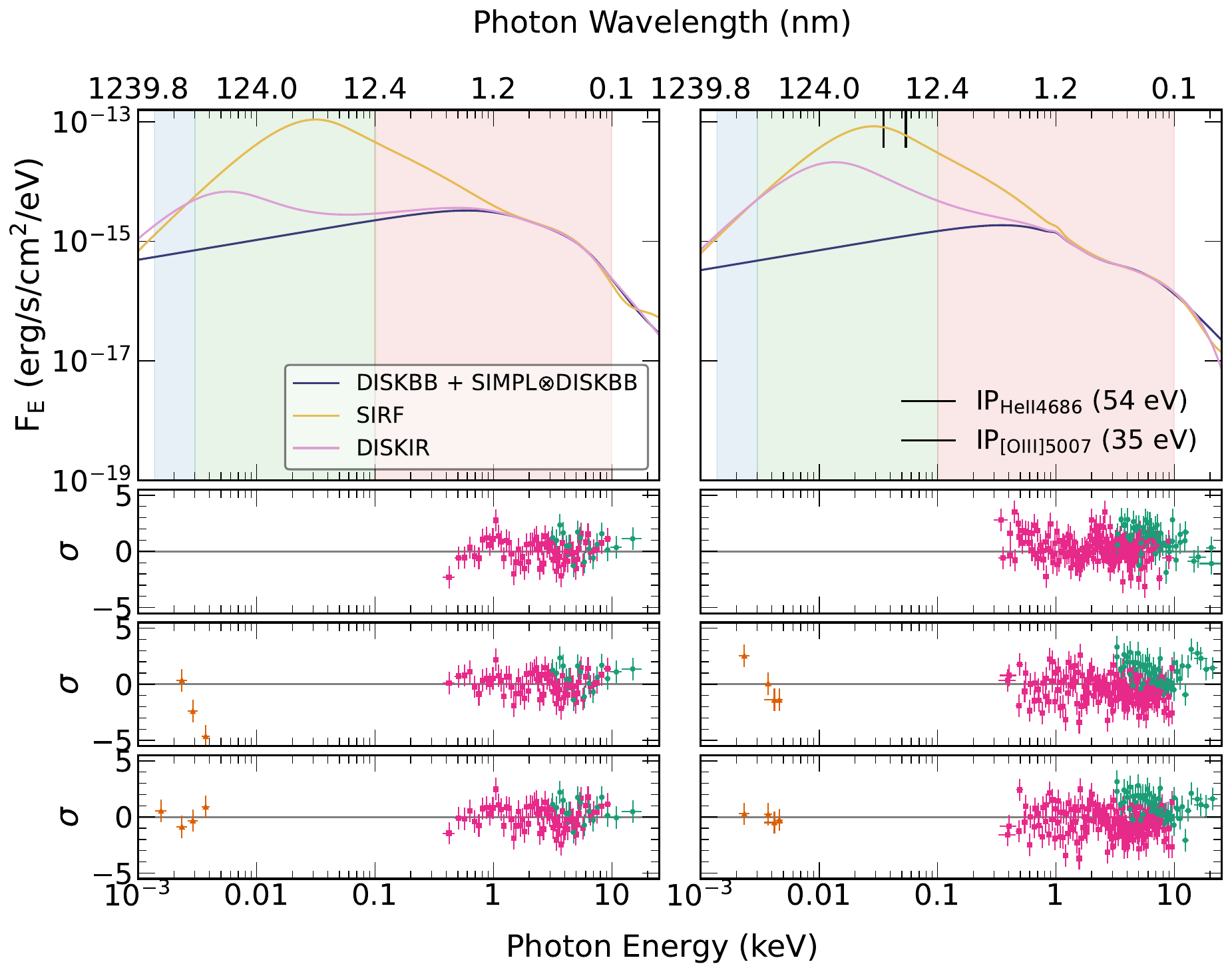}
    \caption{Broadband best-fit spectral models and residuals. (Top) Best-fit absorption-corrected models. The shaded colored areas show the bands used to estimate the fluxes in Table~\ref{tab:sed_modelling}. The two black vertical bars on the right panel show the ionization potential needed to produce two high-excitation optical lines relevant for this work, He$\lambda$4686, [O~{\sc iii}]$\lambda$5007. (Middle top) Best-fit model residuals for the \texttt{phenomenological} model, only fitted to the X-ray data. The pink squares, green circles and orange stars show the \xmm, \nustar\ and \hst\ data respectively. (Middle bottom) As per the above panel for the \texttt{sirf} model. The F814W \hst\ residual redward of $\sim$5555 \AA\ reaching $\sigma\sim7$ has been omitted for clarity. (Bottom) As per the above panel but for the \texttt{diskir} model. The data have been rebinned so that each bin has a significance of at least 10$\sigma$. The left and right panels correspond to the high and low states respectively (notice the change in the X-ray portion of the spectrum).}
    \label{fig:multiband_sed}
\end{figure*}

  Below, we summarise the models employed and the resulting fits to the data. Results from the spectral fitting are reported in Table~\ref{tab:sed_modelling} and Fig.~\ref{fig:multiband_sed} shows the best-fit spectral models and residuals.
\begin{itemize}
    \item \texttt{diskir}: The \texttt{diskir} model is an extension of the multicolour disc blackbody \citep{mitsuda_energy_1984} which takes into account self-irradiation effects by a Compton tail and by emission from the disc inner regions \citep{gierlinski_reprocessing_2009}. The most relevant parameters for the UV/optical data are the reprocessed flux fraction which is thermalized in the disc (\fout) and the outer disc radius (\rout), whereas the rest of the parameters are constrained by the X-ray data. This model has been used to fit the broadband spectrum of ULXs by analogy with XRBs \citep{grise_optical_2012, tao_nature_2012, vinokurov_ultra-luminous_2013, sutton_irradiated_2014, dudik_spitzer_2016} finding satisfactory agreement with the data, although always with some degeneracy with the companion star \citep{grise_optical_2012, tao_nature_2012}. It must be noted that most of these works did not use simultaneous data, which has been shown to be crucial to discriminate between models \citep[e.g.][]{dudik_spitzer_2016}. Compared to existing works \citep{grise_optical_2012, vinokurov_ultra-luminous_2013, dudik_spitzer_2016} where some parameters were frozen (namely those concerning the Compton tail: the electron temperature k$T_\text{e}$ and the ratio of luminosity in the Compton tail to that of the unilluminated disc, $L_\text{C}/ L_\text{D}$) we were able to constrain all parameters simultaneously thanks to the high-energy coverage provided by the \nustar\ data. We found good consistency between the $L_\text{C}/ L_\text{D}$ parameter between the high and low states, which is unsurprising given the lack of variability of \theulx\ above $\sim$10\,keV \citep{gurpide_long-term_2021, walton_unusual_2020}. We therefore tied this parameter between the high and the low states. The fit provides a good description of the continuum with $\chi^2$/dof = 1616/1264 as can be seen from Fig.~\ref{fig:multiband_sed}. We note however, that the reprocessing fraction for the low state ($f_\text{out}\sim4\times10^{-2}$) is about an order of magnitude higher than found in XRBs \citep[see][and reference therein]{urquhart_multiband_2018}. In using this model, we do not necessarily have a physical interpretation in mind, particularly because we expect the accretion flow geometry to deviate substantially from the sub-Eddington accretion flow geometry assumed by the model. We rather consider the \texttt{diskir} as a proxy to obtain a realistic extrapolation to the inaccessible EUV.

    \item \texttt{sirf}: The self-irradiated multi-color funnel can be considered an extension of the \texttt{diskir} to the supercritical regime, in which the disc/wind acquires a cone-shaped geometry altering the self-irradiation pattern \citep{abolmasov_optically_2009}. As a caveat, we note that this model does not include Comptonization or the irradiation of the disc wind onto the outer disc itself \citep[e.g.][]{middleton_thermally_2022}. For simplicity, we fixed some parameters which we considered as nuisance for our analysis to their fiducial values. These were the velocity law exponent of the wind, which we fixed to $-$0.5 (i.e. parabolic velocity law) and the adiabatic index ($\gamma$) which we fixed to 4/3 as the gas is radiation-pressure supported. Leaving the outflow photosphere \rout\ free for both states yielded a loosely constrained and highly degenerate fit. We further noted that this parameter did not strongly affect the fits. We managed to obtain successful fits by leaving \rout\ free for the high state, while for the low state, we fixed \rout\ to 100\,\rsph. We also included self-irradiation effects and found 4--5 iterations were sufficient to reach convergence.
    
    The fit was insensitive to the inclination $i$ so as long as $i$ was smaller than the half-opening angle of the funnel \afunnel\, so we fixed it to 0.5$^\circ$ (i.e. nearly face-on) as expected for a ULX such as NGC~1313~X--1 \citep{middleton_spectral-timing_2015, gurpide_long-term_2021}. Solutions with $i >$ \afunnel\ were statistically excluded by the data.
    
    The model fitted reasonably the continuum (Fig.~\ref{fig:multiband_sed}), although substantially worse than the \texttt{diskir} ($\chi^2$/dof = 1714.2/1268). This model cannot reproduce the fluxes redward of $\sim5555$ \AA\ particularly in the low state (where the residuals $\sigma \gtrsim$7). Perhaps unsurprisingly the parameters are similar to those found for Holmberg~II~X--1 and NGC~5204~X--1 \citep{gurpide_discovery_2021} but with a narrower opening angle of the funnel ($\theta_\mathrm{f} \sim 37^\circ$) and a lower Eddington mass ejection rate $\dot{m}_\mathrm{eje}$ compared to Holmberg~II~X--1.
    
    As can be seen from Fig.~\ref{fig:multiband_sed} the main difference with respect to the \texttt{diskir} model is the predictions in the EUV: the \texttt{sirf} predicts a UV flux more than an order of magnitude higher than the \texttt{diskir}, with the UV flux peaking around the He$^{+}$ ionization potential of 54\,eV (Fig.~\ref{fig:multiband_sed}). Contrary to the \texttt{diskir} model, this model predicts little difference in the UV/optical emission between the high and low states.
        \item \texttt{phenomenological}: Finally, we tested a phenomenological model based on an absorbed dual thermal \texttt{diskbb} including upscattering of the hard \texttt{diskbb} through the empirical \texttt{simpl} model \citep{steiner_simple_2009} \cite[see e.g.][]{gurpide_long-term_2021, walton_unusual_2020}. The complete model in \textsc{XSPEC} was \texttt{tbabs}$\otimes$(\texttt{diskbb} + \texttt{simpl}$\otimes$\texttt{diskbb}). Here too we tied the photon index $\Gamma$ between the high and low states owing to the aforementioned lack of variability above $\sim$10\,keV. Because the model make no account for the optical/near UV emission, we fitted this model to the X-ray data only. Fig.~\ref{fig:multiband_sed} shows this model severely underestimates the UV/optical fluxes compared to the other models. Similar underpredictions of the optical/UV fluxes from direct extrapolation of X-ray spectral models were reported for other ULXs such as Holmberg~IX~X--1 \citep{dudik_spitzer_2016} and NGC~6946~X--1 \citep{abolmasov_optical_2008}. 
\end{itemize}

\begin{table*}
    \begin{center}
     \caption{Best-fit parameters to the multi-band SED of NGC~1313~X--1. All models were fitted to the multi-band data except for the \texttt{phenomenological} without the \texttt{blackbody}, which was fitted to the X-ray data only.}
    \label{tab:sed_modelling}
    \begin{tabular}{cccccc} 
    \hline
    \noalign{\smallskip}
      Parameter &Units& Low State & High State & Low State & High State\\
      \noalign{\smallskip}
    \hline
    \noalign{\smallskip}
                &    &   \multicolumn{2}{c}{\texttt{DISKBB} + \texttt{SIMPL}$\otimes$\texttt{DISKBB}} & \multicolumn{2}{c}{\texttt{DISKBB} + \texttt{SIMPL}$\otimes$\texttt{DISKBB} + \texttt{BBODY}} \\
\noalign{\smallskip}
\hline
\noalign{\smallskip}
\nh\             &10$^{20}$cm$^{-2}$&	\multicolumn{2}{c}{21.7$\pm0.5$}&  \multicolumn{2}{c}{20$\pm0.5$}\\ 
 $T_\text{in}$   & keV & $0.35\pm0.01$ &	0.47$\pm0.03$& 0.38$\pm0.01$  &    0.55$\pm0.03$ \\
$\Gamma$	     &	         &	\multicolumn{2}{c}{3.01$\pm0.09$} & \multicolumn{2}{c}{4.3$\pm0.3$} \\
$f_\text{scatt}$ & \%        & $>$85 & 31$\pm5$ & 60$\pm1$ & 45$\pm11$ \\
$T_\text{in}$    &	keV      & 1.49$_{-0.01}^{+0.10}$ &  1.52$\pm0.07$ & 2.18$\pm0.06$  & 1.63$_{-0.10}^{+0.09}$  \\
$T_\mathrm{*}$              & K         &  &                     &\multicolumn{2}{c}{$33304_{-3133}^{+4061}$} \\
$R_\mathrm{*}$              & $R_\odot$ &    &    & \multicolumn{2}{c}{8$\pm2$} \\
$\chi^2/$dof     &	         & \multicolumn{2}{c}{1629.8/1259} &  \multicolumn{2}{c}{1795.5/1266$^e$} \\
BIC              &           &  \multicolumn{2}{c}{1737.1}     & \multicolumn{2}{c}{1917.2} \\
$L_\mathrm{opt}^a$ & 10$^{37}$\,\luxcgs\ & $0.16\pm0.01$ &  0.23$\pm$0.01  & $1.29\pm0.05$ &  $1.24\pm0.03$ \\ 
$L_\mathrm{UV}^b$ & 10$^{39}$\,\luxcgs\ &  $0.25\pm0.01$ &  $0.38\pm0.02$ & $0.44^{+0.07}_{-0.05}$ &  $0.60^{+0.11}_{-0.07}$\\
$L_\mathrm{X}^c$ & 10$^{40}$\,\luxcgs\ & 1.01$\pm0.01$   & 2.68$\pm$0.02 & $0.977\pm0.006$ &  $2.62\pm0.02$ \\
$L_\mathrm{bol}^d$  & 10$^{40}$\,\luxcgs\ & 1.34$\pm0.03$ & $3.13\pm0.05$ & $1.30\pm0.02$ &  $3.06\pm0.04$ \\
  \noalign{\smallskip}
  \hline
  \noalign{\smallskip}
 &       &  \multicolumn{2}{c}{\texttt{DISKIR}}\ &  \multicolumn{2}{c}{\texttt{DISKIR} + \texttt{BBODY}}\\
  \noalign{\smallskip}
  \hline
    \noalign{\smallskip}
\nh\                    & 10$^{20}$cm$^{-2}$ & \multicolumn{2}{c}{24.1$_{-0.6}^{+0.3}$}&  \multicolumn{2}{c}{24.2$_{-0.7}^{+0.6}$}\\
$kT_\text{disk}$	    &	keV	             & 0.36$\pm0.01$  & 0.46$\pm0.02$       & 0.36$\pm0.01$ & 0.46$_{-0.11}^{+0.03}$  \\
$\Gamma$	            &		             &	1.77$\pm0.3$    &  3.1$_{-0.4}^{+0.1}$& 1.77$_{-0.03}^{+0.04}$ & 3.13$_{-0.35}^{+0.09}$   \\ 
$kT_\text{e}$           &	keV	             &	3.1$\pm0.1$     & $>$23               &	3.1$\pm0.1$  &  $>$12  \\
$L_\text{C}/ L_\text{D}$&                    & \multicolumn{2}{c}{1.67$\pm$0.08}      &   \multicolumn{2}{c}{1.65$_{-0.07}^{+0.09}$}\\
$R_\text{irr}$          &	\rin\            & 1.009$\pm0.001$ &	1.016$_{-0.001}^{+0.002}$ & 1.009$\pm0.001$   & 1.015$\pm0.002$ \\
$f_\text{out}$          &	10$^{-3}$        & 40$_{-10}^{+5}$  & 1.8$_{-0.3}^{+0.5}$ &	 42$\pm7$  &  0.03$\pm0.01$  \\
log(\rout)	            & \rin\              &	3.69$\pm0.03$   & 4.02$\pm0.04$       & 3.23$_{-0.08}^{+0.06}$  &  $>4$ \\
$T_\mathrm{*}$                     &    K               &                  &                     & \multicolumn{2}{c}{$32724_{-5802}^{+8123}$} \\
$R_\mathrm{*}$                     & $R_\odot$          &                  &                     & \multicolumn{2}{c}{$7_{-3}^{+4}$} \\
$\chi^2/$dof            &                    & \multicolumn{2}{c}{1616.0/1264}        & \multicolumn{2}{c}{1613.6/1262} \\
BIC                     &                    &  \multicolumn{2}{c}{1751.9}            & \multicolumn{2}{c}{1763.9}      \\
$L_\mathrm{opt}$        & 10$^{37}$ \luxcgs\ & $1.11\pm0.05$   &  $1.25\pm0.04$          &  $1.16^{+0.05}_{-0.07}$ &  $1.21\pm0.03$ \\ 
$L_\mathrm{UV}$ & 10$^{39}$\,\luxcgs\  & $2.2^{+0.3}_{-0.6}$&  $0.66\pm0.06$          &  $1.9\pm0.3$ &  $0.63^{+0.18}_{-0.05}$\\
$L_\mathrm{X}$  & 10$^{40}$\,\luxcgs\  & $1.12\pm0.03$      &  $2.75^{+0.04}_{-0.03}$ &  $1.13\pm0.03$ & $2.76^{+0.02}_{-0.03}$\\
$L_\mathrm{bol}$&  10$^{40}$\,\luxcgs\ & $1.49\pm0.07$      &  $3.16\pm0.07$          &  $1.48^{+0.07}_{-0.03}$ &  $3.17^{+0.05}_{-0.09}$\\
\noalign{\smallskip}
\hline
\noalign{\smallskip}
&   & \multicolumn{2}{c}{\texttt{SIRF}}  &  \multicolumn{2}{c}{\texttt{SIRF}+ \texttt{BBODY}}\\
\noalign{\smallskip}
\hline
\noalign{\smallskip}
\nh\ & 10$^{20}$cm$^{-2}$ &  \multicolumn{2}{c}{35.2$\pm0.2$} & \multicolumn{2}{c}{34.8$\pm0.3$} \\
$T_\text{in}$	 & keV    & 1.69$\pm0.4$  &  1.18$\pm0.05$             & 1.70$\pm0.05$  &  1.18$\pm0.04$ \\
\rin\            & 10$^{-3}$ \rsph\ & 0.30$\pm0.02$  &  5.40$\pm0.07$  & 0.34$\pm0.03$ & 5.7$_{-0.8}^{+1.1}$\\
\rout\           & $R_\text{sph}$  & 100 & 508$_{-80}^{+115}$          &  100  & 421$_{-70}^{+99}$\\
$\theta_\text{f}$ & $^\circ$ & 36.5$\pm$0.1 & 39.2$\pm0.5$             & 36.5$\pm0.1$  &  39.3$\pm0.5$ \\ 
\mout            & $\dot{M}_\text{Edd}$ &  83$\pm2$  &  16.3$\pm0.5$   &  74$_{-5}^{+4}$  &  4.3$\pm0.6$ \\\
$T_\mathrm{*}$              & K   &     &            &   \multicolumn{2}{c}{$7415_{-1415}^{+2030}$} \\
$R_\mathrm{*}$              & $R_\odot$    &    &    & \multicolumn{2}{c}{$27_{-11}^{+16}$} \\
$\chi^2$/dof     &    & \multicolumn{2}{c}{1714.2/1268} & \multicolumn{2}{c}{1630.2/1266} \\
BIC              &           &  \multicolumn{2}{c}{1821.6}     & \multicolumn{2}{c}{1751.8} \\
$L_\mathrm{opt}$& 10$^{37}$\,\luxcgs\ & $1.03\pm0.04$ &  $1.17\pm0.03$    &  $1.22\pm0.05$ &  $1.21\pm0.04$\\
$L_\mathrm{UV}$ & 10$^{39}$\,\luxcgs\ & $11.3\pm0.2$  &  $15.4\pm0.3$     &  $8.1^{+0.5}_{-0.2}$    &  $10.2\pm0.4$ \\
$L_\mathrm{X}$  & 10$^{40}$\,\luxcgs\ & $1.99\pm0.01$ &  $4.34\pm0.04$    &  $1.80^{+0.04}_{-0.02}$  &  $3.86\pm0.04$ \\
$L_\mathrm{bol}$& 10$^{40}$\,\luxcgs\ & $3.36^{+0.1}_{-0.05}$  &  $6.3^{+0.2}_{-0.1}$&  $2.83^{0.1}_{-0.03}$                & $5.3^{+0.3}_{-0.1}$ \\
  \noalign{\smallskip}
\hline
\hline
\end{tabular}

\begin{minipage}{\linewidth}
\textbf{Notes.} Uncertainties are given at the 1$\sigma$ level. All luminosities are corrected for absorption.\\
$^a$Optical luminosity, defined in the 4000--9000 \AA\ band.\\
$^b$UV luminosity (3\,eV, 4000 \AA\ -- 0.1\,keV).\\
$^c$X-ray luminosity (0.1--10\,keV).\\
$^d$Bolometric luminosity (10$^{-4}-10^{3}$\,keV).\\
$^e$Note the change in number of degrees of freedom here is due to the addition of the \hst\ data in the fit which includes the \texttt{bbody}.\\
\end{minipage}
\end{center}
\end{table*}

\subsubsection{Donor star or irradiated disc?}
As we have seen the only model capable of explaining the broadband data is the \texttt{diskir}, in agreement with earlier works \citep{berghea_spitzer_2012, grise_optical_2012}. Instead, the \texttt{phenomenological} and \texttt{sirf} struggle to describe the optical data. We thus considered an alternative description of the data, in which the optical/near UV emission may have additional contribution from the donor star (in Section~\ref{sec:optical_light} we discuss whether such interpretation holds physically). We approximated its spectrum using a blackbody (\texttt{bbody} in \textsc{XSPEC}) alongside the emission from the accretion flow. We further assumed the putative star parameters remain constant between the high and low states. 

\begin{figure*}
    \centering
    \includegraphics[width=0.8\textwidth]{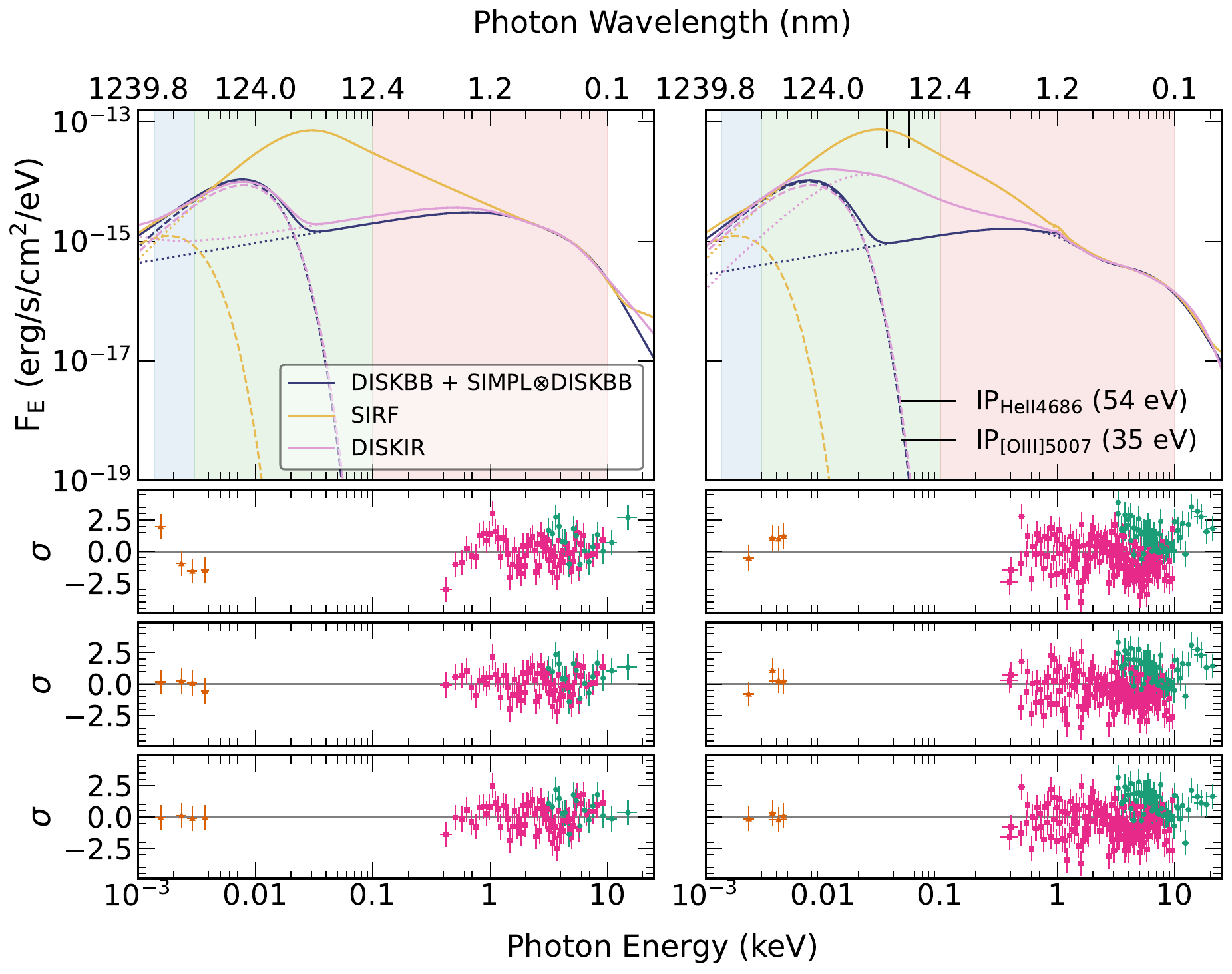}
    \caption{As per Fig.~\ref{fig:multiband_sed} but now including a \texttt{blackbody} for the optical/near UV emission from the putative companion star (indicated by a dashed line). Note the change in scale in the residual panels compared to Fig.~\ref{fig:multiband_sed}.}
    \label{fig:multiband_sed_star}
\end{figure*}

Table~\ref{tab:sed_modelling} shows the resulting best-fit parameter, including the constraints on the star temperature $T_\mathrm{*}$ and radius $R_\mathrm{*}$, while Fig.~\ref{fig:multiband_sed_star} shows the best-fit models and residuals. Both the \texttt{phenomenological} and the \texttt{diskir} would favour $R_\mathrm{*}\sim$7$R_\mathrm{\odot}$ and $T_\mathrm{*} \sim 30,000$K. This would imply an O-type star with a $L\sim 78,000 L_\odot$ and a mass of $>$20$M_\odot$ \citep{ekstrom_grids_2012}. On the other hand, the \texttt{sirf} requires a $\sim$26$R_\mathrm{\odot}$ and $T_\mathrm{*} \sim 8,000$K star, which would correspond to an F0-A star with a mass of $<$9$M_\odot$ \citep{ekstrom_grids_2012}. We discuss the implication of these results in more detail in Section~\ref{sec:optical_light}.

The \texttt{blackbody} provides a significant fit improvement for both the \texttt{phenomenological} and the \texttt{sirf} (for instance a $\Delta \chi^2 = 84$ for 2 degrees of freedom for the \texttt{sirf} model). However, the \texttt{diskir} again provides the best overall fit ($\chi^2$/dof = 1613.6/1262), without much improvement with respect to the starless model, as the optical fluxes were already well described by the disc emission alone. Similar result was found by \citet{berghea_spitzer_2012} in NGC~6946~X--1. The \texttt{sirf} provides only slightly worse fit in terms of $\chi^2$ (1630.2/1266) but as can be seen from Figure~\ref{fig:multiband_sed_star} this model offers a poor description of the high-energy ($>10$ keV) tail. This may be expected as the model does not include Comptonization. However, if we consider the trade off between the likelihood and the number of parameters, in terms of  Bayesian Information Criterion \citep[BIC;][]{schwarz_estimating_1978} it could be argued that the \texttt{sirf} provides a better description of the data, as it offers the lowest BIC ($\Delta $BIC $\simeq$ --12 compared with the \texttt{diskir}). The phenomenological provides a marginal description of the data ($\chi^2_r \approx 1.4$), offering a poor description of the high-energy ($>10$ keV) tail and the optical data. We have verified that the high-energy residuals persists even if we allow $\Gamma$ to vary between the high and low states ($\chi^2_r =$1.42), indicating that the \texttt{phenomenological} model cannot account simultaneously for both the high-energy and optical emission. In terms of overall fit, therefore the best representation of the line-of-sight SED is given by the \texttt{diskir} because it can account simultaneously for both the optical and the high-energy tail. In Section~\ref{sub:front_view} we put further constraints on the line line-of-sight SED, and provide supporting evidence for the UV extrapolation provided by the \texttt{diskir}.

Crucially, these models make different predictions about the UV flux. This is clearly reflected in Figures~\ref{fig:multiband_sed} and ~\ref{fig:multiband_sed_star}. The UV luminosity (defined as the luminosity in the 3\,eV--0.1 keV band) is also reported in Table~\ref{tab:sed_modelling}. The \texttt{phenomenological} predicts the lowest amount, differing in more than an order of magnitude from the prediction made by the \texttt{sirf}. We are now ready to test these predictions against the emission-line nebula.

\section{\cloudy\ modelling: what the nebula `sees'}\label{sec:cloudy}
Having constrained the broadband SED of \theulx, we now turn to examine the effects on its environment through two different sight-lines: sideways, by studying the surrounding, extended nebula reported in \cite{gurpide_muse_2022} (Section~\ref{sub:sideview}) and along the line of sight, by studying the line-of-sight integrated nebular spectrum (Section~\ref{sub:front_view}). In order to select an appropriate ionising SED for our photo-ionization modelling, we noted the numerical calculations presented by \citet{chiang_time-dependent_1996}, who studied nebulae photoionized by variable supersoft X-ray sources. Their calculations considered SEDs and densities to a good degree physically relevant for ULX nebulae. These authors showed that so long as the duty the cycle of the source is much shorter than the recombination timescales, the nebula will effectively 'see' or react to the equivalent of a time-averaged SED of the source. In diffuse nebulae the longest recombination timescale is that of hydrogen \citep{osterbrock_astrophysics_2006}, which based on typical values from our nebular modelling we find of the order of $t($H$^+)\sim$2$\times$10$^5$ yr. Shorter timescales are typically $t$(HeII$^{++})\sim$2,000 yr and $t$(O$^{+++})\sim900$ yr, therefore much longer than the variability timescales of the source. Assuming the \swift-XRT variability is representative of the overall variability of NGC~1313~X--1 \citep[also supported by \xmm\ observations;][]{gurpide_long-term_2021}, the temporal average (black dashed line in Fig.~\ref{fig:swift_hst}) suggest the SEDs from the low state are a good approximation for the ionising SED, as the excursions to the high state are too rapid for any meaningful impact on the overall ionising radiation. We return to this point in Section~\ref{sub:caveats}.

We therefore created a set of \cloudy\ \citep[version C22.02;][]{ferland_2017_2017} models using our best-fit (extinction-corrected) low-state SEDs from Table~\ref{tab:sed_modelling}, totalling 6 different SEDs (3 of them including contribution from the putative companion star). We tested two sets of metallicities for the gas, $Z = 0.15Z_{\sun}$, 0.3$Z_{\sun}$\footnote{Here $Z_{\sun}$ refers to \cloudy's default solar abundances.}, corresponding to \met = 7.86 and \met = 8.17, representative of the nebula around NGC~1313~X--1 \citep{gurpide_muse_2022}, assumed a filling factor of 1 and open geometry. The center of the nebula was taken as the ULX position in the data cube presented in \citet{gurpide_muse_2022}. 

\begin{figure}
    \centering
    \includegraphics[width=0.48\textwidth]{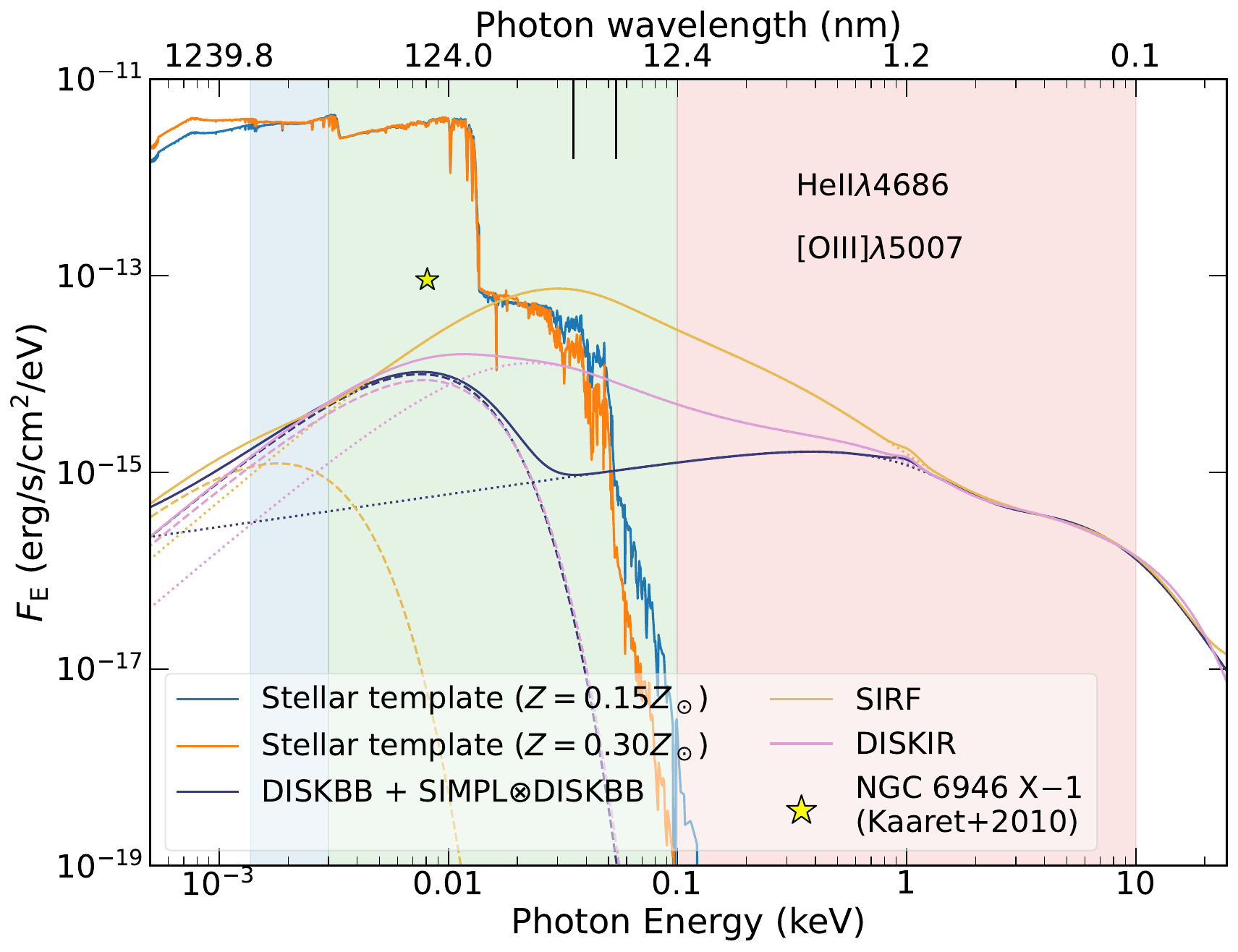}
    \caption{BPASS stellar template alongside the ULX models used for the low state of \theulx. Both stellar templates have an age of 10$^{7.5}$\,yr and were approximately rescaled to the local SFR around \theulx. The shaded bands are as per Fig.~\ref{fig:multiband_sed} and correspond to the bands used to calculate the fluxes in Table~\ref{tab:sed_modelling}.}
    \label{fig:bpass_model}
\end{figure}

We carried out two set of calculations: one where we assumed the ULX was the only source of ionisation and another where we included an ionising stellar-background from the stars in the field. In order to include a realistic stellar ionising background, we obtained a composite stellar spectrum from the Binary Population and Spectral Synthesis (BPASS) v2.1 \citep{eldridge_binary_2017} including masses up to 100\,\msun. We adopted a metallicity matching that of the gas and an age of the stellar population matching that of the nearby stars \citep[$t = 10^{7.5}$\,yr;][]{yang_optical_2011}. We followed \citet{simmonds_can_2021} and rescaled the spectrum based on the star-forming rate (SFR) of NGC~1313. To do so, we noted \citet{suzuki_akari_2013} measured the SFR surface density in NGC~1313 to be $\sim$0.01\,\msun/yr/kpc$^2$. Considering the region of interest here ($\sim$0.04\,kpc$^2$), we found the local SFR = 0.0004\sfr. Using the scaling between UV luminosity and SFR from \citet{kennicutt_star_1998}, a bolometric luminosity for the stellar background of $\sim$9.5$\times$10$^{39}$\luxcgs\ is suggested. This value is only accurate to the order of magnitude owing to uncertainties related to the spatial distribution of the stars in the field, potential contribution from stars outside the photo-ionized region and their exact distance(s) to the nebula. Therefore we have ran our calculations for three different luminosity values for the stellar-background, $L = 2.75\times10^{40}$, 9.5$\times$10$^{39}$ and 2.75$\times10^{39}$ erg/s, in order to consider the effects of varying this parameter. As we show below, independent on the exact luminosity, all calculations strongly support the fact that stars contribute to the line of ratios of the nebula. The best results were found for {$L = 2.75\times10^{40}$ erg/s} and we focus on the results obtained for this luminosity throughout, although we also discuss the less luminous cases below. The stellar templates for $2.75\times10^{40}$ erg/s along with the ULX low-state spectra derived in Section~\ref{sec:multi_band_sed} are shown in Fig.~\ref{fig:bpass_model}. 

To complement the flux maps presented in \citet{gurpide_muse_2022}, we further extracted maps of [Ar~{\sc iii}]$\lambda$7135 and [S~{\sc iii}]$\lambda$9069 from cube 1 in \citet{gurpide_muse_2022}. 
These are shown in Fig.~\ref{fig:fluxmaps}. Other lines such as [O~{\sc ii}]$\lambda$7318,7329 were too faint for any meaningful extraction. 

\begin{figure*}
    \centering
    \includegraphics[width=0.48\textwidth]{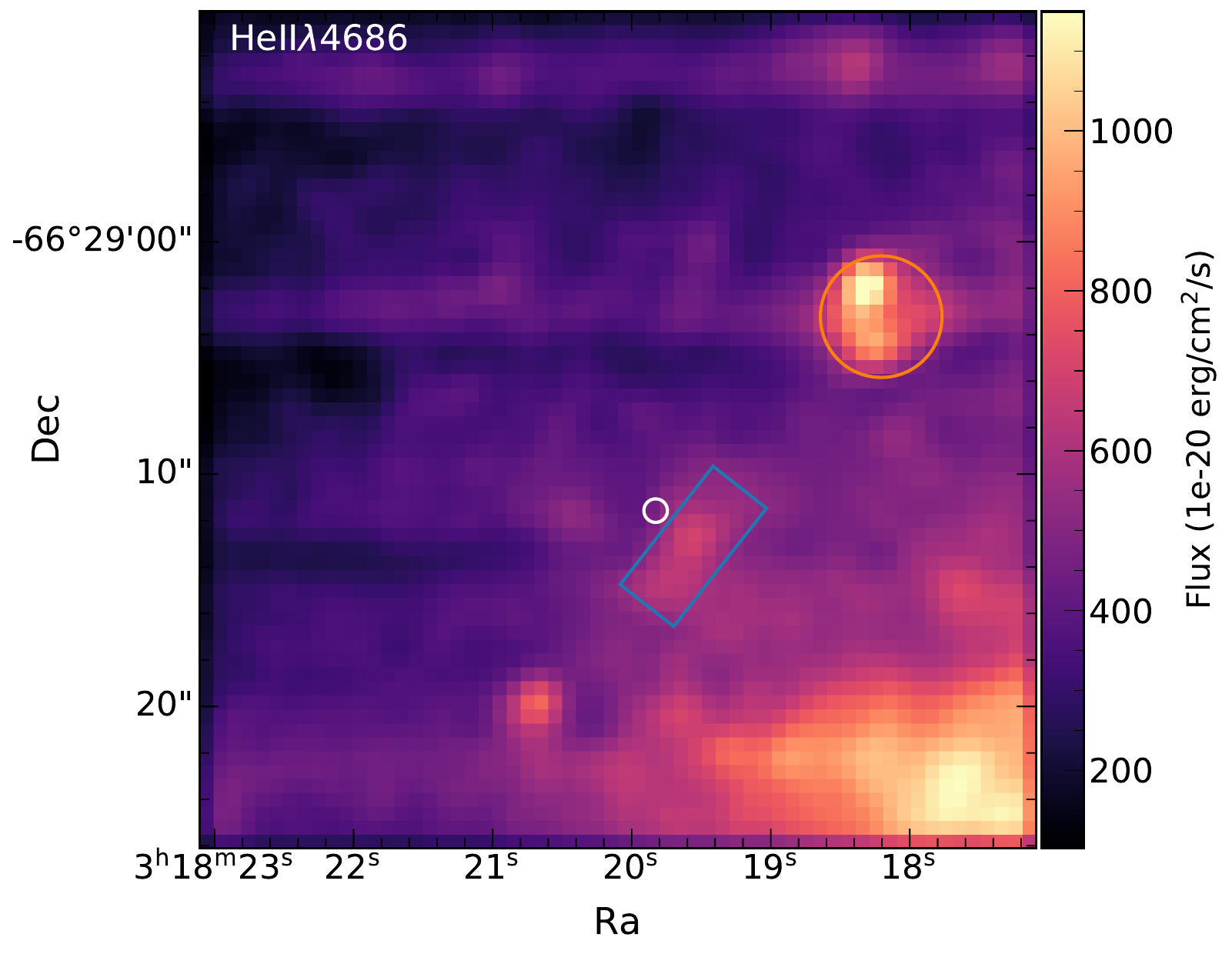}
     \includegraphics[width=0.48\textwidth]{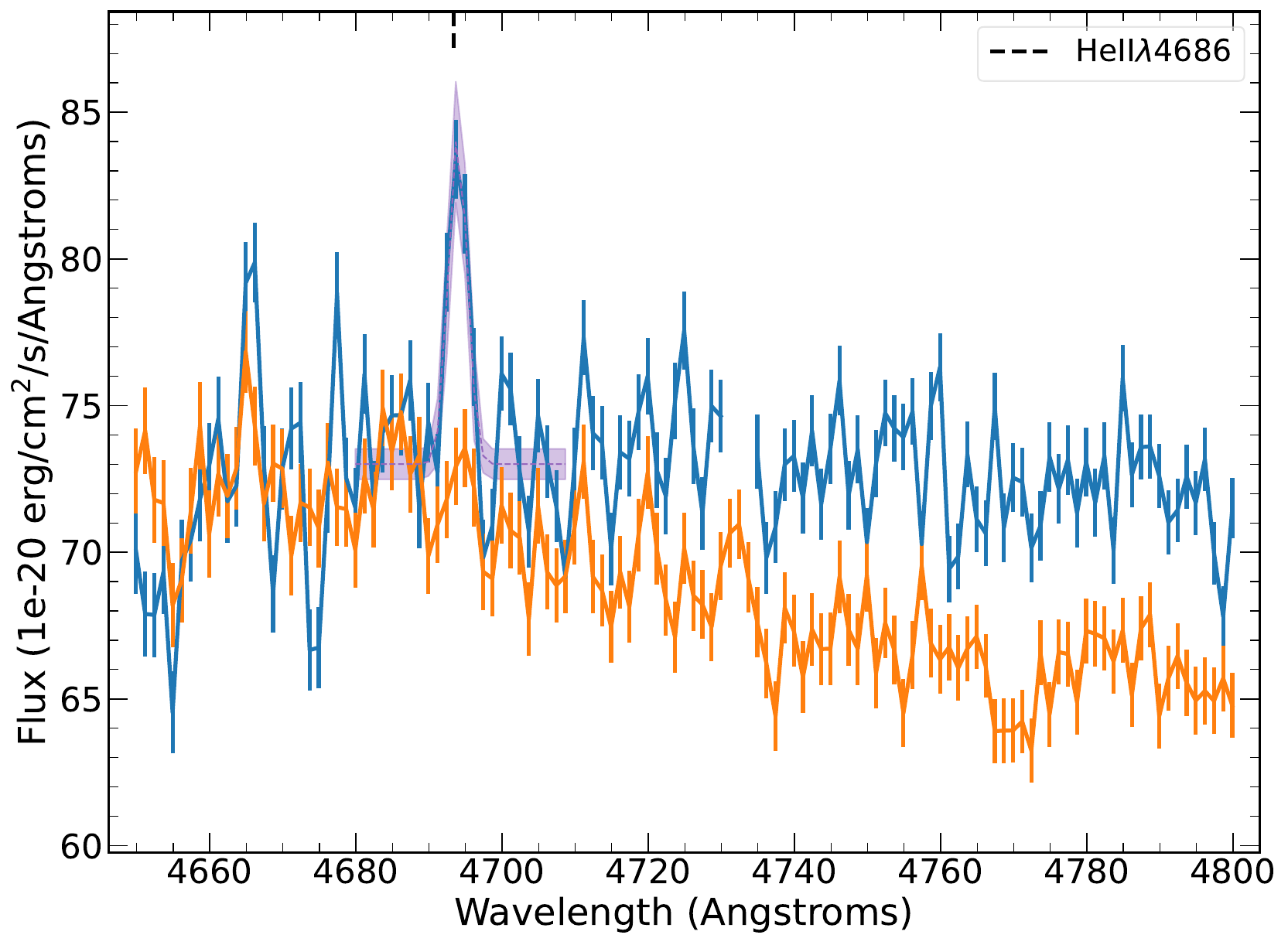}
    \caption{(Left) \heii\ flux map, constructed by integrating the spaxels around the \heii\ line (4688.5--4698.6 \AA) -- accounting for the systemic redshift of NGC1313 -- and subtracting the nearby local continuum \citep[using Cube 2 from ][]{gurpide_muse_2022}. The map has been resampled by a factor 3 compared to the original resolution to enhance a feature (blue rectangle) close to the ULX position (white circle). (Right) Average spectra from the blue and orange regions. The orange spectrum has been divided by 1.4 for visual clarity. The spectrum extracted from the patch close to the ULX shows a clear emission line at the expected position of the \heii\ based on the redshift of NGC~1313 (indicated by a vertical black dashed line). The magenta dashed line and shaded area shows the best-fit with a Gaussian for the \heii\ line and a constant for the local continuum, and its 1$\sigma$ uncertainties.}
    \label{fig:heII}
\end{figure*}

As stated in \citet{gurpide_muse_2022}, the \heii\ was not detected in cube 2 taken in extended mode. The \cloudy\ predictions we make below prompted us to examine carefully the presence of this line or at least derive an upper limit on its flux in order to constrain further our models. Because we found the line too faint for a pixel-by-pixel fit, we constructed a flux map by integrating the spaxels around the expected position of the \heii\ line based on the systemic redshift of NGC~1313 ($z = 0.001568$) -- from 4688.5 \AA\ to 4698.6 \AA\ -- and subtracting the mean value of the nearby continuum. The left panel of Fig.~\ref{fig:heII} shows the resulting map resampled by a factor 3 to highlight a tenuous feature close to the ULX position. We extracted an average spectrum from this region (shown in blue in the right panel of the Figure) along with a spectrum from the nearby stellar cluster (shown in orange) to compare the presence of the \heii\ line. The resulting spectra are shown in Fig.~\ref{fig:heII} (right panel). The line is clearly detected\footnote{$\Delta \chi^2 = 107$ for 3 degrees of freedom with respect to a simple constant model whereas the lower confidence interval on the line flux is inconsistent with 0 at the 3 sigma level.} in the patch next to the ULX at the expected position based on the redshift of NGC~1313 (shown in the Figure by a black dashed vertical tick). Fitting a Gaussian and a constant for the local continuum (shown as a magenta dashed line) we measured an average flux $F$(\heii) = 3.6$\pm$0.7$\times$10$^{-19}$ erg/s/cm$^{2}$. From the \ebv\ map derived in \citet{gurpide_muse_2022}, we measured an average \ebv\ value in the same region of \ebv\ = 0.179$\pm$0.005 mag. With this value and using the \citet{calzetti_dust_2000} extinction curve with $R_\mathrm{v} = 4.05$ we arrived at a \textit{total} \heii\ luminosity of (7.5$\pm$1.4)$\times$10$^{35}$ erg/s. While we do not use this value directly in our model-data comparison, we discuss it further below.

\subsection{The Side View} \label{sub:sideview}
In order to study the extended surrounding nebula discovered in \cite{gurpide_muse_2022}, we converted the 1D \cloudy\ predictions to 2D images and resampled them to the MUSE pixel scale of 0.2". We then blurred the images applying a 2D Gaussian kernel with FWHM matching the datacube's PSF \citep[$\sim$1";][]{gurpide_muse_2022}. Next we compared the line ratios from these set of \cloudy-generated images to the real MUSE-derived line ratio images. We restricted the data-model comparison to the regions where the gas was identified as being EUV/X-ray photoionized, where the ULX contribution dominates (see Figure~\ref{fig:cloudy_2D_diskir} below) and where the influence of shocks is minimized \citep{gurpide_muse_2022}. The choice to work on line ratios rather than in fluxes was in order to reduce uncertainties related to distance, extinction and geometry. 

The geometry above implicitly assumes that the cloud is approximately co-planar with the ULX in the plane of the sky and that the dimension along the line of sight is much smaller than the dimensions on the plane, such that the line ratios along a given line of sight are approximately constant. We are working on extending our modelling to more complex 3D structures and we will present a more refined treatment of the cloud geometry in a future publication, but note our preliminary results assuming a projected, spherical sector give results qualitatively consistent with those presented here.

For each line ratio ([O~{\sc iii}]$\lambda$5007/\hb, [O~{\sc i}]$\lambda$6300/\ha, [N~{\sc ii}]$\lambda$6583/\ha, [S~{\sc ii}]$\lambda$6716/\ha\ and [S~{\sc ii}]$\lambda$6716/[S~{\sc ii}]$\lambda$6730, [Ar~{\sc iii}]$\lambda$7135/\ha and [S~{\sc iii}]$\lambda$9069/\ha) we computed a $\chi^2$ using the model- and data-pixel values and uncertainties on the flux ratio propagated from the measurement error of each line in each spaxel \citep[estimated from the pixel-by-pixel Gaussian-line fitting presented in][]{gurpide_muse_2022}. By adding each individual line ratio $\chi^2$, we were able to select the model with the \textit{overall} lowest \chisq. We did not use [O~{\sc iii}]$\lambda$4959 nor [N~{\sc ii}]$\lambda$6548 as their fluxes were tied by theoretical constraints \citep{storey_theoretical_2000}. Given other sources of uncertainty such as the exact density profile, geometry, clumpiness of the cloud, projection effects, other sources in the field, etc. and the fact that some of the pixels in our models will have no values in them (see below), our $\chi^2$ should be regarded as an heuristic to select the best model, rather than the usual goodness of fit.

\begin{figure}
    \centering
    \includegraphics[width=0.49\textwidth]{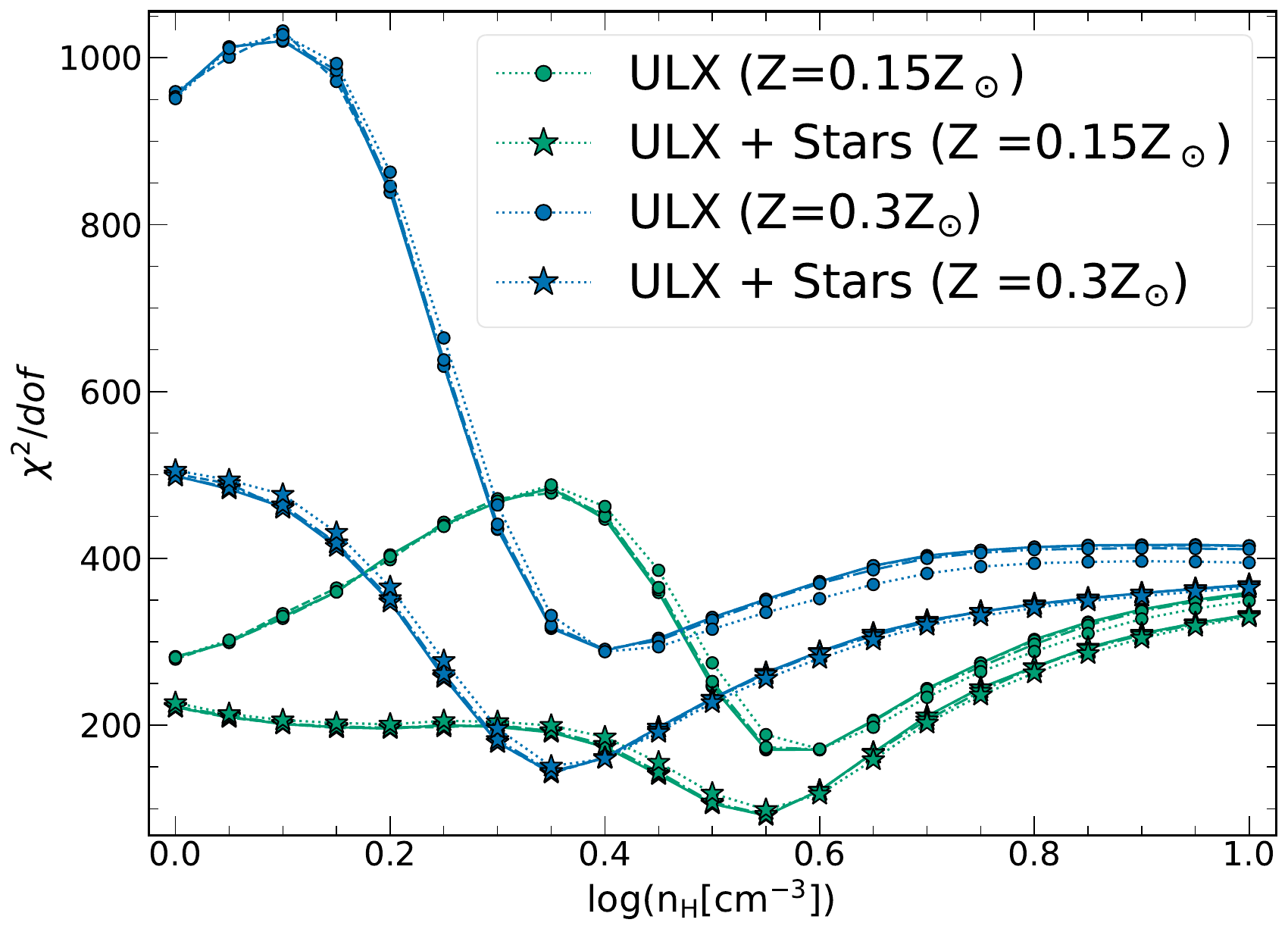}
    \caption{\cloudy\ modelling $\chi^2$/dof as a function of density of the gas for the \texttt{diskir\_star} model without (circular markers) and with the added stellar background contribution (star-shaped markers). Solid, dashed, dot-dashed and dotted lines show the effects of varying $r_\mathrm{in} = 1, 12.48, 22.34, 40$\,pc. Green and blue colors show the results for metallicities Z$ = 0.15$Z$_{\sun}$ and Z$=0.30$Z$_{\sun}$, respectively. We can see that for fixed density, the effects of varying $r_\mathrm{in}$ are negligible and that the inclusion of the stellar background significantly improves the results.}
    \label{fig:chisq}
\end{figure}

We produced models for a range of constant hydrogen number densities $\log(n_\mathrm{H})$, varying from 0.0 to 1.0 in steps of $\Delta \log(n_\mathrm{H}) = $0.05 (noting that the [S~{\sc ii}] lines indicate $n_\mathrm{e} < 100$cm$^{-3}$)\footnote{We ran models with higher densities but they clearly failed to reproduce the extent of the nebula.} and inner radius of the cloud $r_\mathrm{in}=1, 12.5, 22.3, 40$\,pc, setting values inside the cavity or outside of the \cloudy\ calculation to 0. The outer radius was set to 200\,pc, roughly matching the extent of the high [O~{\sc i}]$\lambda$6300/\ha\ ratio and to avoid the emission from a stellar cluster further south (see the bright blob to the far south of the ULX in Figures~\ref{fig:fluxmaps}). Fig.~\ref{fig:chisq} shows an example-$\chi^2$ contours derived for the \texttt{diskir\_star} model (circular markers) for the two metallicity values (green and blue for $Z = 0.15Z_{\sun}$ and $Z = 0.3Z_{\sun}$ respectively), which also illustrates that the effects of varying $r_\mathrm{in}$ are minimal. The Figure also shows the resulting $\chi^2$ contours for the same ULX model when the stellar background is included (star-shaped markers). It is clear that the inclusion of the stellar background improves significantly the fit. Fig.~\ref{fig:cloudy_2D_diskir} shows a 2D data-model comparison for the best-fit \texttt{disk\_star} model for $Z = 0.15Z_{\sun}$ for the ULX and ULX + stellar background runs, to illustrate the region we probed and the type of comparisons we carried out. 

\begin{figure*}
    \centering
    \includegraphics[width=0.98\textwidth]{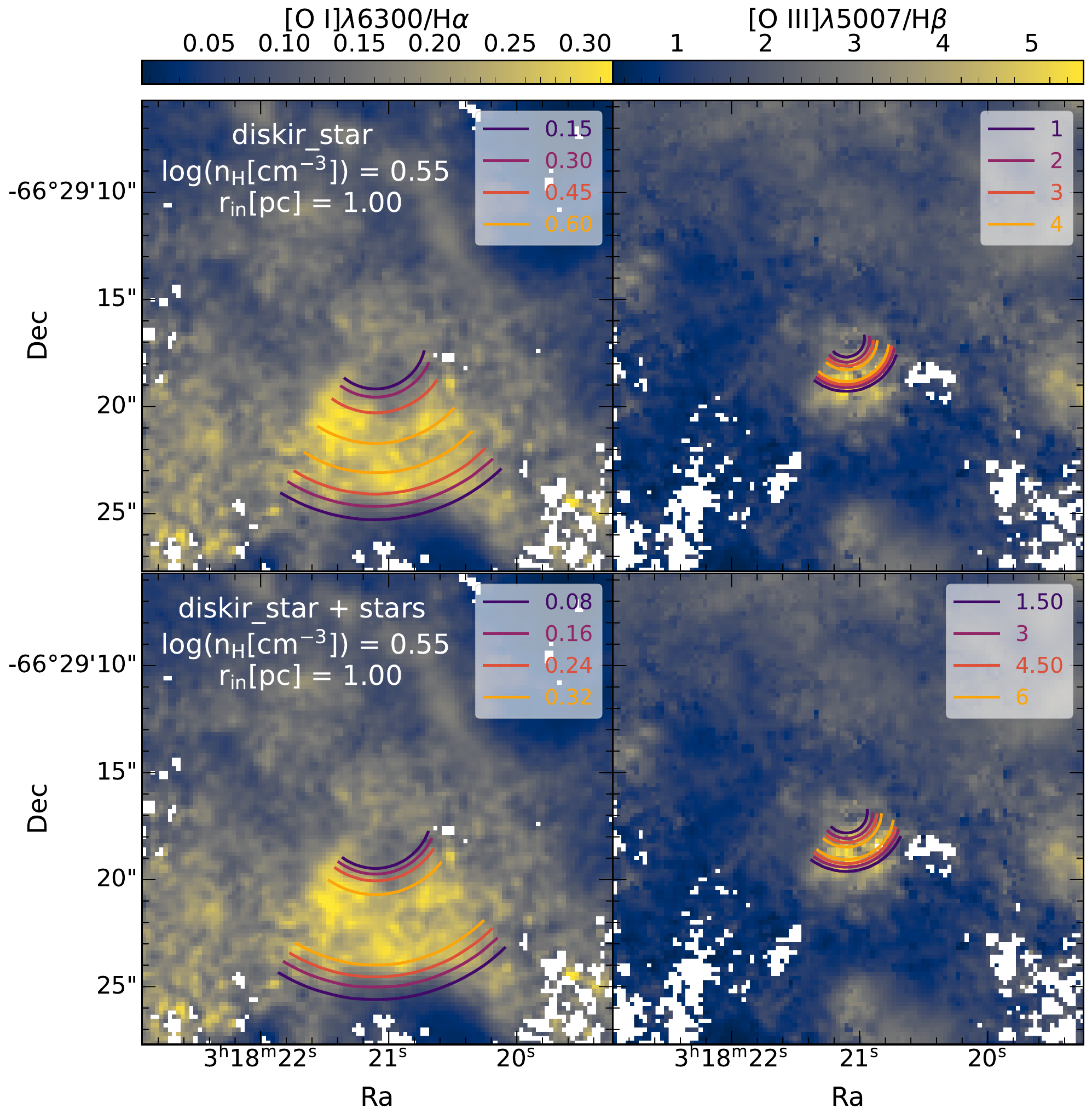}
    \caption{\cloudy-generated 2D best-fit models (contours) and data (background image) comparison for the \texttt{diskir} model with the putative companion star (\textit{top} panels) and same model but now including the stellar background (\textit{bottom panels}), for $Z = 0.15Z_{\sun}$. [OI]$\lambda$6300/\ha\ and [OIII]$\lambda$5007/\hb\ maps are shown on the left and right panels, respectively. The contours show the \cloudy\ model predictions, with the numbers for each colored annuli given in the legend. The contours also show the region used in the comparison, where the EUV/X-ray nebula is located. All images show a 22"$\times22$" region centered around the ULX.}
    \label{fig:cloudy_2D_diskir}
\end{figure*}

Tables~\ref{tab:cloudy_results_Z0004} and~\ref{tab:cloudy_results} show the resulting overall $\chi^2$ alongside the maximum observed line ratios for the best-fit models and the data for $Z = 0.3Z_{\sun}$ and $Z = 0.15Z_{\sun}$, respectively. The models with $Z = 0.3Z_{\sun}$ overpredict all line ratios by a factor of $\sim$2, regardless of whether the stellar background is included or not (Table~\ref{tab:cloudy_results_Z0004}). We therefore were able to reject this metallicity based on the data. This may be surprising as this value matches more closely the metallicity inferred around \theulx. However, as we show below we clearly find that the models with $Z = 0.15Z_{\sun}$ match more reasonably the observed line ratios (see also Fig.~\ref{fig:chisq}). We note that in \citet{gurpide_muse_2022} we were unable to measure the metallicity in the photoionized region itself as metallicity estimators are mostly calibrated for standard HII regions. Hence it may be possible that the nebula has a lower metallicity content than the neighbouring gas. In any case, although such metallicity is low, it is not unrealistic as it is at the lower end of the values measured in \citet{gurpide_muse_2022}. We therefore focused on the results for $Z = 0.15Z_{\sun}$. 

In the case of no stellar background, both \texttt{diskir} and \texttt{phenomenological} offer comparable level of agreement with the data, and superior to that offered by the \texttt{sirf}. The main difference between the \texttt{phenomenological} and the \texttt{diskir} is the peak [O~{\sc iii}]$\lambda$5007/\hb\ ratios, due to the lower UV level of the former model. We note that it is actually possible to reproduce peak [O~{\sc iii}]$\lambda$5007/\hb\ ratios in the 5--6 range with the \texttt{phenomenological} model (cf. 5--7.5 for the \texttt{diskir} and 6--7.5 for the \texttt{sirf}), therefore compatible with the data. However, due to the lower overall values coupled with the smaller region of enhanced [O~{\sc iii}]$\lambda$5007/\hb\ produced by the \texttt{phenomenological} due to its lower UV flux, the spatial smoothing smears these values below 3, which highlights the importance of considering the spatial scale. 

As shown in Fig.~\ref{fig:cloudy_2D_diskir} (upper panels), the extent of the [O~{\sc iii}]$\lambda$5007/\hb\ area is roughly matched although it is too compact with respect to the data, while the peak values are in fair agreement with those observed. For the [O~{\sc i}]$\lambda$6300/\ha\ ratio, although the extent and morphology are well matched, the peak values are overpredicted by a factor 2. This overprediction occurs for most low-ionisation lines (e.g. [S {\sc ii}]$\lambda$6716/\ha) as can be seen in Table~\ref{tab:cloudy_results} and for all models. This can be understood due to a lack of strong soft optical spectra from the background stars, which will excite more readily the Balmer lines and increase them compared to e.g. [O~{\sc i}]$\lambda$6300, which is produced in the outer neutral parts of the nebula via highly-penetrating soft X-rays. The inclusion of the putative companion star does not change this basic conclusion and cannot account for these differences. This prompted us to take into account the contribution from the stellar background.

\begin{figure}
    \centering
    \includegraphics[width=0.49\textwidth]{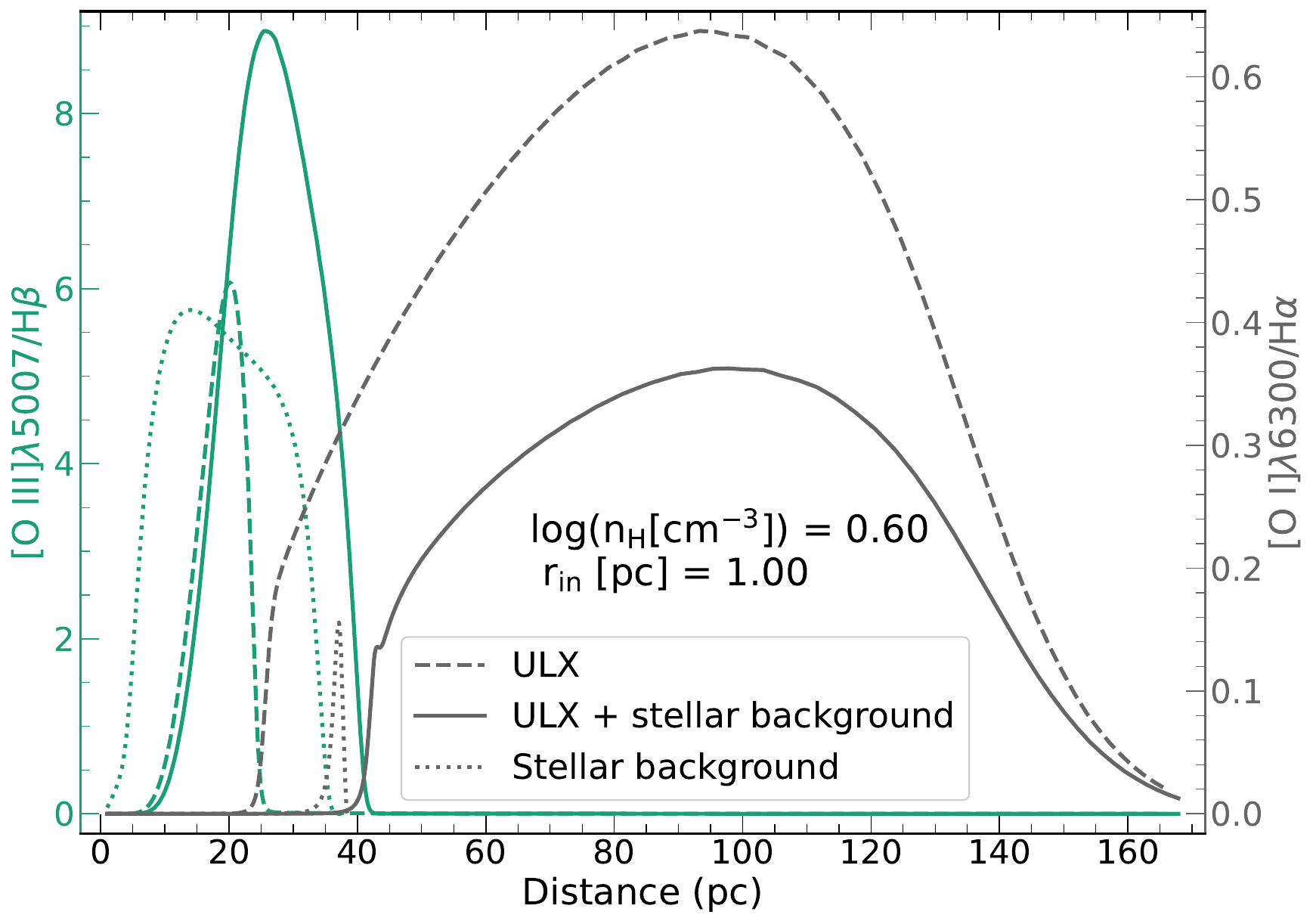}
    \caption{Effects of including the stellar-background on the [O~{\sc iii}]$\lambda$5007/\hb\ (left axis) and [O~{\sc i}]$\lambda$6300/\ha\ (right axis) ratios (shown for the \texttt{phenomenological\_star} model). The stellar background increases the peak [O~{\sc iii}]$\lambda$5007/\hb\ ratio and `widens' the area over which [O~{\sc iii}]$\lambda$5007 is produced, whereas it has the opposite effect on [O~{\sc i}]$\lambda$6300/\ha. These 1D plots are for illustration purposes only, as they are not resampled and smoothed to match MUSE spatial resolution.}
    \label{fig:stellar_background}
\end{figure}

The inclusion of the stellar background not only significantly improved all $\chi^2$ (Fig.~\ref{fig:chisq}; Table~\ref{tab:cloudy_results} for $L = 2.75 \times 10^{40}$ erg/s and Tables~\ref{tab:cloudy_results_lower_stars} for the less luminous stellar-background cases), but now the peak line ratios in low ionisation lines are much closer to the observed values, particularly for the less-UV bright \texttt{diskir} and \texttt{phenomenological} models. To show the effects of adding the stellar background more clearly, Fig.~\ref{fig:stellar_background} shows a 1D comparison between the \texttt{phenomenological} ULX-only model and the same model with the inclusion of the stellar background. The inclusion of the stellar background not only increases the [O~{\sc iii}]$\lambda$5007/\hb\ peak ratio, but also widens the area over which [O~{\sc iii}]$\lambda$5007 is excited. This effect is similar to that observed by \citet{berghea_first_2010, berghea_spitzer_2012} in Holmberg~II~X--1 and NGC~6946~X--1, where it was found that the low-energy photons from the companion star create low-ionisation states which are further ionised by the high-energy ULX photons \citep[Figure 8 in][]{berghea_first_2010}. The effect on the low-ionisation lines is instead the opposite. Both these effects support there is an additional source of ionisation along with the ULX. 

\begin{figure*}
    \centering
    \includegraphics[width=0.98\textwidth]{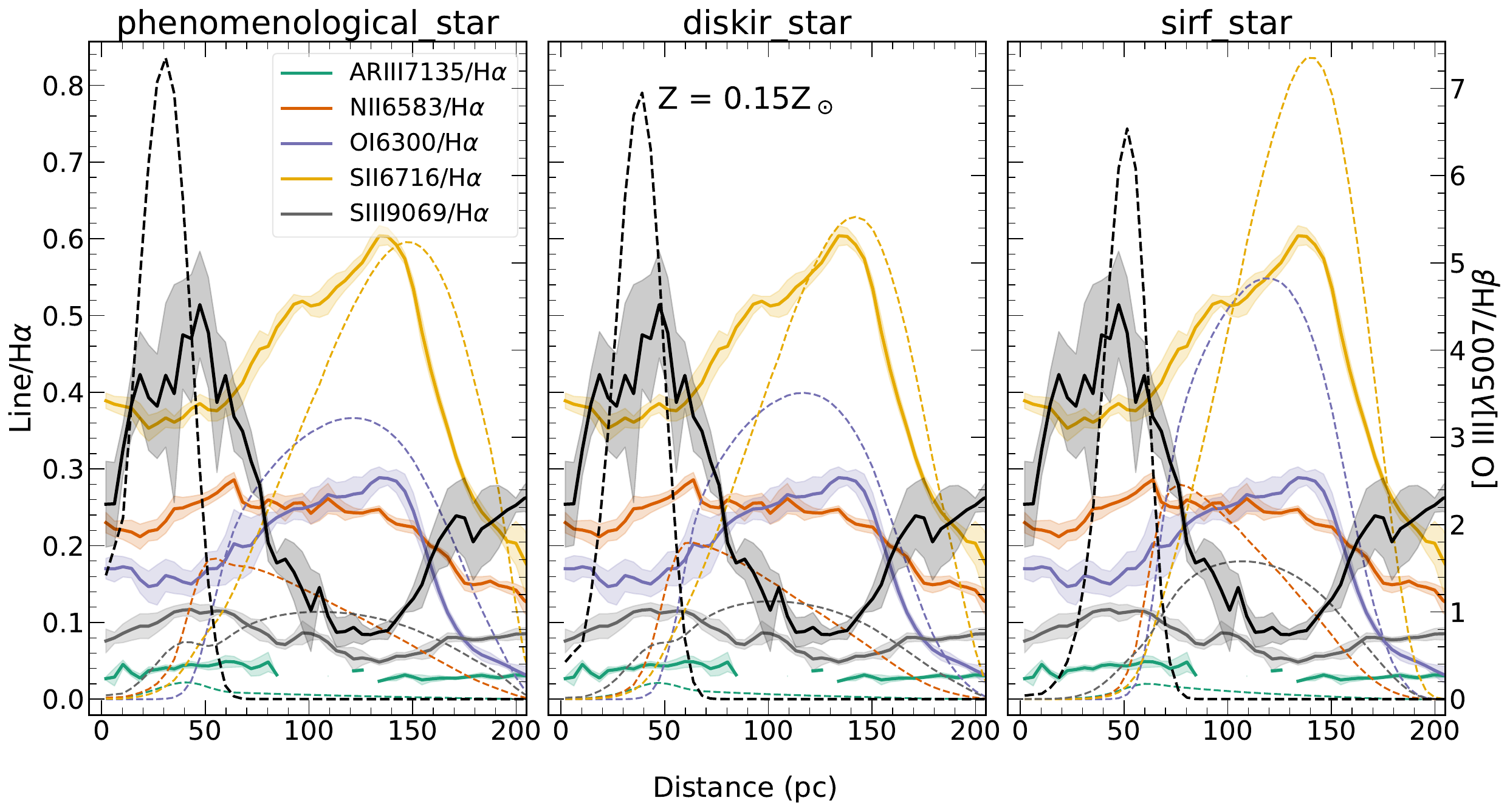}
    \includegraphics[width=0.98\textwidth]{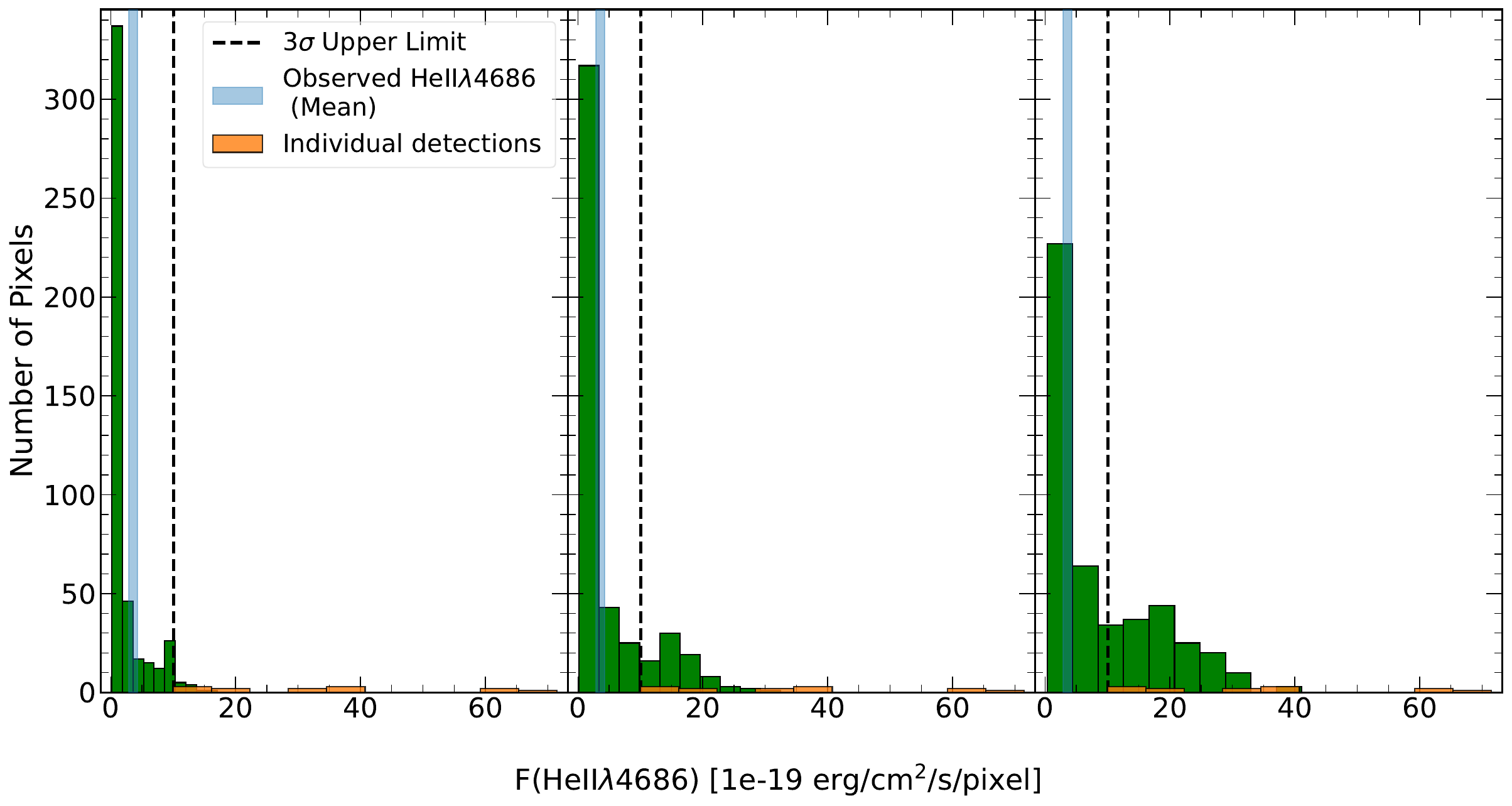}
    \caption{(Top) Comparison of 1D radial profiles extracted from the data (solid and shaded areas) and the best-fit 2D \cloudy\ models (dashed lines), after resampling and smoothing to match MUSE spatial resolution. The profiles were extracted with a width of 10 pixels. For the data, the weighted average and standard deviation are shown as solid lines and dashed regions, respectively. Gaps in the data are due to low signal to noise ratio or bad pixels in those areas. The left axes show the line ratios in low ionisation lines, whereas the right axes (black lines) shows the [OIII]$\lambda$5007/\hb\ ratio. The \texttt{sirf} model overpredicts the low ionization lines in the outer part of the nebula. Note the increase in [O~{\sc iii}]$\lambda$5007/\hb\ towards the end in the data is due to another nearby group of stars. (Bottom) Range of \heii\ flux predicted for each best-fit model (green histograms) averaged over the blue region shown in Figure~\ref{fig:heII}, compared to the average value observed in the data in the same region (blue dashed line showing the $\pm$1$\sigma$ uncertainty) and the 13 individual detections (orange histogram). Both \texttt{diskir} and \texttt{sirf} predict too strong \heii\ compared to the data, particularly considering the 3$\sigma$ upper limit on \heii.}
    \label{fig:1Dcomparison}
\end{figure*}

Despite the widening of the enhanced [O~{\sc iii}]$\lambda$5007/\hb\ region introduced by the stellar background, all models fail to match the width of the observed radial profile. In Fig.~\ref{fig:1Dcomparison} we show a 1D radial profile along the peak of the excitation region comparing the models and the data to illustrate this. The width of the profile was set to 10 pixels to smooth out variations and avoid gaps due low signal-to-noise ratio pixels. Although strictly speaking we compared the 2D generated images with the data, we find these 1D visualization provide an accurate summary of our results. We can see that [O~{\sc iii}]$\lambda$5007/\hb\ is overpredicted in all models while the extent over which [O~{\sc iii}]$\lambda$5007 is produced is underpredicted. We can confirm that part of the reason is due to projection effects. As alluded earlier, here we have assumed that the nebula is thin enough in the line of sight direction such that the line ratios along a given line of sight are approximately constant. Instead, if the nebula has considerable structure in the dimension along the line of sight, for instance in the case of a spherical sector, the line ratios along a given sightline will be averaged over regions with different degree of ionization. This will smooth out the line ratios over a wider region compared to the planar geometry assumed here, hence lowering their peak values and broadening their profiles compared to the line ratio profiles shown in Figure~\ref{fig:1Dcomparison}. We can preliminary confirm these effects from ongoing work, but leave the treatment of more complicated cloud geometries for future work.

Another effect that could be affecting our results is the treatment of the stellar background emission as a point-like source, instead of an extended component. In reality the optical stellar emission will be approximately uniformly distributed over the nebula, both diluting and extending the [O~{\sc iii}]$\lambda$5007/\hb\ region compared to the point-like treatment afforded by \cloudy. This may also explain the wider region of high [N~{\sc ii}]$\lambda$6583/\ha\ in the data compared to the models. 

To inspect whether this is the case, we have run another \cloudy\ simulation with the stellar background alone with a typical $\log(n_\mathrm{H}$[cm$^{-3}$]) = 0.60 to inspect the line ratios it would produce. Fig.~\ref{fig:stellar_background} shows it would produce a ($\sim$30\,pc) with [O~{\sc iii}]$\lambda$5007/\hb\ reaching $\sim$6 dropping to $\sim$5.3 after the PSF smoothing. These values are clearly not typical from HII regions \citep{gurpide_muse_2022} and most likely are an artifact of concentrating all the luminosity in a point-like source. Therefore, we note that our [O~{\sc iii}]$\lambda$5007/\hb\ peak estimates may be partially affected by this and this has to be borne in mind when interpreting the results.

While the fact that the stellar background can produce rather high [O~{\sc iii}]$\lambda$5007/\hb\ ratios may call into question whether the nebula is actually produced by the ULX, we see that the stellar background by itself cannot account for the morphology of the nebula (Fig.~\ref{fig:stellar_background}). In particular, the [O~{\sc iii}]$\lambda$5007/\hb\ emission is concentrated around the point-like source, while in the models including the ULX the gas close to the source is too ionized to produce [O~{\sc iii}]$\lambda$5007. Similarly, the stellar background alone does not produce the extended enhanced [O~{\sc i}]$\lambda6300$/\ha\ region, which can only be produced by the large mean free path of the X-rays. Therefore, our modelling shows undoubtedly that a high-energy source is needed to explain the nebular morphology.

In terms of $\chi^2$, the preferred model is the \texttt{phenomenological}, as it provides the best-overall $\chi^2$. From Fig.~\ref{fig:1Dcomparison} we can see that both the \texttt{diskir} and \texttt{phenomenological} offer comparable level of agreement with the data, while the \texttt{sirf} overpredicts more severely the [O~{\sc i}]$\lambda$6300/\ha\ and [S~{\sc ii}]$\lambda$6716/\ha\ ratios. Despite the differences and the clear more-complex profiles in the data due to the effects mentioned above, the overall structure in most lines is well reproduced. The exception to this is the [S~{\sc iii}]$\lambda$9069/\ha\ (gray line in the plots). While all models peak at around 120\,pc, the observed peak is seen at just 50\,pc from the ULX. We suspect this line might be more strongly affected by the sky subtraction and/or shocks, which may explain this discrepancy. As discussed above, the remaining differences may be attributed to projection effects, the treatment of the stellar background as a point-like source and slight differences in abundances and density profile of the gas. Nevertheless, the peak values in the first two models match those observed in the data (see also Table~\ref{tab:cloudy_results}).

As stated earlier, we have run an additionally set of simulations by lowering the stellar contribution to study the sensitivity of our results to this component. The results are shown in Table~\ref{tab:cloudy_results_lower_stars} and show that the fits worsen in all instances. In particular, we can see that the models now again overpredict most low-ionisation lines by a factor 1.5--2. We conclude that the bright ($L = 2.75 \times$10$^{40}$\,erg) is a better match to the data, with the caveats outlined above. Nevertheless, the \texttt{phenomenological} and \texttt{diskir} continue to be the preferred models regardless of the exact treatment of the background stars. On this basis, we consider the UV flux predicted by the \texttt{sirf} to be too high to match the nebular emission.

To test our final conclusion, based on our best-fit models, we calculated the expected \heii\ flux by multiplying each of the models \heii/H$\beta$ ratios by the observed \hb\ extracted from the datacube \citep{gurpide_muse_2022}. We then extracted the fluxes per spaxel expected from the same blue region shown in Fig.~\ref{fig:heII}. The lower panel of Fig.~\ref{fig:1Dcomparison} shows the range of values expected for each model (averaged over the region), the average $F($\heii) value measured in the data (Figure~\ref{fig:heII}) and the estimated 3$\sigma$ detection limit. To estimate the latter, we inspected the individual spaxels from the blue region in Fig.~\ref{fig:heII} and found the lowest flux at which the 3$\sigma$ error on the flux was consistent with 0. This value was $\sim$10$^{-18}$ \fluxcgs\ and is shown in the Figure as a black dashed line. We further managed to detect the line in 13 individual spaxels although the relative 3$\sigma$ errors are quite high (70--95\%). These detections are also shown in the same Fig.~\ref{fig:1Dcomparison} in the lower panel (orange colored histogram). 

From the histograms in Fig.~\ref{fig:1Dcomparison}, we can see that the \texttt{sirf} model produces $F($\heii) values that are too high with respect the observed values. Instead, the best agreement is again provided by the \texttt{phenomenological} model, because most of the spaxels have values below the detection threshold, as expected based on the lack of strong detections, and the overall histogram shows the best agreement with the data. Therefore, the marginal detection of the \heii\ again reinforces the idea that the predicted UV flux by both the \texttt{sirf} and \texttt{diskir} is overestimated.
\begin{table*}
   \begin{center}
    \caption{\cloudy\ modelling results for the extended nebula for $Z = 0.15Z_{\sun}$.} 
\label{tab:cloudy_results}
\resizebox{\textwidth}{!}{\begin{tabular}{cccccccc} 
\hline 
\hline
Peak Line Ratio & Data & \texttt{phenomenological} & \texttt{phenomenological\_star} & \texttt{diskir} & \texttt{diskir\_star} & \texttt{sirf} & \texttt{sirf\_star}\\ 
\noalign{\smallskip}
\hline
\multicolumn{8}{c}{ULX only}\\
\hline
\noalign{\smallskip}
\oiii/\hb                       & 5.9$\pm$0.3       & 2.18 & 3.48 & 5.82 & 5.32 & 6.32 & 6.03 \\ 
\oi/\ha                       & 0.33$\pm$0.02       & 0.64 & 0.63 & 0.63 & 0.63 & 0.62 & 0.62 \\ 
{[N{\sc ii}]}$\lambda$6583/\ha                       & 0.30$\pm$0.03       & 0.36 & 0.35 & 0.32 & 0.32 & 0.31 & 0.31 \\ 
{[S~{\sc i}]}$\lambda$6716/\ha                       & 0.62$\pm$0.01       & 0.99 & 1.01 & 0.99 & 0.99 & 0.96 & 0.96 \\ 
{[S~{\sc i}]}$\lambda$6716/{[S~{\sc ii}]}$\lambda$6731                         & 1.45$\pm$0.06       & 1.47 & 1.47 & 1.47 & 1.47 & 1.47 & 1.47 \\ 
{[Ar~{\sc iii}]}$\lambda$7135/\ha                       & 00.065$\pm$0.008       & 0.02 & 0.02 & 0.02 & 0.02 & 0.02 & 0.02 \\ 
{[S~{\sc iii}]}$\lambda$9068/\ha                       & 0.135$\pm$0.017       & 0.20 & 0.20 & 0.20 & 0.20 & 0.20 & 0.20 \\ 
\noalign{\smallskip} 
\hline 
\noalign{\smallskip}
$\log(n_\mathrm{H}$[cm$^{-3}$])        & --            &0.55 & 0.50 & 0.55 & 0.55 & 0.75 & 0.75 \\
$r_\mathrm{in}$ (pc)                   & --            &1.00 & 1.00 & 12.48 & 1.00 & 1.00 & 1.00 \\
$\chi^2$ / dof                         & --            &166 & 163 & 169 & 170 & 176 & 175 \\

\noalign{\smallskip}
\hline
\noalign{\smallskip}
\multicolumn{8}{c}{ULX + Stellar background, $L = 2.75 \times10^{40}$\luxcgs} \\
\noalign{\smallskip}
\hline
\noalign{\smallskip}
\oiii/\hb                       & 5.9$\pm$0.3       & 7.13 & 7.50 & 7.41 & 7.10 & 6.83 & 6.65 \\ 
\oi/\ha                       & 0.33$\pm$0.02       & 0.38 & 0.37 & 0.40 & 0.40 & 0.48 & 0.55 \\ 
{[N{\sc ii}]}$\lambda$6583/\ha                       & 0.30$\pm$0.03       & 0.20 & 0.19 & 0.20 & 0.21 & 0.25 & 0.28 \\ 
{[S~{\sc i}]}$\lambda$6716/\ha                       & 0.62$\pm$0.01       & 0.62 & 0.60 & 0.62 & 0.63 & 0.72 & 0.84 \\ 
{[S~{\sc i}]}$\lambda$6716/{[S~{\sc ii}]}$\lambda$6731                       & 1.45$\pm$0.06       & 1.47 & 1.47 & 1.47 & 1.47 & 1.47 & 1.47 \\ 
{[Ar~{\sc iii}]}$\lambda$7135/\ha                       & 0.065$\pm$0.008       & 0.02 & 0.02 & 0.02 & 0.02 & 0.02 & 0.02 \\ 
{[S~{\sc iii}]}$\lambda$9068/\ha                       & 0.135$\pm$0.017       & 0.12 & 0.11 & 0.13 & 0.13 & 0.16 & 0.18 \\ 
\noalign{\smallskip} 
\hline 
\noalign{\smallskip}
$\log(n_\mathrm{H}$[cm$^{-3}$])        & --            &0.50 & 0.45 & 0.55 & 0.55 & 0.75 & 0.75 \\
$r_\mathrm{in}$ (pc)                   & --            &1.00 & 12.48 & 1.00 & 1.00 & 12.48 & 1.00 \\
$\chi^2$ / dof                         & --            &85 & 87 & 92 & 91 & 128 & 148 \\
\noalign{\smallskip}
   \hline
   \hline
   \end{tabular}
}
  \end{center}
\begin{minipage}{\linewidth}
    \textbf{Notes.} $^a$To calculate the reduced $\chi^2$, the number of degrees of freedom is defined as the number of pixels covering the photo-ionized region (11474) minus 2 variables ($n_\mathrm{H}$ and $r_\mathrm{in}$). These $\chi^2$ need to be understood as an heuristic to rank the models, rather than a goodness-of-fit.
    \end{minipage}
\end{table*}

\subsection{The Front View}\label{sub:front_view}

\begin{figure*}
    \includegraphics[width=0.49\textwidth]{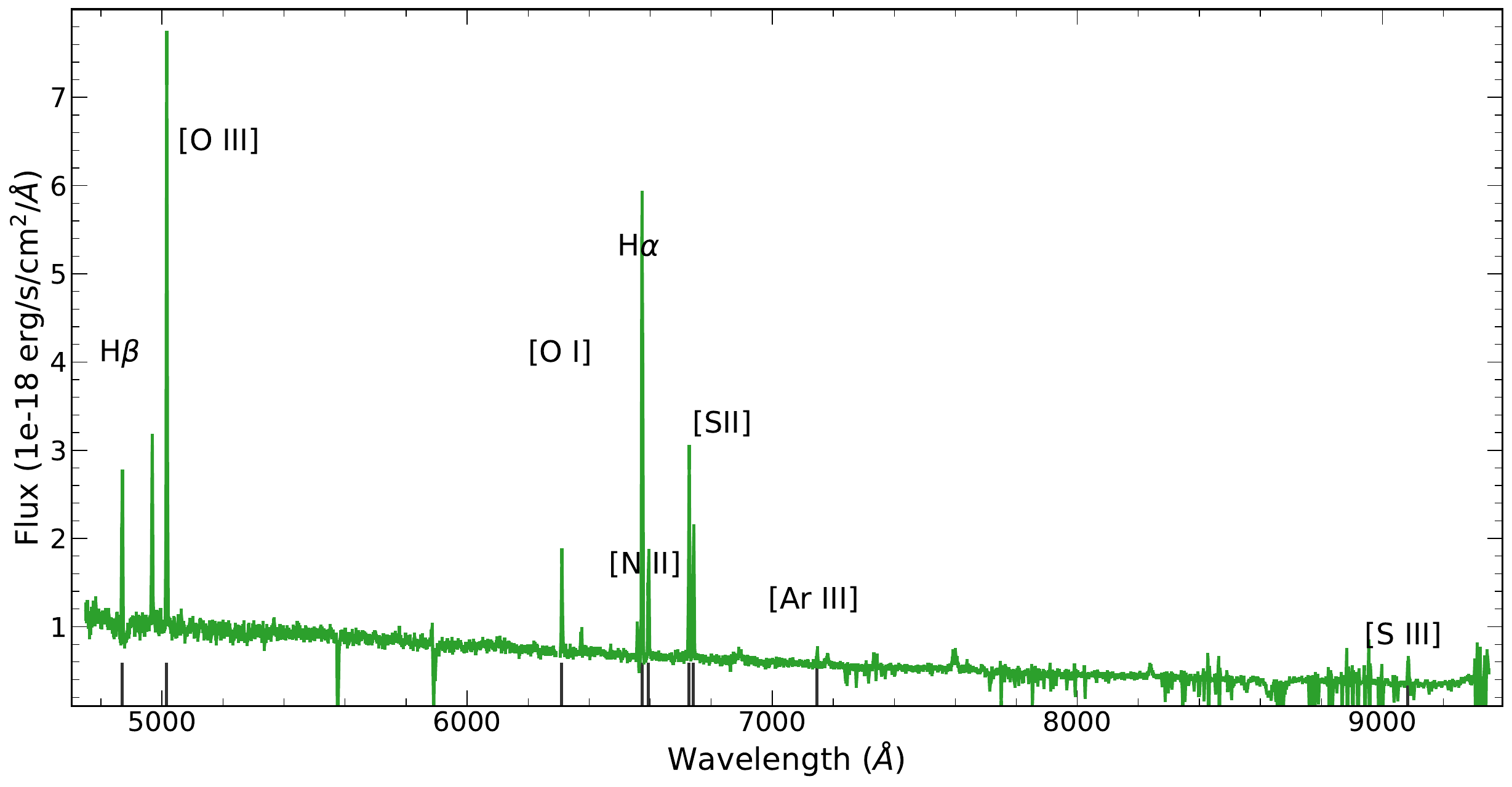}
    \includegraphics[width=0.49\textwidth]{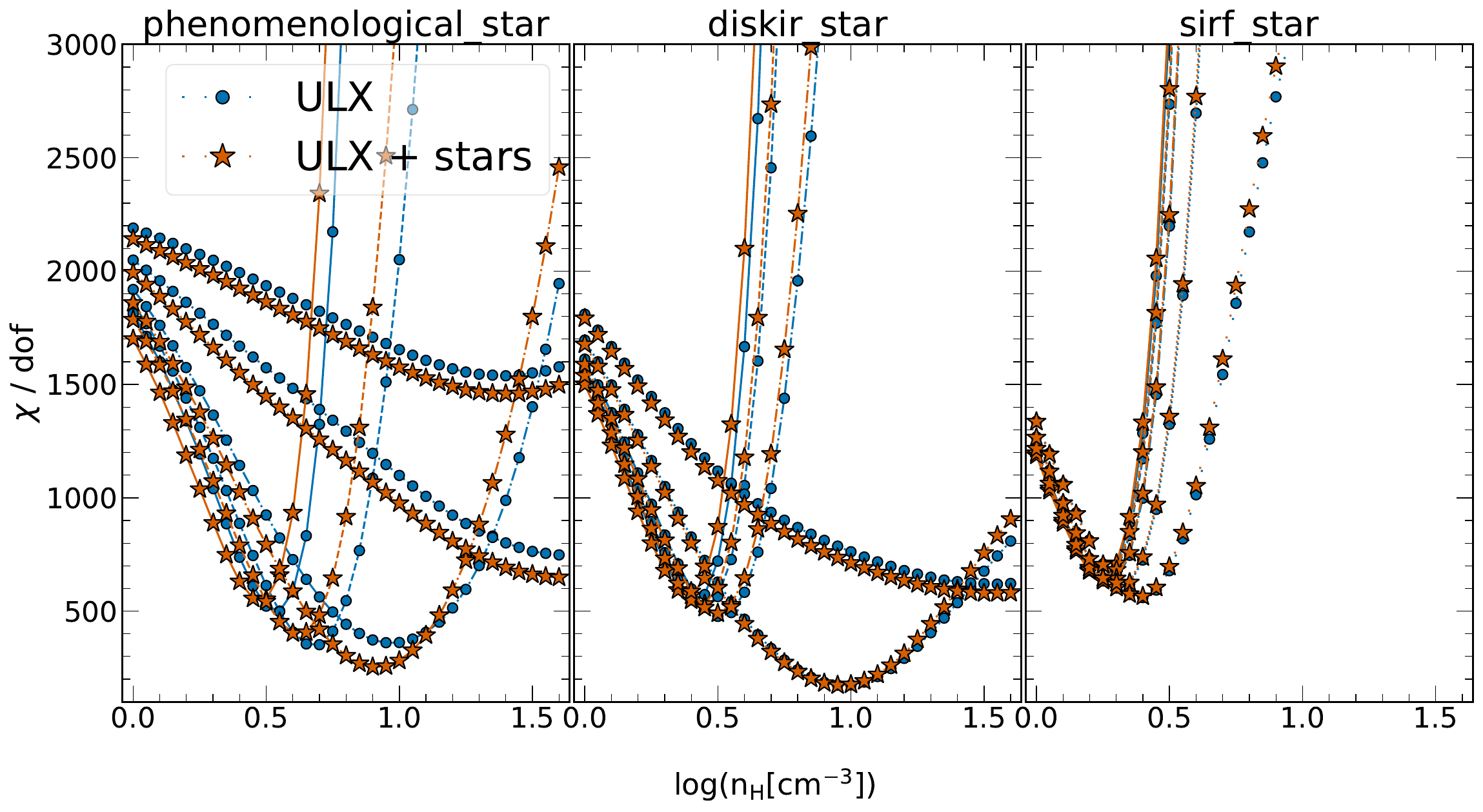}
    \caption{\cloudy\ analysis of the nebular spectrum along the line of sight. (Left) Spectrum extracted around \theulx, averaged over a circular region of 1" in radius roughly matching the cube PSF FWHM. Lines used in the analysis are labelled and their expected position based on the redshift of NGC~1313 are indicated by a black vertical tick. The spectrum has been resampled by a factor 2 compared to MUSE 1.25 \AA\ spectral sampling. (Right) Reduced $\chi^2$ obtained for the three photo-ionization models with and without the stellar background ($L = 2.75\times10^{39}$ erg/s; Z = 0.15Z$_{\sun}$). Lines and symbols as per Figure~\ref{fig:chisq}, including now the results for $r_\mathrm{in} = 60$ pc with loosely dotted lines.} \label{fig:cloudy_los}
\end{figure*}

In \cite{gurpide_muse_2022} we showed there is a lack EUV/X-ray photo-ionization signatures around \theulx\ in other directions, most likely due to a lower-ISM density in those areas. However, as can be appreciated from the BPT diagrams presented in \cite{gurpide_muse_2022} (their Figure 11) the [O~{\sc iii}]$\lambda$5007/\hb\ ratio is slightly enhanced ($\sim$3.2) at the position of \theulx. The enhanced [O~{\sc iii}]$\lambda$5007/\hb\ ratio can also be observed in Figure~\ref{fig:cloudy_los} (left panel), where we show the extinction-corrected spectrum extracted in Section~\ref{sub:extinction} in order to measure the extinction towards \theulx. More specifically, from this spectrum we measured [O~{\sc iii}]$\lambda$5007/\hb\ = 2.93$\pm1.2$, {([S~{\sc ii}]$\lambda$6716 + [S~{\sc ii}]$\lambda$6731)/\ha} = 0.65$\pm$0.01 and \oi/\ha\ = 0.176$\pm$0.006, which again are rather unusual for H{\sc ii} regions. We recall this is the average spectrum per pixel from a circular region of 1" in radius around the source and therefore can be considered to be from a spatial scale of a single MUSE pixel (0.2" or $\sim$4 pc at 4.25 Mpc).

While these ratios are not as extreme as in its vicinity (Section~\ref{sub:sideview}), this is to be expected if we are observing the integrated emission of the photo-ionized nebula along the line of sight. Particularly, should the supercritical funnel in \theulx\ be orientated towards us, then it may be reasonable to consider it as a potential source of ionisation. The lack of clear diagnostics, such as the extent of the \oiii/\hb\ and \oi\/\ha\ regions, makes attributing these enhanced ratios to photo-ionization by the ULX more uncertain (for instance the high ([S~{\sc ii}]$\lambda$6716 + [S~{\sc ii}]$\lambda$6731)/\ha\ ratio could be due to shocks, although we have also seen these ratios are also produced by EUV/X-ray photo-ionization; Figure~\ref{fig:1Dcomparison}). Therefore as a first step we considered whether we could reproduce these ratios as photo-ionization by stellar continua.

To this end, we attempted to reproduce the fluxes from the lines in the spectrum above -- indicated in the Figure with black tick underneath -- using photo-ionization models. As in Section~\ref{sub:extinction}, in order to extract line fluxes we fitted the lines using a Gaussian and a constant for the local continuum. Extinction-corrected fluxes (averaged over the extraction region) for all lines of interest are reported in Table~\ref{tab:cloudy_los}. The flux of \heii\ was derived using the same region but from cube 2. The 3$\sigma$ negative error was consistent with zero, therefore we considered this measurement an upper limit. Its extinction-corrected flux and the 1$\sigma$ uncertainty are also reported in Table~\ref{tab:cloudy_los}. These results were consistent with those obtained using instead a smaller aperture of 0.5\arcsec in radius.


We then compared the observed line fluxes from this spectrum to the fluxes obtained by integrating the \cloudy\ nebulae along the line of sight using again $\chi^2$ statistics. The upper limit on \heii\ was taken into account in the modelling following \cite{hoof_photo-ionization_1997}:
\begin{equation}
    \chi^2 = \left ( \frac{\text{max}(F_\text{obs}, F_\text{model}) - F_\text{obs})} {\sigma_{F_\text{obs}}} \right)^2 
\end{equation}
where $F_\text{obs}$ and $F_\text{model}$ refer to the observed and predicted (\heii) fluxes and $\sigma_{F_\text{obs}}$ is the (1$\sigma$) uncertainty on $F_\text{obs}$.

We assumed we only see nebular emission due to gas located between the ULX and ourselves (i.e. no contribution from an equally-extended nebula behind the ULX). Assuming we only observe the gas in front of the ULX may be reasonable considering much of the emitting nebular gas behind the ULX will be absorbed by the system itself. However, such fraction may be small considering we are observing gas averaged (or integrated) over an angular (physical) diameter of 2\arcsec\ ($\sim$40 pc). Assuming a symmetric case, where the gas is equally distributed in front and behind the ULX, is not expected to affect the results significantly as the effect would be to simply rescale the line fluxes for a given density/inner radius by a factor 2, without altering the predicted line ratios. We have rerun the calculation under this assumption and verified that while there are small numerical differences between the two approaches, our overall conclusions below do not change. We present the results for the former assumption (termed asymmetric case) and discuss the calculations for the latter assumption (symmetric case hereafter) when relevant.

We performed the data-model comparison following two approaches and found very good agreement between them ($\Delta\chi^2 < 6$ for all models), so we only present the results from the latter approach. In the first method, we considered the averaged fluxes over the spatial region and assumed we are observing a column of $p\times p\times r_\mathrm{out}$, where $p$ is the physical size of a MUSE pixel at $D = 4.25$ Mpc and $r_\mathrm{out}$ is the (unknown) outer edge of the nebula along the line of sight. We found best results when considering all models to be ionization-bounded. Therefore the nebula outer radius $r_\mathrm{out}$ was set to the maximum of the \cloudy\ calculation ($r_\mathrm{out} = 200$ pc) or lower in cases of higher density where the hydrogen ionization front was reached before. In the second approach, instead of considering the average spectrum over the 1" aperture, we have instead used the integrated one. We then have compared the integrated line fluxes to those obtained in \cloudy\ by using the \texttt{aperture} command to integrate the simulated nebula over the same spatial region.

We noticed some of the models $\chi^2$ minima were at the limit of our density calculation ($\log(n_\mathrm{H}[$cm$^{-3}$]) = 1), so we extended the calculations to densities up to $\log(n_\mathrm{H}[$cm$^{-3}$]) = 1.6 and $r_\mathrm{in} = 60$ pc to ensure we found the absolute minima for each model.

\begin{table*}
   \begin{center}
    \caption{\cloudy\ modelling results for the emission-line nebula along the line of sight for $Z = 0.15$Z$_{\sun}$. Line fluxes are extinction-corrected and averaged over an extraction region of 1\arcsec.}
    \label{tab:cloudy_los}
    \resizebox{\textwidth}{!}{\begin{tabular}{cccccccc} 
    \hline\hline
    \noalign{\smallskip}
    Line & Data & \texttt{phenomenological} & \texttt{phenomenological\_star} & \texttt{diskir} &
    \texttt{diskir\_star} & \texttt{sirf} & \texttt{sirf\_star} \\
    & 10$^{-20}$ \fluxcgs &  Data/Model & Data/Model & Data/Model& Data/Model& Data/Model & Data/Model \\
    \noalign{\smallskip}
    \hline
    \noalign{\smallskip}
\heii$^a$                    & $125\pm70$  & 0.45  & 0.46  & 0.70  & 0.68  & 0.40  & 0.39  \\
\hb\                         & 776$\pm$29  & 0.99  & 0.92  & 0.91  & 0.98  & 1.04  & 0.99  \\
{[O~{\sc iii}]}$\lambda$5007 & 2279$\pm$42 & 1.51  & 0.96  & 1.29  & 1.42  & 0.82  & 0.78  \\
{[O~{\sc i}]}$\lambda$6300   & 392$\pm$12  & 1.52  & 1.56  & 1.09  & 1.16  & 6.03  & 4.76  \\
\ha\                         & 2220$\pm$22 & 0.88  & 0.84  & 0.86  & 0.93  & 0.99  & 0.94  \\
{[N{\sc ii}]}$\lambda$6583   & 500$\pm12$  & 1.09  & 1.20  & 0.88  & 0.94  & 3.13  & 2.70  \\
{[S~{\sc i}]}$\lambda$6716   &  831$\pm$15 & 4.80  & 4.24  & 1.76  & 1.97  & 9.75  & 8.06  \\
{[S~{\sc ii}]}$\lambda$6731  & 602$\pm$19  & 5.03  & 4.44  & 1.84  & 2.06  & 10.15 & 8.40  \\
{[Ar~{\sc iii}]}$\lambda$7135& 63$\pm$9    & 1.70  & 1.69  & 1.18  & 1.28  & 2.06  & 1.85  \\
{[S~{\sc iii}]}$\lambda$9068 & 128$\pm10$  & 0.79  & 0.78  & 0.52  & 0.56  & 1.11  & 0.98  \\
\noalign{\smallskip}
\hline
\noalign{\smallskip}
\nh[10$^{21}$ cm$^{-2}$]$^c$ & 1.7$^{+0.3}_{-0.6}$ & 2.0   & 2.0   & 2.1   & 2.0   & 1.0   & 1.0 \\
$\log(n_\mathrm{H}[$cm$^{-3}$])  &  --             & 0.65  & 0.70  & 1.00  & 0.95  & 0.35  & 0.40 \\
$r_\mathrm{in}$[pc]   &  --                        & 12.48 & 12.48 & 40.0 & 40.0   & 60.0 & 60.0\\
$r_\mathrm{out}$[pc]  &  --                        & 158   & 141   & 106   & 113   & 200   & 200 \\
$\chi^2$/dof$^b$          &  --                    & 386   & 353   & 164   & 175   & 585   & 561   \\
\noalign{\smallskip}
\hline
\noalign{\smallskip}
\multicolumn{8}{c}{ULX + stellar background ($L = 2.75\times$10$^{39}$ erg/s)} \\
\noalign{\smallskip}
\hline
\noalign{\smallskip}
\heii\                       & $125\pm70$  & 0.82   & 0.92  & 0.68  & 0.67 & 0.40  & 0.39  \\
\hb\                         & 776$\pm$29   & 1.13  & 1.07  & 0.89  & 0.93 & 1.03  & 0.98  \\
{O~{\sc iii}]}$\lambda$5007  & 2279$\pm$43  & 2.69  & 1.87  & 1.08  & 1.27 & 0.82  & 0.77  \\
{[O~{\sc i}]}$\lambda$6300   & 392$\pm$12   & 1.05  & 1.06  & 1.22  & 1.18 & 6.07  & 4.79  \\
\ha\                         & 2220$\pm$22  & 1.02 &  0.98  & 0.85  & 0.89 & 0.99  & 0.93  \\
{[N{\sc ii}]}$\lambda$6583   & 500$\pm12$   & 0.88  & 0.89  & 0.94  & 0.93 & 3.14  & 2.71  \\
{[S~{\sc i}]}$\lambda$6716   &  831$\pm$15  & 2.67  & 2.36  & 1.95  & 1.96 & 9.74  & 8.06  \\
{[S~{\sc ii}]}$\lambda$6731  & 602$\pm$19   & 2.80  & 2.47  & 2.04  & 2.05 & 10.15 & 8.40  \\
{[Ar~{\sc iii}]}$\lambda$7135& 63$\pm$9     & 1.56  & 1.46  & 1.15  & 1.22 & 2.05  & 1.84  \\
{[S~{\sc iii}]}$\lambda$9068 & 128$\pm10$   & 0.62  & 0.59  & 0.53  & 0.55 & 1.11  & 0.98  \\
    \noalign{\smallskip}
\hline
\noalign{\smallskip}
\nh[10$^{21}$ cm$^{-2}$] & 1.7$^{+0.3}_{-0.6}$ & 2.1   & 2.1   & 2.0   & 2.0   & 1.0   & 1.1 \\
$\log(n_\mathrm{H}[$cm$^{-3}$])     &  --      & 0.85  & 0.90  & 0.95  & 0.95  & 0.35  & 0.40  \\
$r_\mathrm{in}$          &  --                 & 22.34 & 22.34 & 40.0  & 40.0  & 60.0  & 60.0 \\
$r_\mathrm{out}$[pc]     &  --                 & 119   & 108   & 113   & 113   & 200   & 200 \\
$\chi^2$/dof             &  --                 & 345   & 254   & 178   & 174   & 587   &  567 \\
\noalign{\smallskip}
\hline
\noalign{\smallskip}
\multicolumn{8}{c}{ULX + stellar background ($L = 2.75\times$10$^{40}$ erg/s)}\\
\noalign{\smallskip}
\hline
\noalign{\smallskip}
\heii\                        & $125\pm70$  & 2.31 & 2.75  & 1.71  & 1.79  & 0.40 & 0.42  \\
\hb\                          & 776$\pm$29  & 1.08 & 1.05  & 1.14  & 1.15  & 0.98 & 1.01  \\
{[O~{\sc iii}]}$\lambda$5007  & 2279$\pm$43 & 3.54 & 3.89  & 4.56  & 7.83  & 0.76 & 0.79 \\
{[O~{\sc i}]}$\lambda$6300    & 392$\pm$12  & 1.23 & 1.26  & 1.40  & 1.24  & 6.47 & 6.30   \\
\ha\                          & 2220$\pm$22 & 0.98 & 0.96  & 1.11  & 1.11  & 0.95 & 0.99 \\
{[N{\sc ii}]}$\lambda$6583    & 500$\pm12$  & 1.04 & 1.04  & 1.10  & 1.09  & 3.22 & 3.21  \\
{[S~{\sc i}]}$\lambda$6716    &  831$\pm$15 & 1.83 & 1.72  & 1.70  & 1.50  & 9.75 & 9.70   \\
{[S~{\sc ii}]}$\lambda$6731   & 602$\pm$19  & 1.91 & 1.79  & 1.77  & 1.57  & 10.15& 10.10   \\
{[Ar~{\sc iii}]}$\lambda$7135 & 63$\pm$9    & 1.74 & 1.75  & 1.90  & 2.05  & 1.99 & 2.03   \\
{[S~{\sc iii}]}$\lambda$9068  & 128$\pm10$  & 0.70 & 0.72  & 0.85  & 0.87  & 1.08 & 1.10  \\
    \noalign{\smallskip}
\hline
    \noalign{\smallskip}
\nh[10$^{21}$ cm$^{-2}$] & 1.7$^{+0.3}_{-0.6}$ & 2.0  & 1.9   & 1.6   & 1.6   & 1.0   & 1.0 \\
$\log(n_\mathrm{H}[$cm$^{-3}$]) &  --          & 1.05 & 1.10  & 1.15  & 1.25  & 0.35  & 0.35 \\
$r_\mathrm{in}$[pc]  &  --                     & 40.0 & 40.0  & 60.0  & 60.0  & 60    & 60.0\\
$r_\mathrm{out}$[pc] &  --                     & 95   & 88    & 97    & 88    & 200   & 200 \\
$\chi^2$/dof         &  --                     & 294  & 294   & 327   & 347   & 614   &  597 \\
\noalign{\smallskip}
\hline \hline
\noalign{\smallskip}
\end{tabular}}
\end{center}
\begin{minipage}{\linewidth}
    \textbf{Notes.} Uncertainties at the 1$\sigma$ level. \\
    $^a$3$\sigma$ upper limit along with the 1$\sigma$ uncertainty. \\
    $^b$Degrees of freedom are defined as 10 (lines) -- 2 variables ($\log(n_\mathrm{H}$) and $r_\mathrm{in}$). \\
    $^c$Extra-galactic neutral absorption column derived from spectral fitting (quoted from the \texttt{diskir} model; Table~\ref{tab:sed_modelling}) i.e. the Galactic contribution along the line of sight \citep[$n_\mathrm{H}^\mathrm{G}$ = 7.07$\times10^{20}$ cm$^{-2}$;][]{hi4pi_collaboration_hi4pi_2016} has been subtracted.
\end{minipage}
\end{table*}

As stated above, we first verified whether we could explain the line fluxes as photo-ionized by a population of stars by running models with the stellar background alone. In particular, we have run models for stellar backgrounds with $L = 2.75\times10^{38,39,40}$ erg/s for the range of densities and radii quoted above. We have found $\chi^2$/dof upwards of 470 in all instances (also for the symmetric case), with unusually lowly extended ($\lesssim$ 30 pc) nebulae due to the hydrogen ionization front being reached very close to the source as a result of the gas being optically thick to the soft radiation. Moreover, sulfur or \oi\ lines were predicted to be $\gtrsim$5 and $\gtrsim$ 15 lower than observed, respectively. This agrees with the fact that these pixels were classified as `AGN' in the BPT diagram presented in \cite{gurpide_muse_2022}. Therefore, we considered whether we could put further constraints on the SED of the ULX along the line of sight by modelling the nebular emission in this direction as photo-ionization by the ULX instead.

Table~\ref{tab:cloudy_los} shows the obtained data/model line ratios and resulting $\chi^2$ for all models, including those for which the stellar contribution discussed in Section~\ref{sub:sideview} is added to the ULX SEDs. Figure~\ref{fig:cloudy_los} (right panel) shows the $\chi^2$ contours obtained for the three models. As opposed to the extended nebula (Section~\ref{sub:sideview}), where the inclusion of the stellar background clearly improves the results, here most models worsen or show little improvement when the background stars are added. It is also clear that best results are obtained for the non-background or moderate ($L = 2.75\times10^{39}$ erg/s) background cases (Table~\ref{tab:cloudy_los} and Figure~\ref{fig:cloudy_los}, right panel). Similar trend was found for the symmetric case. This is reasonable considering here we are modelling a single sight line, where the major (or only) contributor is likely to be the ULX. The best match is obtained for the \texttt{diskir} model alone (or with moderate background stars) and suggest the nebula is denser ($n_\mathrm{H} \sim 9$ cm$^{-3}$) and $\sim$112 pc long along the line of sight. Expectedly, the additional gas implicitly present in the symmetric case instead lowers the required cloud density, to $n_\mathrm{H} \sim$ 5.6 cm$^{-3}$. Note that although the reduced $\chi^2$ is high, almost all lines are predicted within a factor 2 or less with the \texttt{diskir} model. Thus the ULX clearly provides a much better match to the line fluxes than the stellar continua above. The worst agreement is found for the sulfur lines. As alluded in the previous Section, we suspect [S~{\sc iii}]$\lambda$9068 is likely affected by the sky subtraction. The discrepancy between the model predictions and the other sulfur lines is less clear cut, but may be attributed either to some contribution from shocks and/or to a higher abundance of sulfur.

Along with the predicted line fluxes and $\chi^2$, in Table~\ref{tab:cloudy_los} we also report the integrated hydrogen absorption column along the line of sight, together with that derived from spectral fitting (Section~\ref{sec:multi_band_sed}). Most derived \nh\ values from the photo-ionization modelling, although not unrealistic, are higher than those derived from X-ray spectral fitting after subtraction of the galactic contribution to the total \nh. In principle, we would expect \nh\ derived from spectral fitting to be higher than that of the nebula due to additional contribution from the system itself. One possibility is that indeed there is some contribution to the nebular lines from behind the ULX. For the symmetric cases, \nh\ is reduced approximately by $\sim (0.3-0.4)\times10^{21}$ cm$^{-3}$, which would bring them to a level below that derived from spectral fitting, arguably in more reasonable agreement with the X-ray data. It is also likely that there are additional sources of uncertainty due to the exact underlying model and the fact that we have not included the exact (unknown) abundances into account when deriving \nh\ from spectral fitting. Therefore these values should be taken with caution. Nevertheless, our analysis suggest most contribution to the absorption column may be due to the photo-ionized nebula itself.

While our photo-ionization modelling here is more uncertain due to additional contribution from possible shocks and the unknown extent of the nebula, we can confidently rule out the \texttt{sirf} model as a good line-of-sight SED, as it cannot account for the low-ionisation lines (as they are all underestimated by a factor 3 or more) while producing too strong \heii. At the same time, it suggest our best-fit line of sight SED (the \texttt{diskir}; Section~\ref{sec:multi_band_sed}) is not unrealistic based on the nebular line fluxes observed along the line of sight. Therefore, while we cannot rule some contribution from shocks, we can at least ascertain that a high-energy source (namely the ULX) is a good match to the line fluxes if these are to be explained by a photo-ionization and that these are best explained by the \texttt{diskir} model.

\section{Discussion}\label{sec:discussion}

Through multi-band spectroscopy and modelling of the nebular emission we have been able to confirm our previous assertion \citep{gurpide_muse_2022} that \theulx\ powers an extended $\sim$200\,pc EUV/X-ray photo-ionized region, with additional contribution from the stars in the field. Using state-resolved multi-band spectroscopic data, we have constrained the SED of the ULX along the line of sight. The best-fit model, capable of explaining both the high-energy tail ($>$10 keV) and the UV/optical data simultaneously is the \texttt{diskir} model (regardless of whether we add any contribution from the putative stellar counterpart; Section~\ref{sec:multi_band_sed}). This result agrees with earlier works that have found that such model, despite not describing the accretion flow geometry envisioned for a supercritically accreting compact object \citep[e.g.][]{lipunova_supercritical_1999, poutanen_supercritically_2007, abolmasov_optically_2009}, fits ULX broadband data satisfactorily \citep[e.g.][]{kaaret_photoionized_2009, grise_optical_2012, berghea_spitzer_2012, tao_nature_2012}. We do not however take this as evidence for \theulx\ being powered by a standard self-irradiated disc -- an unlikely scenario owing to the unusual X-ray spectral shape \citep{bachetti_ultraluminous_2013}, the presence of relativistic outflows \citep{pinto_resolved_2016} and the $\sim$400 pc shock-ionized bubble surrounding the source \citep{gurpide_muse_2022}. Instead, we consider the \texttt{diskir} as a physically reasonable extrapolation to the unaccessible EUV. Such extrapolation to the UV -- although uncertain -- is supported by our modelling of the nebula along the line of sight (Section~\ref{sub:front_view}). We have further shown that direct extrapolation of X-ray only models (the \texttt{phenomenological} model; Section~\ref{sec:multi_band_sed}) fail to account for the full SED when the optical data is considered, an issue highlighted also in previous works \citep{abolmasov_optical_2008, dudik_spitzer_2016}.

Through photo-ionization modelling of the extended emission-line nebula (Section~\ref{sub:sideview}) we have instead attempted to constrain the SED seen by the nebula through a different sight line. The fact that the extended [O ~{\sc iii}]$\lambda$5007 and [O~{\sc i}]$\lambda$6300 emission do not overlap (Figures~\ref{fig:cloudy_2D_diskir} and ~\ref{fig:stellar_background}) is a strong indication that the nebula sees the accretion flow in \theulx\ sideways. Therefore NGC~1313~X--1 and its EUV/X-ray excited nebula offers us an opportunity to study super-Eddington accretion flows effectively from two different sightlines. Here we have shown that the nebular lines are best described with a model with a lower UV flux that than constrained along the line of sight. In particular, the best match to the nebular lines is provided by the \texttt{phenomenological} model, whose UV flux is about a factor 4 lower compared to the \texttt{diskbb} (Table~\ref{tab:sed_modelling}). This model not only provides the best match to the nebular lines (Table~\ref{tab:cloudy_results}) but also accounts for the lack of strong nebular \heii\ detection (Fig.~\ref{fig:1Dcomparison}). One may argue that similar results were found by \citet{berghea_first_2010, berghea_spitzer_2012} in Holmberg~II~X--1 and NGC~6946~X--1, wherein it was found that the models with the lowest UV levels were a better match to the nebular lines (their Figures 4 and 3 respectively or their PLMCD and MBC models respectively)-- although we caution that these works did not use spatially-resolved data, which is needed if the degree of beaming is to be determined.

Below we elaborate on how these findings, that is, the discrepancy between the best-fit line-of-sight SED (with $L_\text{UV} \sim 2\times10^{39}$ erg/s) and the best-fit nebular model ($L_\text{UV} \sim 0.5 \times10^{39}$ erg/s) allows us to put constraints on the degree of anisotropy of the UV emission in the ULX \theulx.

\subsection{Quasi-isotropic UV emission}

Our results may suggest the extended nebula is not seeing the same SED as that derived from the line-of-sight. That is, our observations may be interpreted as evidence for a mild degree of anisotropy in the emission of \theulx. Because of the nature of the emission lines used in this work, with the highest ionisation potentials falling in the UV band (Figures~\ref{fig:multiband_sed} and ~\ref{fig:multiband_sed_star}), our observations constrain the degree of anisotropy mostly in the EUV, and any extrapolation to the soft/hard X-rays remains more speculative. This can be observed, for instance, by the fact that the \heii\ line is insensitive to the X-rays ($\gtrsim$0.1 keV), as the three models predict vastly different \heii\ fluxes (Fig.~\ref{fig:1Dcomparison}) despite all having similar X-ray luminosities constrained by the data (Table~\ref{tab:sed_modelling}). Observations probing higher excitation lines found in the IR \citep{berghea_first_2010,berghea_spitzer_2012} will allow to put tighter constrains and extend our measurements to higher energies. 

Although below we provide quantitative calculations for the degree of beaming based on our results, we would like to highlight that there are inevitably additional sources of uncertainty we cannot account for, potential contribution of shocks, other SED extrapolations we have not tested for and the treatment of the stellar background (we discuss these in more detail below in Section~\ref{sub:caveats}). Nevertheless, while the quantitative details may be uncertain, we believe our results can be confidently interpreted as a lack of strong anisotropy in the UV emission. 

The differences in UV luminosity between the best-fit line-of-sight SED (\texttt{diskir}) and the nebular one (\texttt{phenomenological}) may be used to constrain the beaming factor $b$ proposed by \citet{king_ultraluminous_2001}. \citet{king_ultraluminous_2001} defines the beaming factor as:
\begin{equation}
    b = \frac{L}{L_\mathrm{sph}}
\end{equation}
where $L$ is the true emitted radiative luminosity and $L_\mathrm{sph}$ is the observed luminosity under the assumption of isotropic emission. Crucially, this factor must also depend on the inclination of the system and energy. The dependence of $b$ with the inclination complicates estimating this value quantitatively from an observational point of view, particularly due to the uncertain inclination of the ULX with respect to our line of sight and the nebula. Nevertheless, we may approximate its value as:
\begin{equation}
    b \simeq \frac{L^i_\mathrm{nebula}}{L_\mathrm{sph}}
\end{equation}
where for $L_\mathrm{sph}$ we assume that NGC~1313~X--1 is observed close to face on ($i = 0^\circ$) and where $L^i_\mathrm{nebula}$ is the luminosity observed by the nebula at an unknown inclination angle $i$. We may get a crude estimate of $i$ from the \texttt{diskir} normalization, which is $\propto \cos(i)$. We have re-ran our analysis in Section~\ref{sub:sideview} by `inclining' the \texttt{diskir\_star} to inclination angles $i = 45^\circ, 60^\circ, 80^\circ$, finding the best-fit $n, r_\mathrm{in}$ in each case. All calculations were carried out with the stellar background stars and assuming the flux of the putative companion is isotropic. We have found the best match to the nebular lines is given for $i =45^\circ$ with $\log(n_\mathrm{H}[$cm$^{-3}$]) = 0.45, $r_\mathrm{in} = 1$ pc and $\chi^2$/dof = 89 (cf. $\chi^2$/dof = 92 for the \texttt{diskir\_star}). However, this inclined version not only provides a worse fit than the \texttt{phenomenological} model, but also still produces \heii\ in comparable numbers to the nominal \texttt{diskir\_star}. This suggests the differences between line-of-sight and the ionising SED are more complex than a simple scaling factor. 

Alternatively, we may find $i$ by considering what value of $i$ is needed to reduce the UV luminosity of the \texttt{diskir} to a level comparable to that of the \texttt{phenomenological} (i.e. a factor 4 dimmer). This would imply $i = 80^\circ$, although this model gives worse fits to the nebular emission than the nominal \texttt{diskir}. These values are obviously uncertain due to the uncertainties related in using the \texttt{diskir} to describe a super-Eddington accretion flow, but may suggest the nebula sees an inclined $i = 45^\circ-80^\circ$ version of the line-of-sight SED. Nevertheless, we have already argued that the non-overlapping [O~{\sc iii}]$\lambda$5007/\hb\ and [O~{\sc i}]$\lambda$6300/\ha\ regions strongly suggest the nebula sees the emission sideways. We additionally note that the \texttt{sirf} instead predicts a nearly isotropic UV flux and no amount of inclination can yield the necessary reduction in UV flux. 

Therefore considering the differences in UV luminosity between the line-of-sight SED (\texttt{diskir}) and that inferred from the extended nebula (\texttt{phenomenological}), we constrain the beaming factor in the EUV of NGC~1313~X--1 to about 0.3--0.15 and suggest the nebula sees an inclined $i = 45^\circ-80^\circ$ version of the line-of-sight SED. Such estimates are in agreement with measurements of the photo-ionized nebula around Holmberg~II~X--1 \citep{kaaret_high-resolution_2004} who found beaming factors $b >> 0.1$ (their Table~1) -- with the caveat of the extrapolation of the X-ray spectrum (see the Introduction). Such estimates are also consistent with detailed analytical calculations by \citet{abolmasov_optically_2009}. If such results can be extrapolated to the X-ray band, then we may be able to rule out the strong beaming invoked to explain the emission from some PULXs \citep{king_pulsing_2020} and would support the constraints derived on $b$ from modelling of the observed PULX pulse-fractions \citep{mushtukov_bright_2023, mushtukov_pulsating_2021}. 

Alternatively, the beaming factor may indeed be more pronounced in the X-ray band as alluded in the Introduction, as these photons emanate from within the wind funnel from the supercritical disk. Instead, the wind photosphere, which is expected to dominate the UV emission, is expected to emit quasi-isotropically \citep[see e.g.][]{shakura_black_1973, weng_evidence_2018}. Hence our results would be consistent with this picture. In this regard, our results also seem broadly consistent with the general relativistic radiation-magnetohydrodynamic simulations presented by \citet{narayan_spectra_2017}, although it is hard to derive a quantitative comparison. Broadly speaking, their post-processed spectra (e.g. their Figure 13) show the UV is reduced by about a factor $\sim$6 when going from $i = 10^\circ$ to 60$^\circ$. This would be broadly consistent with our reasoning based on the \texttt{diskir} model and the increase in inclination needed to match a reduction in flux of about 4.

\subsection{ULXs in the UV}
Regardless of whether we consider the intrinsic UV flux in \theulx\ as the line-of-sight value or that inferred from the extended nebula, both estimates are significantly lower than those measured in NGC~6946~X--1 \citep{abolmasov_optical_2008, kaaret_direct_2010}. The measurement in the F140LP reported by \citet{kaaret_direct_2010} is shown in Fig.~\ref{fig:bpass_model} for comparison and is about a factor 5 than predicted by any of our models. Similarly, we have integrated the \heii\ line predicted by the best-nebular model (\texttt{phenomenological} with the background stars) over the whole nebula (assuming isotropy), and found $L($He{\sc ii}$\lambda$4686) $\sim$9$\times$10$^{35}$ erg/s, slightly above our observed value of (3.4$\pm$0.6)$\times$10$^{35}$ erg/s (Section~\ref{sec:cloudy}). The value derived from the modelling may be considered an upper limit as we have considered the same \heii\/\hb\ ratios everywhere around the source. Still, such value is significantly lower than the measured $L($He{\sc ii}$\lambda$4686) = (2$\pm0.2)\times$10$^{37}$ erg/s in the MF16 nebula surrounding NGC~6946~X--1 reported by \citet{abolmasov_optical_2008}. If these UV differences were due to an inclination effect, then we should have seen evidence for strong UV (comparable to that of NGC~6946~X--1) either along the line of sight or in the nebular lines. Such differences suggest the UV emission in NGC~6946~X--1 and NGC~1313~X--1 is \textit{intrinsically} different, suggesting we are probing differences in the mass-transfer rate.

Because the UV emission is thought to be linked to the wind photosphere \citep[$\propto \dot{m_0}^{-3/4}$;][where $\dot{m_0}$ is the mass-transfer rate at the companion in Eddington units]{poutanen_supercritically_2007} our analysis suggest a lower mass-transfer rate in \theulx\ compared to NGC~6946~X--1, despite the stronger X-ray luminosity of the former \citep{gurpide_long-term_2021}. We suggest the inclination of these two systems might be comparable, but NGC~6946~X--1 might possess a narrower funnel due to higher accretion-rate, creating a strong soft X-ray/EUV source and mimicking a highly inclined source. In NGC~1313~X--1 instead we peer down the funnel most of the time owing to the wider opening angle of the funnel, except in the extremely soft and unusual `obscured state' \citep{gurpide_long-term_2021}, where we showed the source becomes even softer than NGC~6946~X--1. A brighter UV and softer X-ray spectrum in NGC~6946~X--1 due to a higher mass-transfer rates would be fully consistent with predictions from R(M)HD simulations \citep{kawashima_comptonized_2012, narayan_spectra_2017} and arguments made by \citet{abolmasov_optical_2007} based on the observed nebular lines in a sample of ULXs. 

\begin{table*}
    \centering
    \caption{Comparison of ULX spectral properties and their high-excitation emission-line nebulae.} \label{tab:heII_ULXs}
    \resizebox{\textwidth}{!}{\begin{tabular}{lccccccc}
    \hline \hline
    \noalign{\smallskip}
    Source & Spectral Regime & $L($\heii$^a$) & \heii/\hb & [O~{\sc iii}]$\lambda$5007/\hb$^b$& $L_\mathrm{UV}^c$ & $P_\mathrm{mec}$ & Ref\\

            &                &  10$^{36}$ erg/s &    &    & 10$^{39}$ erg/s &  10$^{39}$ erg/s  & \\
    \noalign{\smallskip}
    \hline
    \noalign{\smallskip}
NGC~6946~X--1  & Soft & 13--20$\pm2$ & 0.17$\pm$0.02; 0.22$\pm$0.01 & Yes & 12 & -- &\citet{abolmasov_optical_2007, abolmasov_optical_2008} \\
Holmberg~II~X--1   & Soft (Variable) & 2.2--2.7; 3.6 & 0.22$\pm$0.02; 0.33 &  Yes & -- &  2$^{+8}_{-2}$ & \citet{kaaret_high-resolution_2004, lehmann_integral_2005, abolmasov_optical_2007,moon_large_2011, cseh_unveiling_2014} \\
NGC~5408~X--1 & Soft & $>$1.2 & 0.27--0.38 & Yes & --  & -- & \citet{kaaret_photoionized_2009} \\
M51 ULX-1 & Soft &  0.82$\pm$0.03 & 0.22$\pm$0.01 & Yes & -- & 2$\pm$0.5 & \citet{urquhart_two_2016, urquhart_multiband_2018} \\
IC~342~X--1 & Hard & Undetected, 0.01$\pm$0.002 & 0.036$\pm$0.02 & Yes & -- &  3.4 & \citet{roberts_unusual_2003, ramsey_optical_2006, abolmasov_optical_2007, cseh_black_2012} \\
NGC~1313~X--2 & Hard & Undetected; $<$6 $\times$10$^{35}$ erg/s & $<0.05$--0.07 &  Yes & $<$1 & 6.5$\pm$0.3 & \citet{ramsey_optical_2006, zhou_very_2022, gurpide_absence_2024}\\
M81~X--6 & Hard & $\sim$0.09; $<$4.4 & $\sim$0.19 &  -- & -- & -- & \citet{moon_large_2011} \\
Holmberg~IX~X--1 & Hard & 0.007$\pm$0.002, $\sim$0.37; $<$12 & 0.06$\pm$0.02;$\sim$0.12 & Yes & 1.9 &  3.5--7 & \citet{abolmasov_optical_2007,abolmasov_kinematics_2008, moon_large_2011} \\
NGC~1313~X--1 & Hard & 0.34--0.9 & 0.10--0.20 & Yes & 0.55 & 30$\pm$20&  This work; \citet{gurpide_muse_2022} \\
       \noalign{\smallskip}
       \hline \hline
       \noalign{\smallskip}
    \end{tabular}
    }
    \begin{minipage}{\linewidth}
   \textbf{Notes.} $^a$Nebular \heii\ luminosity. Different values correspond to different measurements reported in the references.\\
    $^b$Whether unusually high [O~{\sc iii}]$\lambda$5007/\hb\ ratios indicative of high-excitation have been reported.\\
    $^c$UV luminosity inferred from the nebula when available.
    \end{minipage}
\end{table*}

In Table~\ref{tab:heII_ULXs} we have collated literature results regarding the properties of high-excitation nebulae surrounding ULXs for which the X-ray spectral regime is known, which we have taken from \citet{sutton_ultraluminous_2013, urquhart_two_2016, gurpide_long-term_2021}. Notice the apparent differences in nebular emission between hard ULXs such as NGC~1313~X--1 and M81~X--6 and soft ULXs such as NGC~5408~X--1 \citep{kaaret_photoionized_2009}, Holmberg~II~X--1 \citep{kaaret_high-resolution_2004} or NGC~6946~X--1 \citep{abolmasov_optical_2008}. These differences are not trivial as we have already stressed the fact that the \heii\ is insensitive to the X-rays (Fig.~\ref{fig:1Dcomparison} lower panel) as its ionization cross-section falls roughly as $\nu^{-3}$ \citep{osterbrock_astrophysics_2006}. The \heii\ is obviously sensitive to the density or distribution of material around the ULX. For this reason in the Table we also report the \heii/\hb\ ratio, which should be less sensitive to density differences. As can be seen, the ratios are generally lower in hard ULXs, indicating that the differences in \heii\ luminosity are not due to a density difference. Thus the differences in nebular \heii\ around hard and soft ULXs are a telltale that hard and soft ULXs are not only distinct in their X-ray spectral properties, but also in their UV/EUV. Because strong EUV is linked to the mass-transfer rate according to RMHD simulations \citep{narayan_spectra_2017} and analytical estimates \citep{poutanen_supercritically_2007}, bright \heii\ around soft ULXs signals these systems must possess higher mass-transfer rates compared to their hard counterparts. Instead, hard ULXs, due to their faint EUV, can produce strong [O~{\sc iii}]$\lambda$5007/\hb\ ratios but dim or no detectable \heii\ (e.g. as it is the case in NGC~1313~X--1, Holmberg~IX~X--1 and NGC~1313~X--2). Our findings are consistent with earlier assertions made by \citet{abolmasov_optical_2007} and suggest the mass-transfer rate may be more relevant parameter distinguishing hard and soft ULXs, as opposed to their inclination \citep{sutton_ultraluminous_2013}. 

In Table~\ref{tab:heII_ULXs} we have also collated mechanical powers $P_\text{mec}$ inferred from observations of optical/radio bubbles surrounding ULXs. Because we expect the outflow power to increase with $\dot{m}_\mathrm{0}$ \citep[e.g.][]{kitaki_origins_2021}, we should expect soft ULXs to possess higher $P_\text{mec}$. From the current limited sample, it does not appear that soft ULXs show higher $P_\mathrm{mec}$, but certainly a systematic study is needed here, which is beyond the scope of this paper. It is also unclear whether some of these nebulae are comparable: for instance, the bubble around Holmberg~IX~X--1 shows a nearly spherical morphology \citep{abolmasov_kinematics_2008}, while Holmberg~II~X--1 or M51~ULX-1 show instead bipolar bubbles associated with collimated jets \citep{cseh_unveiling_2014, urquhart_multiband_2018}. Soft ULXs seem also to be less likely associated with shock signatures \citep[e.g. the case of Holmberg II X--1 or NGC~5408~X--1][]{kaaret_high-resolution_2004, cseh_black_2012}, which may suggest hard ULXs are more likely to clear out the material around them, hindering the detectability of the \heii\ line. Other factors such as the mass, or even the nature of the accretor are also likely to play a role in explaining such differences. In particular, hard ULXs have been systematically shown to be more likely to host NSs \citep{pintore_pulsator-like_2017, walton_evidence_2018, gurpide_long-term_2021, amato_ultraluminous_2023}. Whether the nature of the accretor could also explain such differences remains to be seen, but from the limited sample it would seem that differences in the X-ray spectral are translating into differences in the interaction with the environment.

\subsection{The nebular \heii\ problem} 

Whether ULXs can produce He{\sc ii} emission in enough numbers to account for the \heii\ line observed in the integrated spectra of metal-poor galaxies (the \heii\ problem) has been recently examined in \citet{simmonds_can_2021} and \citet{kovlakas_ionizing_2022}, reaching contradictory results. \citet{simmonds_can_2021} found that a the multi-band \texttt{diskir} model presented by \citet{berghea_spitzer_2012} could produce \heii\ in enough numbers to explain it. \citet{kovlakas_ionizing_2022} on the other hand, built empirical models based analytical descriptions of super-Eddington accretion discs \citep{shakura_black_1973, lipunova_supercritical_1999, poutanen_supercritically_2007} and found ULXs do not produce enough ionising UV photons to explain the \heii\ line. 

The main uncertainty on these works was the lack of knowledge about the UV emission (see Figure 6 in \citet{kovlakas_ionizing_2022}), to which the \heii\ line is most sensitive. Based on our discussion above and our analysis, we suggest that there should be a dichotomy between hard and soft ULXs in terms of their capacity to excite \heii, with hard ULXs ruled out as potential candidates to explain the nebular \heii\ in metal-poor galaxies. Finally, we also note that the relatively isotropic EUV found here would render uncertainties related to beaming and its dependency with $\dot{m}$ in the study of the nebular \heii\ problem nearly unimportant \citep[e.g.][]{kovlakas_ionizing_2022}. We aim to pursue similar studies in other ULXs with different X-ray spectral hardness to reliable confirm these results. 

\subsection{The origin of the optical/near-UV light}\label{sec:optical_light}

In Section~\ref{sec:multi_band_sed} we have included a blackbody component to model the optical/near-UV fluxes from the \hst. While this component is not required by the \texttt{diskir}, it is strongly required by the \texttt{sirf} and \texttt{phenomenological} models to fit the \hst\ data. We initially presented this component as a proxy for the contribution from the companion star. However, based on its luminosity and temperature, we now investigate whether a stellar origin is physically plausible and consider alternative explanations.

We have seen both the \texttt{phenomenological} and the \texttt{diskir} give similar parameters for this blackbody. The temperature and luminosity of this component imply an O-type star, requiring a star $>$20\,\msun. This would be consistent with the donor OB-types inferred from direct optical modelling of the ULX spectra \citep{tao_compact_2011}. However, here we cannot rule out this type of star as it would be at odds with constraints from the population and age of the nearby stars presented in \citet{yang_optical_2011}, which suggest masses under 10\,\msun\ instead. Therefore for \theulx, we can rule out the optical/near-UV fluxes being dominated by the donor star, which cast doubts on the OB-type stars inferred in other ULXs \citep[e.g.][]{grise_optical_2012}. This is consistent with the optical short-term variability observed by \citet{yang_optical_2011} (and confirmed here Section~\ref{sec:data_reduction}) and the strong optical variability observed during the nascent ULX in M83 \citep{soria_birth_2012}.

\citet{yang_optical_2011} and \citet{tao_compact_2011} presented additional diagnostics that can help shed light on the nature of the emission. Based on our reanalysis of the \hst\ data and the state-resolved optical and X-ray data we updated the Johnson-Cousin Vega extinction-corrected magnitude $B_\mathrm{0} = 23.28\pm0.07$ and the optical to X-ray ratio diagnostic from \citet{van_paradijs_average_1981} \textbf{$\xi = B_\mathrm{0} + 2.5 \log(F_\mathrm{X}) = 21.8\pm0.7$}, reported in \citet{yang_optical_2011}. As stated in \citet{yang_optical_2011}, such values are typical for low-mass X-ray binaries and suggest the emission is dominated by the disc itself (or reprocessed emission from it) rather than the companion star \citep{kaaret_ultraluminous_2017}. Indeed, \citet{tao_compact_2011} found through careful analysis of the \hst\ data of a handful of ULXs counterparts that the optical emission is not consistent with any stellar type.

An alternative explanation is that this component represents the emission from the wind photosphere. The temperature of this component ($T\sim3\times10^4$\,K) is comparable to that measured in SS433 \citep[$T = (7\pm2)\times10^4$;][]{dolan_ss_1997} and in NGC~6946~X--1 \citep[$T = 3.1 \times 10^4$\,K;][]{kaaret_direct_2010} both using similar photometric filers and a single blackbody model. We note however that the value on SS433 is highly uncertain as relies on a likely overestimated level of extinction, as noted by the detailed X-SHOOTER analysis on SS433 \citep{waisberg_collimated_2019}. 

Nevertheless, the radii inferred for both SS433 and NGC6946~X--1 are of the order of 10$^{12}$\,cm, while for NGC~1313~X--1 we measured about half that value. Whether we can associated any predictive power to this component is questionable owing to the uncertainty on the exact modelling \citep[and in fact most likely additional components are present/needed;][]{kaaret_direct_2010, soria_birth_2012}, however given the photosphere radius is expected to scale with $\dot{m}_\mathrm{0}^{3/2}$, this is indeed in line with our previous suggestion of a lower mass-transfer rate in NGC~1313~X--1. Such interpretation is consistent by the lower overall UV luminosity in NGC~1313~X--1 ($\lesssim10^{39}$\,erg/s) compared to NGC~6946~X--1 \citep{abolmasov_optical_2008} or SS433 \citep{dolan_ss_1997, waisberg_collimated_2019}, which both have a UV luminosities in excess of 10$^{40}$\,erg/s. 

In this regard, while we have shown the level of UV predicted by the \texttt{sirf} is too high to explain the nebular emission (Section~\ref{sec:cloudy}), it may offer a more accurate description of the optical/near-UV data. This is because when employing the \texttt{sirf} model, it is the (supercritical) disc that dominates the optical/near-UV fluxes, with the putative companion adding a negligible contribution (Fig.~\ref{fig:multiband_sed_star}). Such description of the SED would fit more accurately our expectation of the optical/near-UV fluxes based on the reasoning above. Given the putative star parameters, we infer an F0-A-type star with a mass of $\lesssim9$\msun\ \citep{ekstrom_grids_2012} which would be consistent with the population of stars around NGC~1313~X--1 \citep{yang_optical_2011}. We further note that this is the stellar-type inferred for the companion star of SS433 \citep[see][and references therein]{goranskij_photometric_2011}, which may add supporting evidence for the link between SS433 and ULXs. However, we must stress that this ignores binary evolutionary effects and irradiation by the X-rays from the disc/wind, which will distort the spectrum of the companion star \citep[see discussion in][and references therein]{ambrosi_modelling_2022,sathyaprakash_multi-wavelength_2022}. In any case, our analysis suggest that the optical/UV emission in NGC~1313~X--1 is \textit{not} dominated by the companion star. Similar situation was found in Holmberg~II~X--1 \citep{tao_nature_2012} and may suggest the same is true in NGC~5408~X--1 \citep{grise_optical_2012}.

Finally, we consider the results presented by \citet{vinokurov_optical_2018}, who pointed out that the brightest ULXs in the optical (approximately $M_\mathrm{V} < -5.5$ where $M_\mathrm{V}$ is the absolute magnitude in the Johnson $V$ band) show a powerlaw-like spectra, whereas dimmer ULXs instead have blackbody-like optical spectra. \citet{vinokurov_optical_2018, fabrika_ultraluminous_2021} argue that the dimmer appearance of some ULXs is due to a lower contribution of the wind photosphere, linked to the mass-accretion rate. Such interpretation would reinforce our conclusion that indeed the $\dot{m}_\mathrm{0}$ in NGC~1313~X--1 is lower compared to softer ULXs such as e.g. NGC~6946~X--1 or NGC~5408~X--1, which instead have $M_\mathrm{V} < $-6 \citep{tao_compact_2011}. \citet{vinokurov_optical_2018} argues that the blackbody-like spectra in the dimmer systems may represent emission from the donor. While the low optical luminosity of NGC~1313~X--1 ($M_\mathrm{V} \sim -4.9$) may suggest the latter scenario applies here as well, we have already seen that the companion star inferred (O-type) from blackbody fits is at odds with the stellar population around NGC~1313~X--1. Ignoring geometrical and irradiation effects \citep{abolmasov_optically_2009}, assuming the wind has the virial velocity at the spherization radius \citep{poutanen_supercritically_2007}, and that the wind photosphere radiates as a spherical blackbody, one can show that its luminosity is roughly equal to a third of the Eddington luminosity and independent of $\dot{m}_\mathrm{0}$ \citep{poutanen_supercritically_2007}:
\begin{equation}
    L_\mathrm{ph} \sim \frac{1}{3} L_\mathrm{Edd} \sim 4 \times 10^{37} \left(\frac{M}{M_\odot} \right) \mathrm{erg/s}
\end{equation} 
Thus the luminosity of the optical blackbody ($\sim 2 \times$10$^{38}$ erg/s) can be easily explained a by a $\sim$5 \msun\ BH. However we note that we could not match its temperature ($T\sim$3$\times$10$^{-3}$ keV) for a reasonable value of $\dot{m}_\mathrm{0}$.

To illustrate this, consider the temperature of the wind photosphere may be expressed as \citet{poutanen_supercritically_2007}:
\begin{equation}
    T_\mathrm{ph} \sim 0.8 \epsilon_\omega^{-1} m^{-\frac{1}{4}} m_\mathrm{0}^{-3/4} \text{keV}
\end{equation}
where $\epsilon_\omega$ = $L_\mathrm{wind}$/($L_\mathrm{wind}$ + $L_\mathrm{rad}$), with $L_\mathrm{wind}$ the kinetic luminosity of the wind and $L_\mathrm{rad}$ the observed radiative luminosity \citep{poutanen_supercritically_2007}. Assuming $\epsilon_\omega = 0.5$ \citep{pinto_resolved_2016, gurpide_muse_2022}, in order to match the observed blackbody temperature we would need an abnormally high $\dot{m}_0 \sim 1000$ for the aforementioned BH mass. Thus, as alluded above, it is unlikely that we can associate this component solely to the wind photosphere and the picture may be more complex due to geometrical, irradiation effects \citep{abolmasov_optical_2008} and deviations from a blackbody due to scattering \citep{lipunova_supercritical_1999}. 

\subsection{Caveats} \label{sub:caveats}

There are several limitations of our study that are worth highlight and need to be borne in mind when interpreting the results. The first is the timescales involved in the nebular emission with respect to the variability of the source, which will unavoidably affect any study of this kind. As alluded in Section~\ref{sec:cloudy}, the recombination timescales of the nebula are of the order of thousands of years, while the observation baseline of \theulx\ is only of dozens of years, if we consider \xmm\ observations prior to the \swift-XRT. Hence the time-averaged spectrum could be higher or lower depending on the (unaccessible) activity history of \theulx. However, the fact that the \texttt{diskir} provides a reasonable description of the line of sight nebular lines (all lines accurately predicted within a factor 2) -- which are obviously subject to similar recombination timescales -- suggests that the time-average SED cannot deviate substantially from the present-day estimate. In fact, we can rule out a brighter time-average spectra as the \texttt{sirf} substantially overpredicts the line fluxes both along the line of sight and in the extended nebula (Tables~\ref{tab:cloudy_results} and ~\ref{tab:cloudy_los}). We have explicitly confirmed that the high-state SEDs did not offer better match to the nebular lines than the low state SEDs. There remains the possibility that the time-average spectrum is dimmer than the \texttt{diskir} but still brighter than the \texttt{phenomenological}, because the level of UV provided by the \texttt{phenomenological} already provides a worse match to the line-of-sight line fluxes than the \texttt{diskir} (Table~\ref{tab:cloudy_los}). It may thus be possible to explain both the line of sight and extended nebular line fluxes better by a time-average spectrum whose UV brightness sits between these two spectra, which may suggest the UV is closer to being isotropic than we have estimated.  

Another possibility is that some of the line ratios (or fluxes) are slightly affected by shocks. In our modelling of the extended emission (Section~\ref{sub:sideview}) we do not expect this to be an issue affecting significantly our results. The first reason is that we already showed in \cite{gurpide_muse_2022} that shocks are mainly concentrated in the edges or rim of the bubble, whereas the inner parts where instead dominated by EUV/X-ray excitation. Secondly, the morphology of the nebula, that is the observed ionization gradient, most remarkable in the \oiii\ and \oi\ lines, can only be produced by EUV/X-ray photo-ionization. Shocks would instead produce rather co-spatial \oi\ and \oiii\ regions. Moreover, we already showed in \cite{gurpide_muse_2022} that to produce the observed levels of \oiii\ would require shock velocities $>$300 km/s \citep[see also][]{berghea_first_2010}, clearly ruled out by the kinematic data. Therefore, if anything, we may expect a slight contribution from shocks in low-ionization lines in the outermost parts of the photo-ionized region. However, winds can still alter the distribution of the gas, by depleting and compressing it onto a thin shell \citep[e.g.][]{siwek_optical_2017, garofali_modeling_2023, gurpide_absence_2024}. To some extent this has been taken into account, since we have marginalized our results over a range of radii and density. In practice however, it is often hard to derive an accurate description of the geometry of the nebulae, due to projection effects. We are also working on extending our modelling to 3D structures to provide potentially a more accurate description of the cloud geometry. However, preliminary results suggest that extending to 3D, at least for the present work, is not likely to have a strong impact on the results.

Shocks may be important however in our photo-ionization modelling of the nebula along the line of sight. As stressed in Section~\ref{sub:front_view}, the lack of clear diagnostics such as the extent of the regions with enhanced oxygen to Balmer line ratios makes the attribution of the observed line ratios to photo-ionization by the ULX less certain. A more refined modelling, self-consistently accounting for shocks and photo-ionization may provide a more accurate picture. Nevertheless, once again we do not expect this to strongly affect lines such as \oiii\ and \heii\ and therefore we can confidently ascertain that the line of sight SED is best described by the \texttt{diskir} or the \texttt{phenomenological} models. Thus we consider our assertion that beaming in the UV must be small to be robust to these caveats. We leave for future work a more detailed treatment self-consistently accounting for photo-ionization and shocks.

Lastly, moving forward it may be advantageous to include the stars in the field in a spatially resolved manner, here and for other ULXs in crowded fields such as Holmberg~II~X--1 \citep{pakull_optical_2002}. At present, this cannot be done in \cloudy, but may be explored in the future with three-dimensional photo-ionization codes such as \texttt{MOCASSIN} \citep{ercolano_mocassin_2003}.

\section{Conclusions}\label{sec:conclusion}
Coupling multi-band spectroscopy with detailed modelling of the EUV/X-ray excited emission-line nebula surrounding NGC~1313~X--1, we have attempted to constrain the degree of anisotropy of the UV emission in this archetypal ULX. Our results suggest that the UV emission is mildly beamed, by a about a factor $\sim$4 at most. We have also shown that the optical emission in NGC~1313~X--1 is unlikely to be dominated by the companion star.

We have also discussed the weak detection of \heii\ in the nebula surrounding NGC~1313~X--1, which seems to be a common finding around other hard ULXs. Instead, bright nebular \heii\ seems to be a common finding around soft ULXs. We suggest differences in mass-transfer rate may explain such dichotomy, since according to analytical calculations and numerical simulations only at high-mass transfer rates a ULX will become an extreme EUV source. This implies that only a subset of the whole ULX population may excite \heii\ in high enough numbers to account for the observed \heii\ line in metal-poor galaxies.

Moving forward, a better understanding of the ULX SED or observations targeted at reducing the uncertainty on the line-of-sight SED by probing the FUV emission would be of great interest to reduce the uncertainties in our work. While the lines probed here do not allow us to constrain the degree of anisotropy of the soft X-ray emission, probing the high-excitation lines found in the IR with \textit{James Webb Space Telescope} would enable to constrain the degree of anisotropy of the ULX emission at higher energies, allowing us to test existing super-Eddington accretion theories and improving our understanding of ULXs and the feedback on their environments.

\section*{Acknowledgements}
We would like to thank the anonymous referee for their thoughtful comments which helped improved our manuscript. N.C.S. and A.G. acknowledge support from the Science and Technology Facilities Council (STFC) consolidated grant ST/V001000/1. We are grateful to P. Abolmasov for useful commentary which helped improved the paper, to C. Knigge and O. Godet for stimulating discussion and to P. van Hoof for assistance in the use of \cloudy. This research has made use of NASA/ESA Hubble Space Telescope obtained from the Hubble Legacy Archive, which is a collaboration between the Space Telescope Science Institute (STScI/NASA), the Space Telescope European Coordinating Facility (ST-ECF/ESA), and the Canadian Astronomy Data Centre (CADC/NRC/CSA), the NuSTAR Data Analysis Software (NuSTARDAS), jointly developed by the ASI Space Science Data Center (SSDC, Italy) and the California Institute of Technology (Caltech, USA) and the data supplied by the UK Swift Science Data Centre at the University of Leicester. Software used: Python v3.8, mpdaf \citep{piqueras_mpdaf_2017}, pyCloudy \citep{morisset_pycloudy_2013}, CAMEL \citep{epinat_massiv_2012}. 

\section*{Data Availability}

All data used in this work is publicly available in the corresponding archives.



\bibliographystyle{mnras}
\bibliography{references.bib, refs.bib}




\appendix

\section{\chandra\ data analysis} \label{sec:chandra}

We reduced \chandra\ using the \texttt{CIAO} software v.4.12 and \texttt{CALDB} 4.9.3 and reprocessed the events using the script \texttt{chandra\_repro}. Extraction regions for the source were determined by running the detection algorithm \texttt{wavdetect}, whereas the background was selected from a large nearby $\gtrsim$10"-radius circular ``source-free'' region. We inspected each observation for pile up using the \texttt{pile\_map} tool. Obsids 4747/8 were severely affected by pile up ($P_\text{f} > 20\%$) as the source was placed on the aimpoint whereas for obsid 4750 the effects of pileup were mitigated owing to the relatively large axis position of the source ($\sim$6.9'). We therefore discarded 4748 for further analysis as it was taken on the same date as obsid 4750. 

To obtain \swift-XRT equivalent count rates in the 0.3 -- 10\,keV band, we fit these observations in order to determine the source spectral state (using \swift-XRT as a reference), at the time of the \chandra\ observations. We extracted spectra using the task \texttt{specextract} and rebinned the spectra using the scheme proposed by \cite{kaastra_optimal_2016}. We adopted $\chi^2$ statistics, as the spectra had 20 counts/bin for most of the range, and restricted the fit to the $\sim$0.35--8\,keV band, where the source count rate was well above the background in order to avoid introducing negative counts in the $\chi^2$ computation. We then folded the best-fit models through the \swift-XRT response files to determine the \swift-XRT equivalent count-rate. The same conversion factor was used to transform the flux upper and lower confidence interval into uncertainties on the count rates. 

All spectral fits were performed with \texttt{XSPEC} software 12.12.1. Uncertainties in this Section are given at the 90\% confidence level for one parameter of interest. Fluxes were estimated using the pseudomodel \texttt{cflux}. In all our fits we included a \texttt{tbabs} \citep{wilms_absorption_2000} model component with abundances from \cite{verner_atomic_1996} to account for neutral absorption along the line sight. Owing to the low data quality we restricted the spectral models to a single spectral component: either a \texttt{diskbb} or a \texttt{powerlaw}.

To model the 2003 observation severely affected by pile up we convolved all models with the \texttt{pileup} model \citep{davis_event_2001} to attempt to correct for pile up effects. We first attempted to fit these data with an absorbed \texttt{powerlaw} along with the grade morphing parameter ($\alpha_\text{G}$) and the PSF fraction ($f_\text{PSF}$) from the \texttt{pileup} model free to vary. While the fit was statistically acceptable ($\chi^2$/dof = 69.2/67), we found that both $\alpha_\text{G}$ and $f_\text{PSF}$ had unconstrained upper confidence intervals, and we were unable to reliable estimate the flux. We therefore froze $f_\text{PSF}$ to the typical value of 95\% for a point-like source but the fit was once again unstable during the flux estimation mainly due to $\alpha_\text{G}$ being once again unconstrained ($\alpha_\text{G} < $ 0.58). Similar results were found if we froze the neutral hydrogen absorption column to the average value determined in \cite{gurpide_long-term_2021} (\nh\ = 2.63$\pm$0.03 $\times$10$^{21}$ \text{cm}$^{-2}$). The data was instead well fitted with an absorbed \texttt{diskbb} with a $\chi^2$/dof = 68.7/68 with $\alpha_\text{G}$ = 0.47$_{-0.05}^{+0.04}$, $f_\text{PSF}$ = 0.94$\pm$0.01 and $T_\text{in} = 0.55_{-0.07}^{+0.01}$\,keV and \nh\ frozen to the aforementioned value. We note the $f_\text{PSF}$ was in good agreement with the fraction of piled-up photons inside the PSF expected for an aimpoint source. The best-fit model observed flux was {(8$\pm$1)$\times10^{-12}$ \fluxcgs} which translated to 0.38$^{+0.06}_{-0.05}$ \swift-XRT equivalent count rates. Unabsorbed fluxes and luminosities for this model are given in Table~\ref{tab:chandra_data}. Both the \swift-XRT lightcurve and the unabsorbed luminosity of $\sim$2.9 $\times$10$^{40}$\,\luxcgs suggest the \chandra\ observation caught \theulx\ at its peak brightness \citep[Fig.~\ref{fig:swift_hst} ; cf.][]{kobayashi_new_2019, walton_unusual_2020}. 

Given the uncertainties associated with the pile up model, we carried a series of tests to assess the reliability of this flux estimate and additional sources of uncertainty. To do so, we inspected whether the \chandra\ data was compatible with either the dimmest or the brightest states of \ulx, but distorted by pile up. To obtain a model for the brightest state, we followed \cite{walton_new_2020} and extracted the first 45\,ks from \xmm\ obsid 0803990101 (see their Figure 2 and Table~2). Since we were only interested in a rough estimation of the continuum we fitted the three EPIC detectors with a simple absorbed \texttt{cutoffpl} which provided a satisfactory fit ($\chi^2$/dof = 451.5/320), with most of the contribution to the $\chi^2$ coming from the well-known residuals at 1\,keV \cite[e.g.][]{middleton_diagnosing_2015}. However, the best-fit model flux of (6.68$\pm0.05)\times10^{-12}$ \fluxcgs\ translated only to 0.188$\pm$0.001 \swift-XRT equivalent count rates, slightly below the brightest state of the source (Fig.~\ref{fig:swift_hst}). We therefore re-extracted the spectra using only the first 10\,ks of the observation to really probe the peak of the rapid flaring. The same model yielded ($\chi^2$/dof = 317.4/284) with an observed flux (9.8$\pm0.1)\times10^{-12}$ \fluxcgs\ which translated into 0.26$\pm$0.003 \swift-XRT equivalent ct/s, consistent with the high state (Fig.~\ref{fig:swift_hst}). For the low state we used \xmm\ obsid 0106860101 \cite[cf.][]{gurpide_long-term_2021}, where we employed a simple absorbed \texttt{powerlaw} as the fit was insensitive to the high-energy cutoff. Here we obtained $\chi^2$/dof = 516.4/341 with 2.84$^{+0.05}_{-0.05}\times10^{-12}$\,\fluxcgs\ equivalent to $0.075\pm$0.001 \swift-XRT ct/s, in agreement with the dimmest states seen in the \swift-XRT lightcurve (Fig.~\ref{fig:swift_hst}).
\begin{table*}
    \centering
        \caption{\xmm\ datasets used to verify the fits to the \chandra\ obsid 4747 (see text for details). The model used was an absorbed \texttt{cutoffpl} powerlaw and a simple \texttt{powerlaw} for the high and low states respectively. Uncertainties are given at the 1$\sigma$ level.}
    \label{tab:xmm_chandra}
    \begin{tabular}{cccccccc}
    \hline
     \noalign{\smallskip}
        ObsID & Date & State &  \nh\ & $\Gamma$ & $E_\text{cutoff}$ & $\chi^2$/dof  & \swift-XRT\\
        & & & $10^{22}$cm$^{-2}$ & & keV &  & ct /s \\
        \noalign{\smallskip}
        \hline
         \noalign{\smallskip}
       0803990101(-10ks)  & 2000-10-17 &  High &  0.28$\pm$0.02 & 1.4$\pm$0.1 & 5.3$_{-0.7}^{+0.9}$ & 317.4/284 & 0.38$^{+0.06}_{-0.05}$\\
       0106860101 &  2017-06-14 & Low & 0.22$\pm$0.01 & 1.88$\pm$0.03 & -- & 516.4/341 & $0.075\pm$0.001\\
       \noalign{\smallskip}
       \hline
    \end{tabular}
\end{table*}

We then took these two best-fit models from the high and low states and convolved them with the obsid 4747 \chandra\ response, froze all the source-model parameters to their best \xmm\ values and fit for the $\alpha_\text{G}$ and $f_\text{PSF}$ from the \texttt{pileup} model. The model from the low state not only can almost statistically be excluded ($\chi^2$/dof = 117.5/70), but required $\alpha_\text{G}>$ 0.95 and $0.75 < f_\text{PSF}<$ 0.86 which are values beyond those expected for the \texttt{pileup} model\footnote{\url{https://cxc.harvard.edu/ciao/download/doc/pileup_abc.pdf}}, implying that the data is not consistent with the low state being distorted by pile up. The model from the bright state is instead in excellent agreement with the data ($\chi^2$/dof = 71.3/70) and yielded reasonable values for $\alpha_\text{G} = 0.20\pm0.04$ and $f_\text{PSF} = 0.91\pm0.01$. Such good agreement with the peak spectrum implies that the \chandra\ observation is indeed consistent with \theulx\ being at its peak brightness. 
We finally inspected how our model choice for the \chandra\ data (the \texttt{diskbb}) probing the peak of \theulx\ affected our best-fit \swift-XRT equivalent count rate. We did so by leaving the photon index $\Gamma$ from the best-fit \xmm\ \texttt{cutoffpl} powerlaw free to vary and recalculating the observed flux/\swift-XRT equivalent count rate. In brief we found the \swift-XRT count rate measured in this manner is slightly lower ($0.25^{+0.25}_{-0.09}$ ct/s) than our estimate with the \texttt{diskbb} (0.38$^{+0.06}_{-0.05}$ ct/s) but still fully consistent with the high state. We therefore considered this estimate as a lower limit on the flux probed by the \chandra\ data. We conclude that despite the data being affected by pile up we can confidently say that the \chandra\ probed the high state of \theulx\ (see Fig.~\ref{fig:swift_hst}) and thus we can place the simultaneous \hst\ observations from 2003 in this state.  

For obsid 4750 we initially attempted to fit the data with the \texttt{pileup} model to verify whether our estimate of low pile up fraction of $\sim$2.5--3\% was reasonable. An absorbed \texttt{diskbb} with \nh\ frozen to the average value quoted above yielded a marginally acceptable fit ($\chi^2$/dof = 92/57), with an unabsorbed luminosity $>5\times10^{40}$ \luxcgs, which translates to an equivalent \swift-XRT count of $\sim$0.76 ct/s, which we deem implausibly high based on the source long-term behaviour (Fig.~\ref{fig:swift_hst}). The fit improves if \nh\ is left free to vary ($\chi^2$/dof = 79.8/56), but the luminosity is even more implausible ($8^{+2}_{-1}\times 10^{40}$\,\luxcgs). For the \texttt{powerlaw} fits we found that both $f_\text{PSF}$ and $\alpha_\text{G}$ were consistent with 0 regardless of our treatment of \nh, indicating that this component is not necessary. Pile up solutions are often `double-valued' i.e. an incident count rate can be modelled as heavily or marginally piled up due to the degeneracy between incident and detected count rates \footnote{\url{https://cxc.harvard.edu/ciao/download/doc/pileup_abc.pdf}}. Given the observed counts per frame ($<$0.06 ct/frame) it is most likely that our spectrum falls in the `low' count rate solution rather than the heavily piled up one. We therefore concluded that the effects of pile up are indeed negligible, in agreement with the earlier estimates. 

Removing thus the \texttt{pileup} component, we found that a single absorbed \texttt{diskbb} was statistically excluded ($\chi^2$ = 134.5/58). The data was broadly consistent with a simple absorbed \texttt{powerlaw} ($\chi^2$ = 76.6 for the same degrees of freedom) -- in agreement with $\alpha_\text{G}$ converging towards 0 when employing the \texttt{pileup} model -- with an unabsorbed luminosity of ($1.8\pm0.1)\times10^{40}$ \luxcgs\ and an equivalent \swift-XRT count rate of $0.128^{+0.009}_{-0.008}$ ct/s, placing NGC~1313~X--1 in the low state during the February 2004 observations. 

\section{Flux maps for the [Ar~{\sc iii}]$\lambda$7135 and [S~{\sc iii}]$\lambda$9069 lines}
\begin{figure*}
    \centering
    \includegraphics[width=0.485\textwidth]{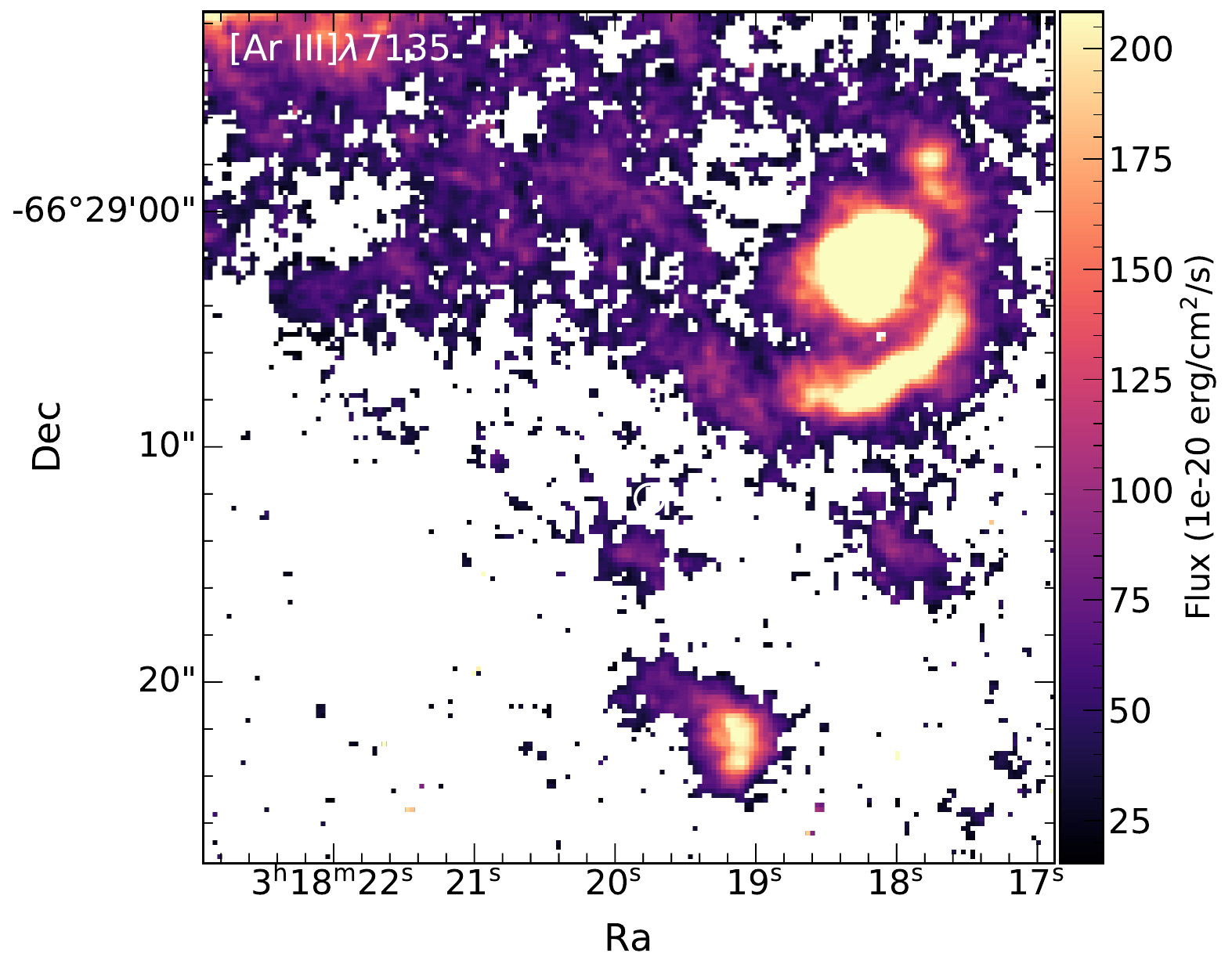}
    \includegraphics[width=0.485\textwidth]{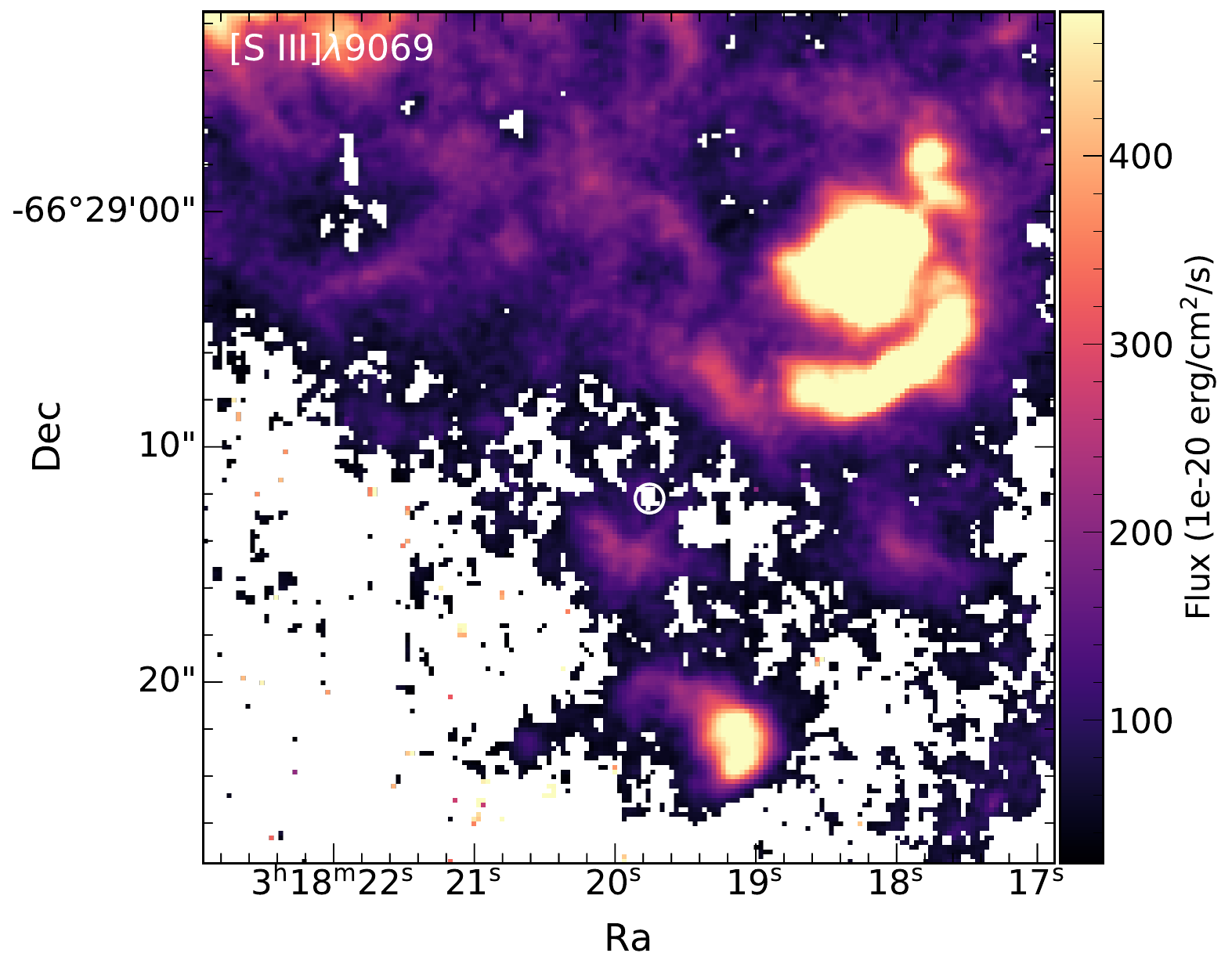}
    \caption{Flux maps for [Ar~{\sc iii}]$\lambda$7135 (\textit{left}) and [S~{\sc iii}]$\lambda$9069 (\textit{right}) derived as per \citet{gurpide_muse_2022} (see their Figure~5). Pixels below a S/N of 5 have been masked. The brighter blob to the far south of the ULX is due to a nearby stellar cluster.}
    \label{fig:fluxmaps}
\end{figure*}

Here we present the line flux maps for the [Ar~{\sc iii}]$\lambda$7135 and [S~{\sc iii}]$\lambda$9069 lines (Figure~\ref{fig:fluxmaps}). These were extracted using the Gaussian line pixel-by-pixel fitting presented in \citet{gurpide_muse_2022}, where we used the routine \texttt{CAMEL} \citep{epinat_massiv_2012}.

\section{\cloudy\ modelling results for variations in luminosity of the stellar background and metallicy.}

In this Appendix we present the \cloudy\ modelling of the extended nebulae for variations in the stellar background luminosity (Table~\ref{tab:cloudy_results_lower_stars}) and metallicity ($Z = 0.30Z_{\sun}$; Table~\ref{tab:cloudy_results_Z0004}). Both these calculations yielded significantly worse fits than our main result, which was obtained for a stellar background luminosity of $L = 2.75\times10^{40}$ erg/s and $Z = 0.15Z_{\sun}$ (Table~\ref{tab:cloudy_results}).

\begin{table*}
\centering    
\caption{As per Table~\ref{tab:cloudy_results} but with different contribution from the background stars. $Z = 0.15Z_{\sun}$.}\label{tab:cloudy_results_lower_stars}
\resizebox{\textwidth}{!}{\begin{tabular}{lccccccc} 
\hline 
\hline
Peak Line Ratio & Data & \texttt{phenomenological} & phenomenological\_star & \texttt{diskir} & \texttt{diskir\_star} & \texttt{sirf} & \texttt{sirf\_star}\\ 
\noalign{\smallskip}
\hline
\noalign{\smallskip}
\multicolumn{8}{c}{ULX + Stellar background, $L = 9.5\times10^{39}$\luxcgs} \\
\noalign{\smallskip}
\hline
\noalign{\smallskip}
\oiii/\hb                      & 5.9$\pm$0.3       & 4.94 & 5.68 & 6.49 & 6.07 & 6.50 & 6.26 \\ 
\oi/\ha                       & 0.33$\pm$0.02       & 0.52 & 0.51 & 0.52 & 0.53 & 0.56 & 0.60 \\ 
{[N{\sc ii}]}$\lambda$6583/\ha                          & 0.30$\pm$0.03       & 0.28 & 0.27 & 0.27 & 0.27 & 0.29 & 0.30 \\ 
{[S~{\sc i}]}$\lambda$6716/\ha                     & 0.62$\pm$0.01       & 0.84 & 0.82 & 0.83 & 0.83 & 0.86 & 0.92 \\ 
{[S~{\sc i}]}$\lambda$6716/{[S~{\sc ii}]}$\lambda$6731                      & 1.45$\pm$0.06       & 1.47 & 1.47 & 1.47 & 1.47 & 1.47 & 1.47 \\ 
{[Ar~{\sc iii}]}$\lambda$7135/\ha                      &  0.065$\pm$0.008       & 0.02 & 0.02 & 0.02 & 0.02 & 0.02 & 0.02 \\ 
{[S~{\sc iii}]}$\lambda$9068/\ha                       & 0.135$\pm$0.017       & 0.16 & 0.16 & 0.17 & 0.17 & 0.18 & 0.20 \\ 
\noalign{\smallskip} 
\hline 
\noalign{\smallskip}
$\log(n_\mathrm{H}$[cm$^{-3}$])        & --            &0.50 & 0.50 & 0.55 & 0.55 & 0.75 & 0.75 \\
$r_\mathrm{in}$ (pc)                   & --            &1.00 & 1.00 & 12.48 & 1.00 & 1.00 & 1.00 \\
$\chi^2$ / dof                         & --            &117 & 109 & 120 & 120 & 151 & 165 \\
\noalign{\smallskip}
\hline
\noalign{\smallskip}
\multicolumn{8}{c}{ULX + Stellar background, $L = 2.75 \times10^{39}$\luxcgs} \\
\noalign{\smallskip}
\hline
\noalign{\smallskip}
\oiii/\hb                       & 5.9$\pm$0.3       & 3.10 & 4.20 & 6.08 & 5.52 & 6.38 & 6.08 \\ 
\oi/\ha                       & 0.33$\pm$0.02       & 0.60 & 0.59 & 0.59 & 0.60 & 0.60 & 0.61 \\ 
{[N{\sc ii}]}$\lambda$6583/\ha                       & 0.30$\pm$0.03       & 0.33 & 0.32 & 0.30 & 0.31 & 0.30 & 0.31 \\ 
{[S~{\sc i}]}$\lambda$6716/\ha                       & 0.62$\pm$0.01       & 0.95 & 0.95 & 0.94 & 0.94 & 0.93 & 0.95 \\ 
{[S~{\sc i}]}$\lambda$6716/{[S~{\sc ii}]}$\lambda$6731                       & 1.45$\pm$0.06       & 1.47 & 1.47 & 1.47 & 1.47 & 1.47 & 1.47 \\ 
{[Ar~{\sc iii}]}$\lambda$7135/\ha                       & 0.065$\pm$0.008       & 0.02 & 0.02 & 0.02 & 0.02 & 0.02 & 0.02 \\ 
{[S~{\sc iii}]}$\lambda$9068/\ha                       & 0.135$\pm$0.017       & 0.19 & 0.19 & 0.19 & 0.19 & 0.20 & 0.20 \\ 
\noalign{\smallskip} 
\hline 
\noalign{\smallskip}
$\log(n_\mathrm{H}$[cm$^{-3}$])        & --            &0.55 & 0.50 & 0.55 & 0.55 & 0.75 & 0.75 \\
$r_\mathrm{in}$ (pc)                   & --            &1.00 & 12.48 & 1.00 & 12.48 & 1.00 & 12.48 \\
$\chi^2$ / dof                         & --            &153 & 143 & 151 & 153 & 167 & 171 \\
\noalign{\smallskip}
   \hline
   \hline
\end{tabular}
}
\end{table*}

\begin{table*}
\centering    
\caption{As per Table~\ref{tab:cloudy_results} but for Z = 0.30Z$_{\sun}$.}\label{tab:cloudy_results_Z0004}
\resizebox{\textwidth}{!}{\begin{tabular}{lccccccc} 
\hline 
\hline
Peak Line Ratio & Data & \texttt{phenomenological} & phenomenological\_star & \texttt{diskir} & \texttt{diskir\_star} & \texttt{sirf} & \texttt{sirf\_star}\\ 
\noalign{\smallskip}
\hline
\multicolumn{8}{c}{ULX only}\\
\noalign{\smallskip}
\hline
\noalign{\smallskip}
{[O~{\sc iii}]}$\lambda$5007/\hb\                       & 5.9$\pm$0.3       & 0.87 & 2.24 & 11.70 & 9.35 & 11.66 & 12.33 \\
{[O~I]}$\lambda$6300/\ha\                               & 0.33$\pm$0.02     & 0.92 & 0.92 & 0.91  & 0.91 & 0.90  & 0.90 \\
{[N~{\sc ii}]}$\lambda$6583/\ha\                        & 0.30$\pm$0.03     & 0.54 & 0.54 & 0.51  & 0.51 & 0.49  & 0.49 \\
{[S~{\sc ii}]}$\lambda$6716/\ha                         & 0.62$\pm$0.01     & 0.73 & 0.74 & 0.72  & 0.72 & 0.69  & 0.69 \\
{[S~{\sc ii}]}$\lambda$6716/{[S~{\sc ii}]}$\lambda$6731 & 1.45$\pm$0.06     & 1.47 & 1.47 & 1.47  & 1.47 & 1.47  & 1.47  \\
{[Ar~{\sc iii}]}$\lambda$7135/\ha                       & 0.065$\pm$0.008   & 0.04 & 0.04 & 0.04  & 0.04 & 0.04  & 0.04 \\
{[S~{\sc iii}]}$\lambda$9069/\ha                        & 0.135$\pm$0.017   & 0.26 & 0.26 & 0.26  & 0.26 & 0.27  & 0.27 \\
\noalign{\smallskip}
\hline
\noalign{\smallskip}
$\log(n_\mathrm{H}$[cm$^{-3}$])        & --            &0.35 & 0.35 & 0.40 & 0.40 & 0.65 & 0.60 \\
$r_\mathrm{in}$ (pc)                   & --            &40.00 & 40.00 & 1.00 & 40.00 & 40.00 & 1.00 \\
$\chi^2$ / dof                         & --            &304 & 303 & 286 & 288 & 300 & 297 \\
\noalign{\smallskip}
\hline
\noalign{\smallskip}
\multicolumn{8}{c}{ULX + Stellar background, $L = 2.75\times10^{40}$\luxcgs} \\
\noalign{\smallskip}
\hline
\noalign{\smallskip}
{[O~{\sc iii}]}$\lambda$5007/\hb                       & 5.9$\pm$0.3       & 13.16 & 13.49 & 13.36 & 13.10 & 13.41 & 13.13 \\ 
{[O~I]}$\lambda$6300/\ha                      	 & 0.33$\pm$0.02       & 0.57 & 0.54 & 0.60 & 0.61 & 0.74 & 0.83 \\ 
{[N~{\sc ii}]}$\lambda$6583/\ha                        & 0.30$\pm$0.03       & 0.41 & 0.41 & 0.43 & 0.44 & 0.51 & 0.56 \\ 
{[S~{\sc ii}]}$\lambda$6716/\ha                        & 0.62$\pm$0.01       & 0.70 & 0.67 & 0.71 & 0.72 & 0.82 & 0.93 \\ 
{[S~{\sc ii}]}$\lambda$6716/{[S~{\sc ii}]}$\lambda$6731& 1.45$\pm$0.06       & 1.47 & 1.47 & 1.47 & 1.47 & 1.47 & 1.47 \\ 
{[Ar~{\sc iii}]}$\lambda$7135/\ha                      & 0.065$\pm$0.008       & 0.04 & 0.04 & 0.04 & 0.04 & 0.04 & 0.04 \\ 
{[S~{\sc iii}]}$\lambda$9069/\ha                       & 0.135$\pm$0.017       & 0.22 & 0.21 & 0.23 & 0.23 & 0.29 & 0.33 \\ 
\noalign{\smallskip} 
\hline 
\noalign{\smallskip}
$\log(n_\mathrm{H}$[cm$^{-3}$])        & --            &0.30  & 0.25 & 0.35  & 0.35  & 0.60  & 0.60 \\
$r_\mathrm{in}$ (pc)                   & --            &12.48 & 1.00 & 12.48 & 12.48 & 12.48 & 1.00 \\
$\chi^2$ / dof                         & --            &123   & 118  & 141   & 142   & 220   & 259 \\
\noalign{\smallskip}
   \hline
   \hline
\end{tabular}
}
\end{table*}
\end{document}